\documentclass[12pt,preprint]{aastex}
\shorttitle{Observations of NGC 4038/39 with Herschel}
\shortauthors{Schirm $\&$ Wilson}

\usepackage{graphicx}
\usepackage{epstopdf}
\usepackage{amssymb}
\usepackage{hyperref}

\newcommand{\unit}[1]{\ensuremath{\, \mathrm{#1}}}
\newcommand{\mol}[1]{\ensuremath{\mathrm{#1}}}
\newcommand{\vm}{\ensuremath{\mathrm{v}}}

\begin{document}
\title{\emph{Herschel\footnote{Herschel is an ESA space observatory with science instruments provided by European-led Principal Investigator consortia and with important participation from NASA}} SPIRE-FTS Observations of Excited $\unit{CO}$ and $\mol{[CI]}$ in the Antennae (NGC 4038/39): Warm and Cold Molecular Gas}
\author{Maximilien R.P. Schirm\altaffilmark{1},  Christine D. Wilson\altaffilmark{1}, Tara J. Parkin \altaffilmark{1}, Julia Kamenetzky\altaffilmark{2}, Jason Glenn\altaffilmark{2}, Naseem Rangwala\altaffilmark{2}, Luigi Spinoglio\altaffilmark{3}, Miguel Pereira-Santaella\altaffilmark{3}, Maarten Baes\altaffilmark{4}, Michael J. Barlow\altaffilmark{5}, Dave L. Clements\altaffilmark{6}, Asantha Cooray\altaffilmark{7}, Ilse De Looze\altaffilmark{4}, Oskar \L. Karczewski\altaffilmark{8}, Suzanne C. Madden\altaffilmark{9}, Aur\'elie R\'emy-Ruyer\altaffilmark{9}, Ronin Wu\altaffilmark{9}}
\altaffiltext{1}{Department of Physics and Astronomy, McMaster University, Hamilton, ON L8S 4M1 Canada; schirmmr@mcmaster.ca, wilson@physics.mcmaster.ca}
\altaffiltext{2}{Center for Astrophysics and Space Astronomy, 389-UCB, University of Colorado, Boulder, CO, 80303}
\altaffiltext{3}{Istituto di Astrofisica e Planetologia Spaziali, INAF-IAPS, Via Fosso del Cavaliere 100, I-00133 Roma, Italy}
\altaffiltext{4}{Sterrenkundig Observatorium, Universiteit Gent, Krijgslaan 281 S9, B-9000 Gent, Belgium}
\altaffiltext{5}{Dept. of Physics \& Astronomy, University College London, Gower Street, London WC1E 6BT, UK}
\altaffiltext{6}{Astrophysics Group, Imperial College, Blackett Laboratory, Prince Consort Road, London SW7 2AZ, UK}
\altaffiltext{7}{Dept. of Physics \& Astronomy, University of California, Irvine, CA,92697, USA}
\altaffiltext{8}{Department of Physics and Astronomy, University of Sussex, Brighton, BN1 9QH, UK}
\altaffiltext{9}{CEA, Laboratoire AIM, Irfu/SAp, Orme des Merisiers, 91191 Gif-sur-Yvette, France}
\begin{abstract}
We present \emph{Herschel} SPIRE-FTS observations of the Antennae (NGC 4038/39), a well studied, nearby ($22 \unit{Mpc}$) ongoing merger between two gas rich spiral galaxies. The SPIRE-FTS is a low spatial ($\unit{FWHM} \sim 19''-43''$) and spectral ($\sim 1.2 \unit{GHz}$) resolution mapping spectrometer covering a large spectral range ($194 - 671 \unit{\mu m}$, $450 \unit{GHz} - 1545 \unit{GHz}$). We detect 5 $\mol{CO}$ transitions ($J=4-3$ to $J=8-7$), both $\mol{[CI]}$ transitions and the $\unit{[NII]205\mu m}$ transition across the entire system, which we supplement with ground based observations of the $\mol{CO}$ $J=1-0$, $J=2-1$	 and $J=3-2$ transitions, and Herschel PACS observations of $\mol{[CII]}$ and $\unit{[OI]63\mu m}$. Using the $\mol{CO}$ and $\mol{[CI]}$ transitions, we perform both a LTE analysis of $\mol{[CI]}$, and a non-LTE radiative transfer analysis of $\mol{CO}$ and $\mol{[CI]}$ using the radiative transfer code RADEX along with a Bayesian likelihood analysis. We find that there are two components to the molecular gas: a cold ($T_{kin}\sim 10-30 \unit{K}$) and a warm ($T_{kin} \gtrsim 100 \unit{K}$) component.  By comparing the warm gas mass to previously observed values, we determine a $\mol{CO}$ abundance in the warm gas of $x_{\mol{CO}} \sim 5\times 10^{-5}$. If the $\mol{CO}$ abundance is the same in the warm and cold gas phases, this abundance corresponds to a $\mol{CO}$ $J=1-0$ luminosity-to-mass conversion factor of $\alpha_{\mol{CO}} \sim 7 \ M_{\odot}\unit{pc^{-2} \ (K \ km \ s^{-1})^{-1}}$ in the cold component, similar to the value for normal spiral galaxies. We estimate the cooling from $\mol{H_2}$, $\mol{[CII]}$, $\mol{CO}$ and $\unit{[OI]63\mu m}$ to be $\sim 0.01 L_{\odot}/M_{\odot}$. { We compare PDR models to} the ratio of the flux of various $\mol{CO}$ transitions, along with the ratio of the $\mol{CO}$ flux to the far-infrared flux in NGC 4038, NGC 4039 and the overlap region. We find that the densities recovered from our non-LTE analysis are consistent with a background far-ultraviolet field of strength $G_0\sim 1000$.  Finally, we find that a combination of turbulent heating, due to the ongoing merger, and supernova and stellar winds are sufficient to heat the molecular gas.  
\end{abstract}

\keywords{galaxy: individual(NGC 4038, NGC 4039), infrared: galaxy}
\section{Introduction}

Luminous and ultra luminous infrared galaxies are a well studied class of infrared (IR) bright galaxies whose excess IR emission comes from dust heated by enhanced star formation activity \citep{cluver2010}.  Their enhanced star formation rate directly correlates with a star formation efficiency (SFE) $\sim 4-10$ times greater than in normal galaxies \citep{genzel2010} typically seen in the form of starbursts throughout the galaxy \citep{sanders1996}. Many luminous infrared galaxies (LIRGs, $L_{IR}>10^{11} L_\odot$) and almost all ultra luminous infrared galaxies (ULIRGs, $L_{IR}>10^{12} L_\odot$) are found to be in an advanced merging state between two or more galaxies \citep{clements1996,sanders1996}; these starbursts are likely the result of the ongoing merger \citep{hibbard1997}, triggered by the redistribution of material throughout the merging galaxies \citep{toomre1972}.  It has been suggested that all mergers undergo a period of ``super starbursts" \citep{joseph1985}, where $\gtrsim 10^9 M_\odot$ of stars form in $\sim 10^8 \unit{years}$ \citep{gehrz1983}. 

The Antennae (NGC 4038/39, Arp 244) is a young, nearby, ongoing merger between two gas rich spiral galaxies, NGC 4038 and NGC 4039. At a distance of only $22 \pm 3 \unit{Mpc}$ \citep{schweizer2008}, the Antennae may represent the nearest such merger.  Its infrared brightness of $L_{IR} = 7.5 \times 10^{10} L_\odot$ \citep{gao2001} approaches that of a LIRG and is likely the result of merger triggered star formation. The majority of the star formation is occurring within the two nuclei (NGC 4038, NGC 4039), along with a third region where the two disks are believed to overlap \citep{stanford1990}, also sometimes referred to as the interaction region (IAR, e.g. \citealt{schulz2007}). A fourth star forming region of interest is located to the west of NGC 4038 and is known as  the ``western loop". The X-ray luminosity in the Antennae is enhanced relative to normal spiral galaxies, with $50\%$ of the X-ray emission originating from hot diffuse gas outside of these four regions \citep{read1995}; however, the presence of an active galactic nucleus (AGN) has been ruled out \citep{brandl2009}.

As a result of the star formation process, most of the star clusters in the Antennae are in the two nuclei and the overlap region, with the overlap region housing the youngest ($\lesssim 5 \unit{Myr}$) of these star clusters, and the second youngest found in the western loop ($5-10\unit{Myr}$) \citep{whitmore1995,whitmore1999}. Some of the star clusters found near the overlap region are super star clusters with masses of a few $10^6 \unit{M_\odot}$ \citep{whitmore2010}. Comparisons of \emph{Spitzer}, GALEX, HST and 2MASS images by \cite{zhang2010} showed that the star formation rate itself peaks in the overlap region and western loop, agreeing with previous results by \cite{gao2001}.  The overlap region exhibits characteristics of younger, more recent star formation \citep{zhang2010}. Far-infrared spectroscopy of the Antennae with the ISO-LWS spectrometer, together with ground-based Fabry-Perot imaging spectroscopy, has been used to constrain the age of the instantaneous starburst to $(7-8) \times 10^6 \unit{yr}$, producing a total stellar mass of $(2.5 \pm 1.5) \times 10^8 \unit{M_\odot}$ with a luminosity of $(7 \pm 4) \times 10^{10} \unit{L_\odot}$ \citep{1996A&A...315L..97F}.

The star forming molecular gas in the Antennae is well studied, with observations of the molecular gas tracer $\mol{CO}$ in the $J=1-0$ \citep{wilson2000,gao2001,wilson2003,zhu2003,schulz2007}, $J=2-1$ \citep{zhu2003,schulz2007,2012ApJ...750..136W}, $J=3-2$ \citep{zhu2003,schulz2007,2012ApJ...745...65U}, $J=6-5$ \citep{bayet2006} and $J=7-6$ \citep{bayet2006} transitions.  Interferometric observations of the ground state transition by \cite{wilson2000} and \cite{wilson2003} found $\sim$ 100 super-giant molecular complexes (SGMCs) scattered throughout the Antennae.  The 7 most massive of these SGMCs consist of the two nuclei (NGC 4038 and NGC 4039), and five others located in the overlap region. Assuming a $\mol{CO}$-to-$\mol{H_2}$ conversion factor of $3\times10^{20} \unit{H_2} \unit{cm^{-2}} \unit{(K \ km \ s^{-1})^{-1}}$, the molecular gas { masses} of the SGMCs are on the order of $10^8 \unit{M_\odot}$. In addition, recent interferometric observations of the $\mol{CO}$ $J=3-2$ transitions by \cite{2012ApJ...745...65U} show that half of the $J=3-2$ emission originates from the overlap region. Only $30\%$ of the giant molecular clouds (GMCs) resolved in the $J=3-2$ map coincide with star clusters which have been detected in the optical and near infrared, suggesting that the $J=3-2$ emission may in fact be tracing future star forming regions. 

Single-dish observations of the $\mol{CO}$ $J=1-0$ line by \cite{zhu2003} suggest that, assuming the same conversion factor as \cite{wilson2000}, the total molecular gas mass of the system is $1.2\times 10^{10} \unit{M_\odot}$, with about $\sim 40\%$ of the total gas mass in the overlap region ($4.5\times 10^{9} \unit{M_\odot}$). Large Velocity Gradient (LVG) models by \cite{zhu2003} suggest that there are at least two phases to the molecular gas: a cold phase ($T_{kin} \sim 40 \unit{K}$) and a warm phase ($T_{kin} \sim 100 \unit{K}$). \cite{bayet2006} performed a single component LVG analysis of NGC 4038 and the overlap region, finding that in the overlap region the molecular gas of their single component is warm ($T_{kin} \sim 140 \unit{K}$).  

Launched in 2009, the Herschel Space Observatory (\emph{Herschel}; \citealt{pilbratt2010}) explores the largely unobserved wavelength range of $55-671 \unit{\mu m}$. The Spectral and Photometric Imaging Receiver (SPIRE; \citealt{griffin2010}) spectrometer is the Fourier Transform Spectrometer (FTS; \citealt{naylor2010}), an imaging spectrometer covering a total spectral range from $194 \unit{\mu m}$ to $671 \unit{\mu m}$ ($\sim 450 \unit{GHz}$ to $\sim 1545 \unit{GHz}$). The SPIRE-FTS allows us to simultaneously observe all molecular and atomic transitions which lie within its spectral range. A total of 10 $\mol{CO}$ transitions, from $J=4-3$ to $J=13-12$, and both $\mol{[CI]}$ transitions { at $492\unit{GHz}$ ($609 \unit{\mu m}$) and $809 \unit{GHz}$ ($370 \unit{\mu m}$),} lie within this spectral range, making the SPIRE-FTS ideal for studying both cold and warm molecular gas in extra-galactic sources (e.g. see \citealt{panuzzo2010,vanderwerf2010,rangwala2011,kamenetzky2012,
spinoglio2012,meijerink2013,pereirasantaella2013}). 

We have obtained observations of the Antennae using the SPIRE-FTS as part of the guaranteed time key project ``Physical Processes in the Interstellar Medium of Very Nearby Galaxies" (PI: Christine Wilson), which we supplement with ground based observations of the $\mol{CO}$ $J=1-0$, $J=2-1$ and $J=3-2$ transitions.  In Section \ref{obs} we present the observations and method used to reduce these data. In Section \ref{radtrans} we present a radiative transfer analysis used to constrain the physical properties of the gas.  We discuss the implications in Section \ref{discuss}, where we investigate the possible heating mechanisms of the molecular gas, including modeling of Photon Dominated Regions (PDRs). 

\section{Observations} \label{obs}

We observed both the nucleus of NGC 4038 (hereafter NGC 4038) and the overlap region using the SPIRE FTS in high spectral resolution (FWHM $=0.048 \unit{cm^{-1}}$),  full-sampling mode on December 12th, 2010 (OD 572).  The observation of NGC 4038 is centered at ($12\unit{h}01\unit{m}53.00\unit{s}, -18^{\circ}52'01.0''$) while the observation of the overlap region is centered at ($12\unit{h}01\unit{m}54.90\unit{s},-18^{\circ}52'45.0''$). The observation IDs for the observations of NGC 4038 and the overlap region are 1342210860 and 1342210859, respectively, and the total integration time for each observation is 17,843 seconds, for a total integration time of 35,686 seconds ($\sim 10 \unit{hours}$).  In addition to the SPIRE observations, we also present here ground-based $\mol{CO}$ $J=1-0$ from the Nobeyama Radio Observatory \citep{zhu2003}, along with $\mol{CO}$ $J=2-1$ and $J=3-2$ maps from the James Clerk Maxwell Telescope (JCMT). We also include observations of $\mol{[CII]}$ and $\unit{[OI]}63$ from the \emph{Herschel} Photodetecting Array Camera and Spectrometer (PACS) instrument. 

\subsection{FTS Data reduction}

\begin{figure}[ht] %%%Beam Image
\centering
\includegraphics[width = \linewidth]{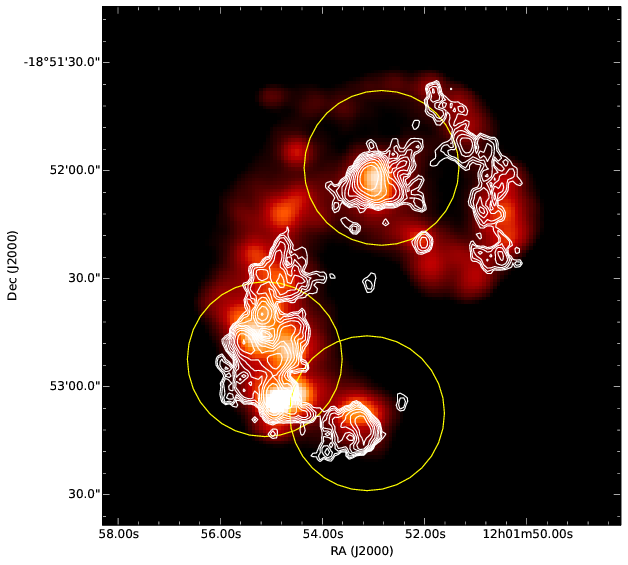} 
\caption[]{{ $\mol{CO}$ $J=1-0$ contours from \cite{wilson2003} overlaid on the} PACS $70 \unit{\mu m}$ observations of the Antennae from \cite{Klaas2010}. { The white contours correspond to $1\%$, $1.6\%$, $2.5\%$, $4\%$, $6\%$, $9.5\%$, $15\%$, $23\%$, $37\%$, and $57 \%$ the peak intensity.} The yellow circles indicate the approximate region observed for NGC 4038 (\emph{north-west}), the overlap region (\emph{south-east}), and NGC 4039 (\emph{south-west}) for the FTS beam at $\sim 460 \unit{GHz}$ ($\sim 43 ''$).}
\label{PACS70circles}
\end{figure}

We reduce the FTS data using a modified version of the standard Spectrometer Mapping user pipeline and the Herschel Interactive Processing Environment version 9.0, and SPIRE calibration context version 8.1.  \cite{fulton2010} and \cite{swinyard2010} described an older version of the data reduction pipeline and process.  The standard mapping pipeline assumes that the source is extended enough to fill the beam uniformly. Interferometric observations of NGC 4038/39 show that the molecular gas is partially extended \citep{wilson2000} and does not fill the beam (Figure \ref{PACS70circles}). To account for this, we apply a point-source correction to all of the detectors in both of our bolometric arrays in order to calibrate the flux accurately across the entire mapped region. { Furthermore, by applying this point source correction, we obtain a cube with the same calibration scale as our ground based observations.} This point-source correction is { calculated} from models and observations of Uranus, { and is the product of the beam area and of a point source coupling efficiency (see Chapter 5 of the SPIRE Observers Manual version 2.4\footnote{Available at \url{http://herschel.esac.esa.int/Docs/SPIRE/html/spire_om.html}}). The correction itself varies with frequency and is unique for each individual detector.}

\begin{figure}[ht] %%%Beam Image
\centering
\includegraphics[width = \linewidth]{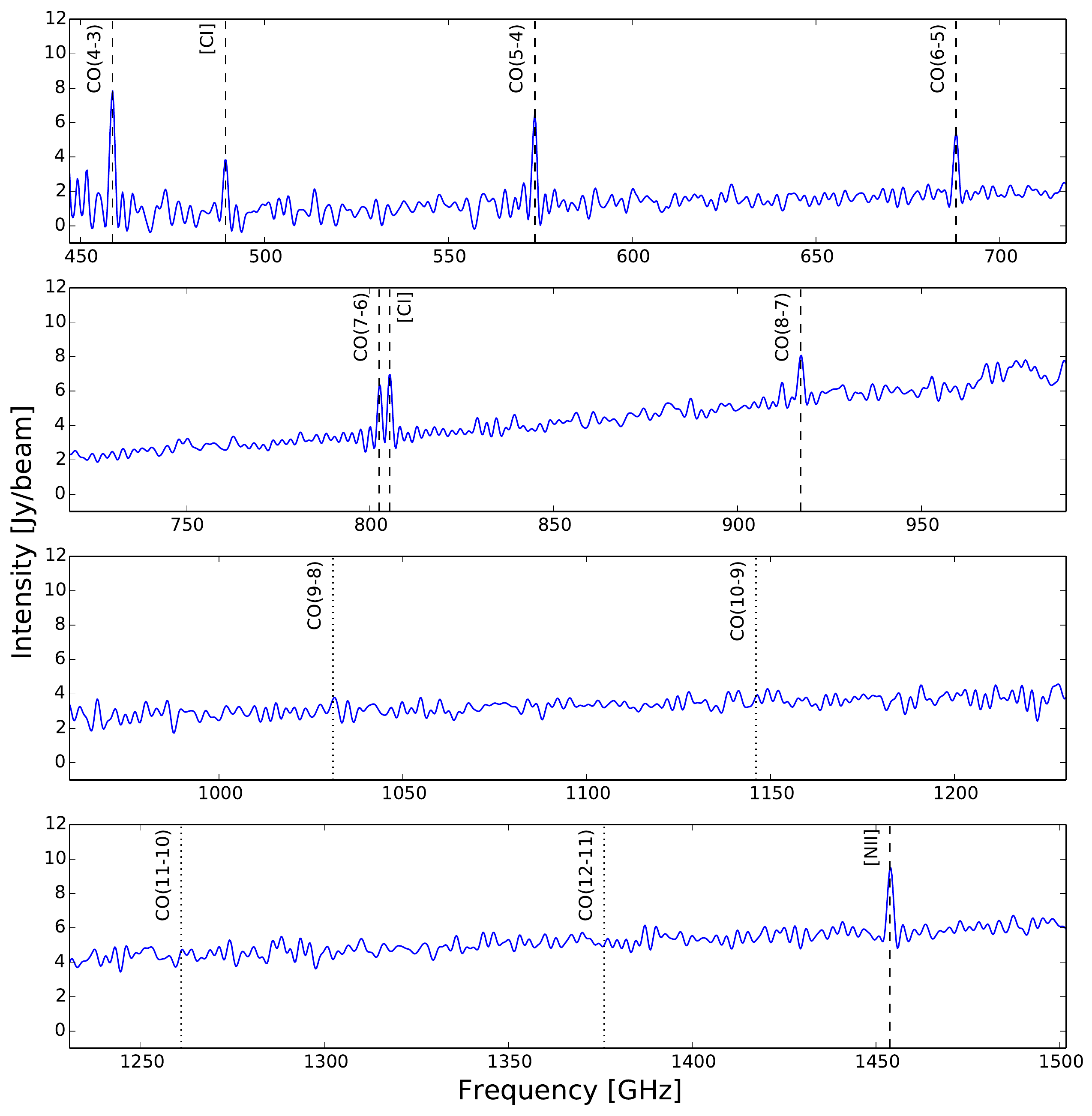} \caption[]{FTS spectrum for the overlap region covering the entire spectral range of the FTS. This position corresponds to the blue box in Figure \ref{IMconv}. { The top two panels correspond to the SLW and the bottom two panels correspond to the SSW, while the jump in the continuum between the second and third panels from the top is due to the large difference in beam size between the SLW ($\sim 37''$) and the SSW ($\sim 21 ''$) at the junction.}  The dashed lines indicate the detected and identified molecular and atomic transitions while the dotted lines indicate where we would expect to find the undetected higher $J$ $\mol{CO}$ transitions.}
\label{OSpec}
\end{figure}

After applying the point-source correction, we combine the observations using the \emph{spireProjection} task into two data cubes : one for the spectrometer long wave (SLW) bolometric array, the other for the spectrometer short wave (SSW) bolometric arrays. We use a pixel size of $15''$ for both data cubes, which we determined empirically as a balance between a sufficient number of detector hits per pixel and small pixel sizes. In addition, we create a data cube for the SSW with $3''$ pixels for the purposes of correcting the $\mol{[CII]}$ ionized gas fraction (see Section \ref{cooling}). The FTS spectrum for the overlap region is shown in Figure \ref{OSpec}.

\subsubsection{Line Fitting}

\begin{figure}[ht] %%%Unconvolved maps
\centering
$\begin{array}{@{\hspace{-0.2in}}c@{\hspace{0in}}c@{\hspace{0in}}c}
\includegraphics[width=0.33\linewidth]{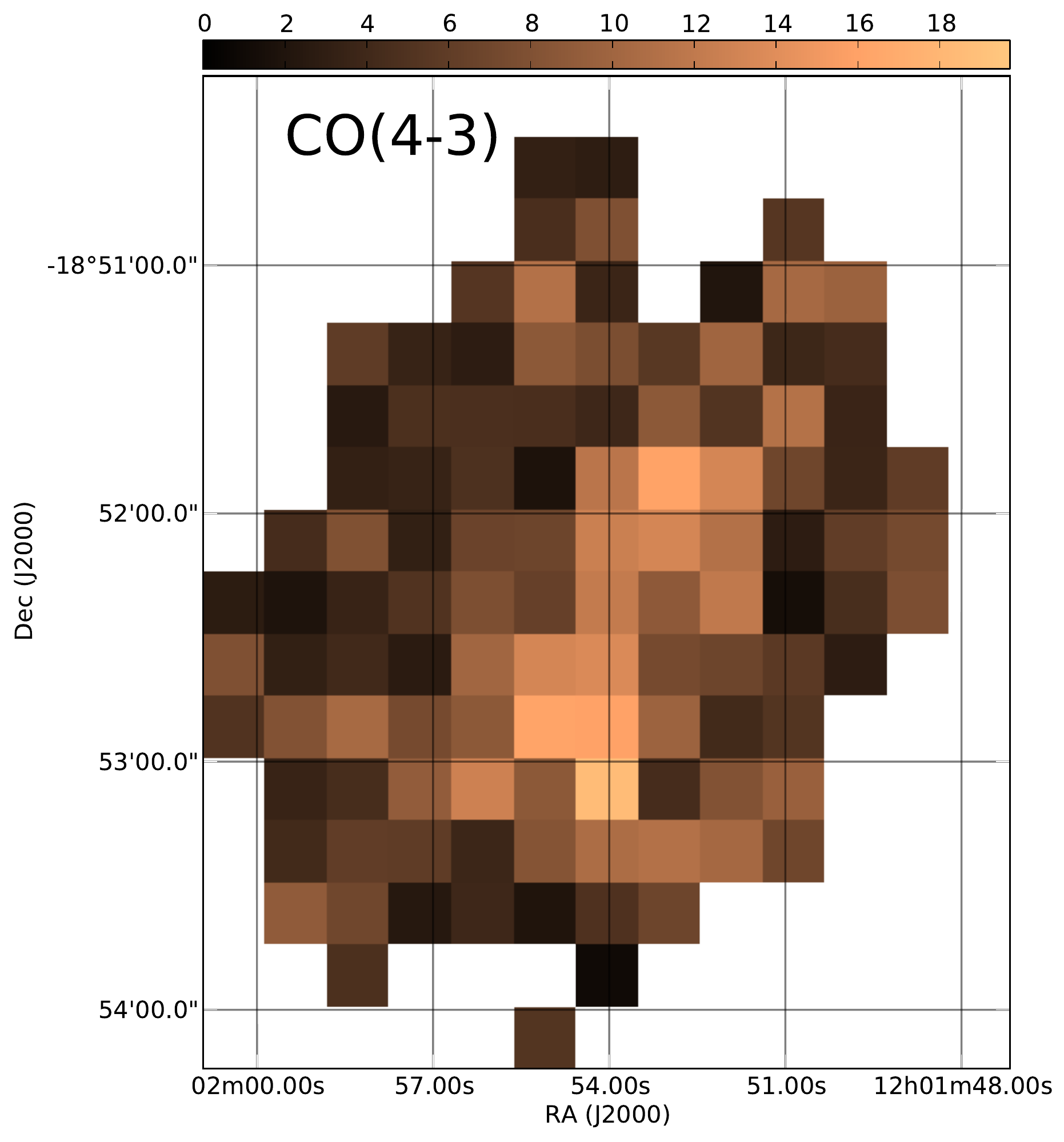} &
\includegraphics[width=0.33\linewidth]{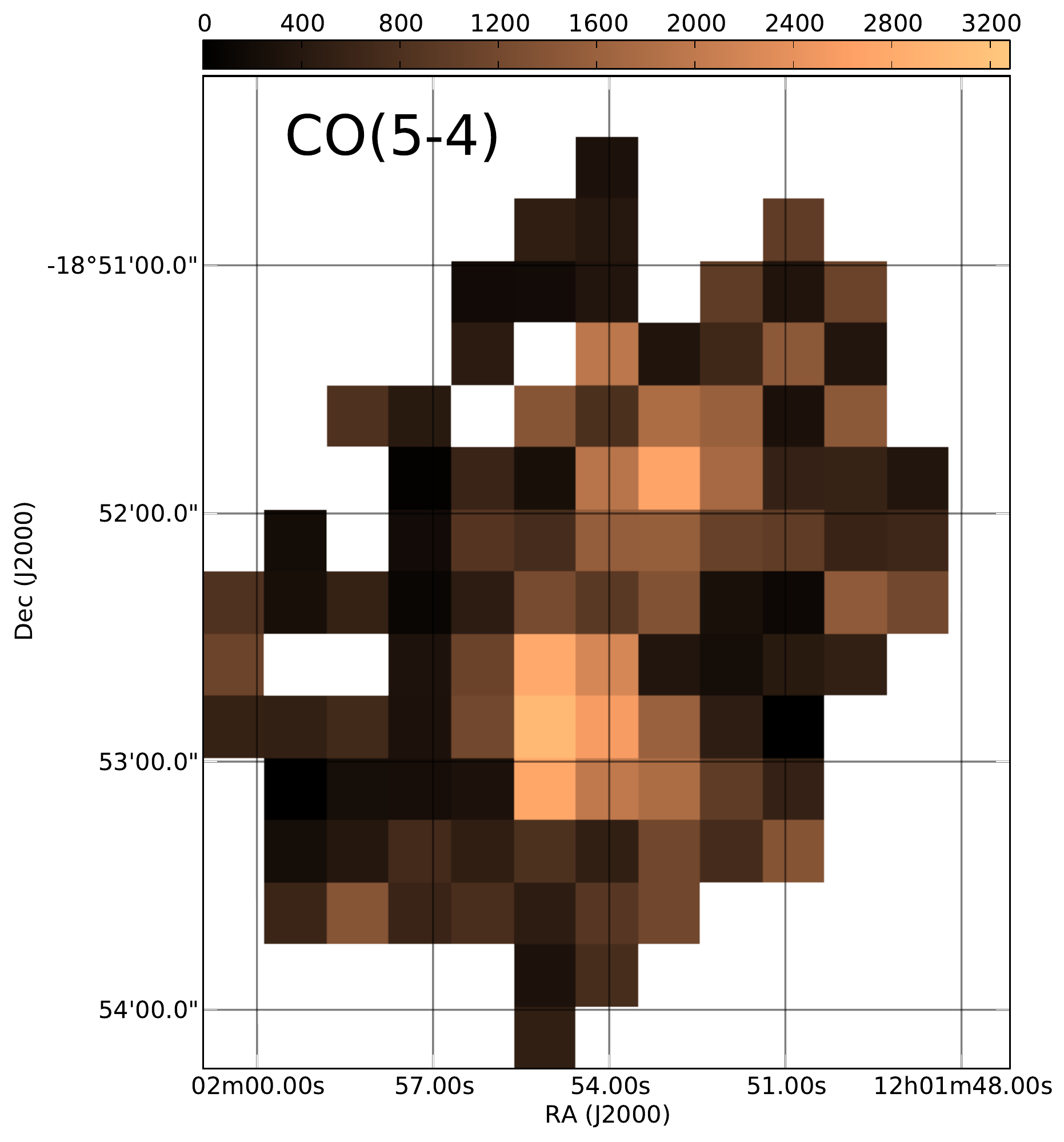} &\includegraphics[width=0.33\linewidth]{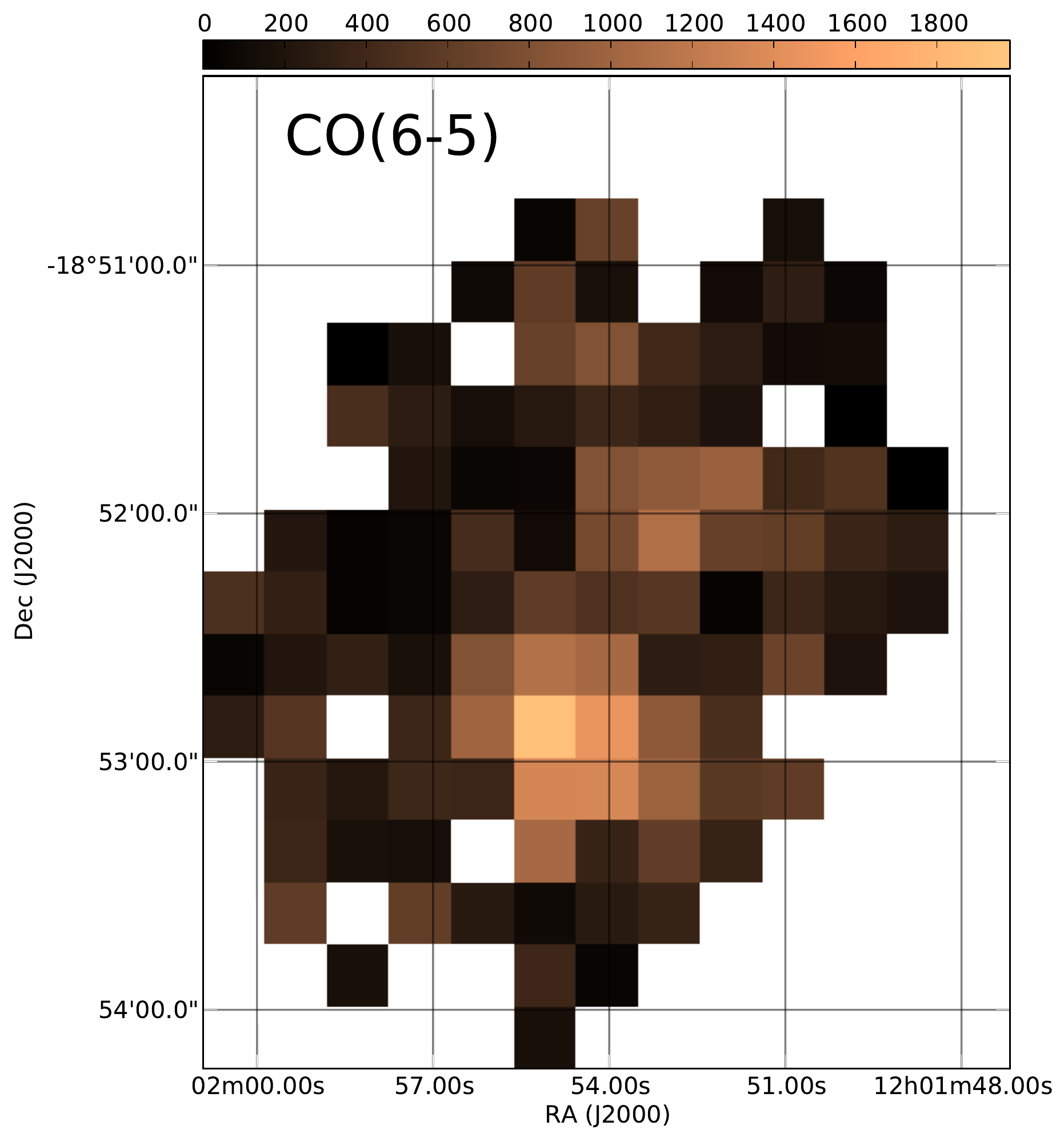} \\
\includegraphics[width=0.33\linewidth]{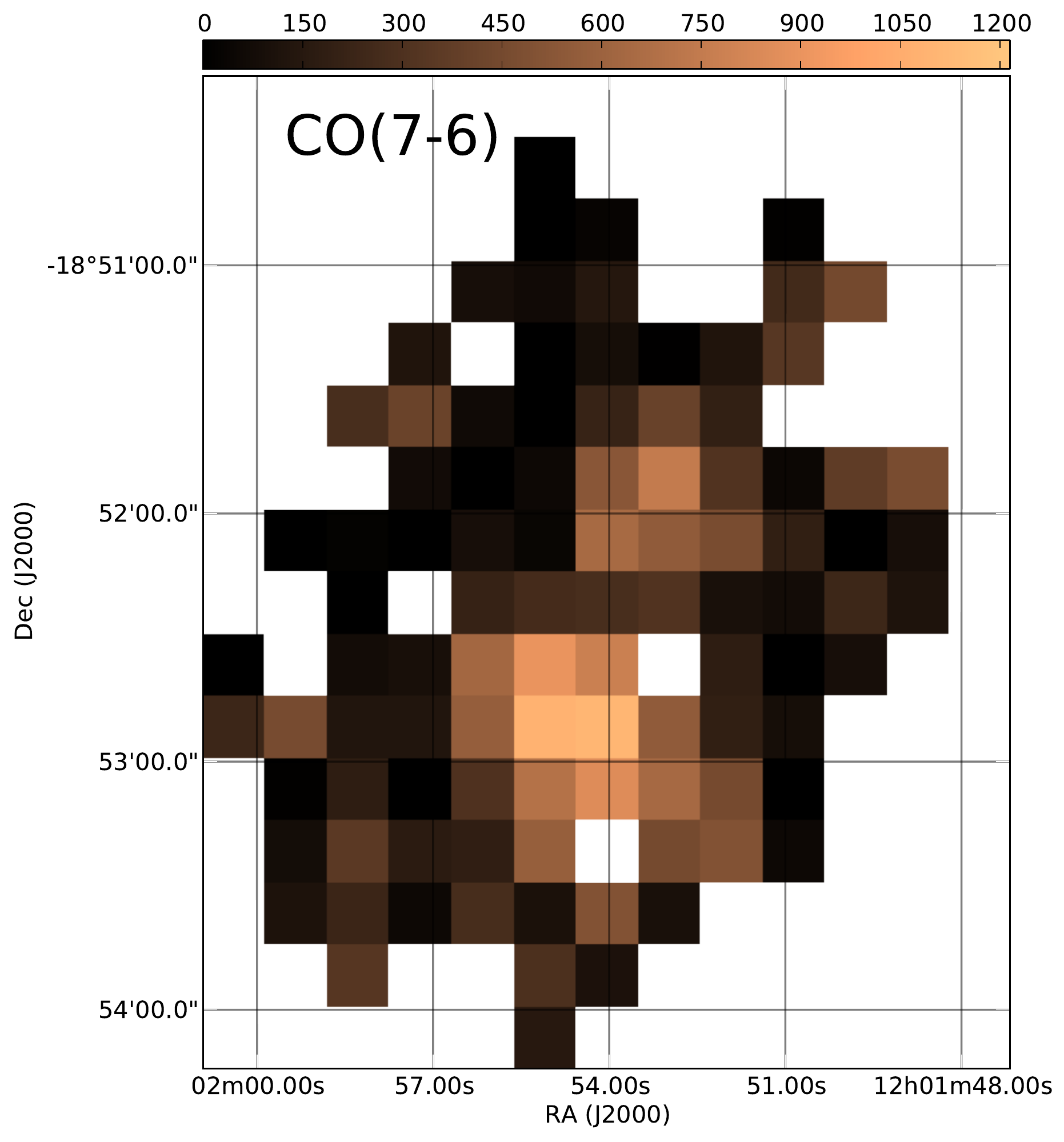} &\includegraphics[width=0.33\linewidth]{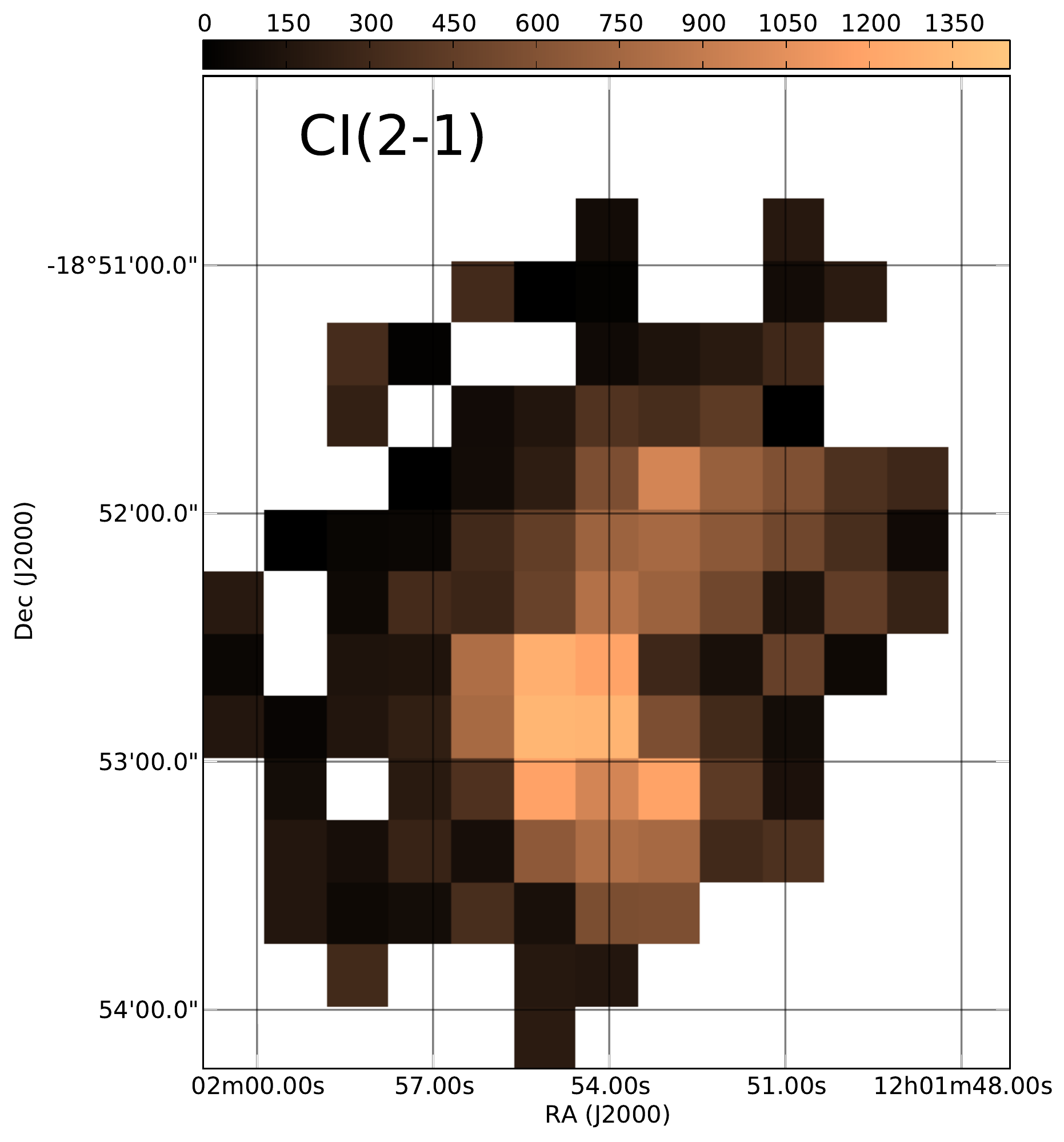}
&\includegraphics[width=0.33\linewidth]{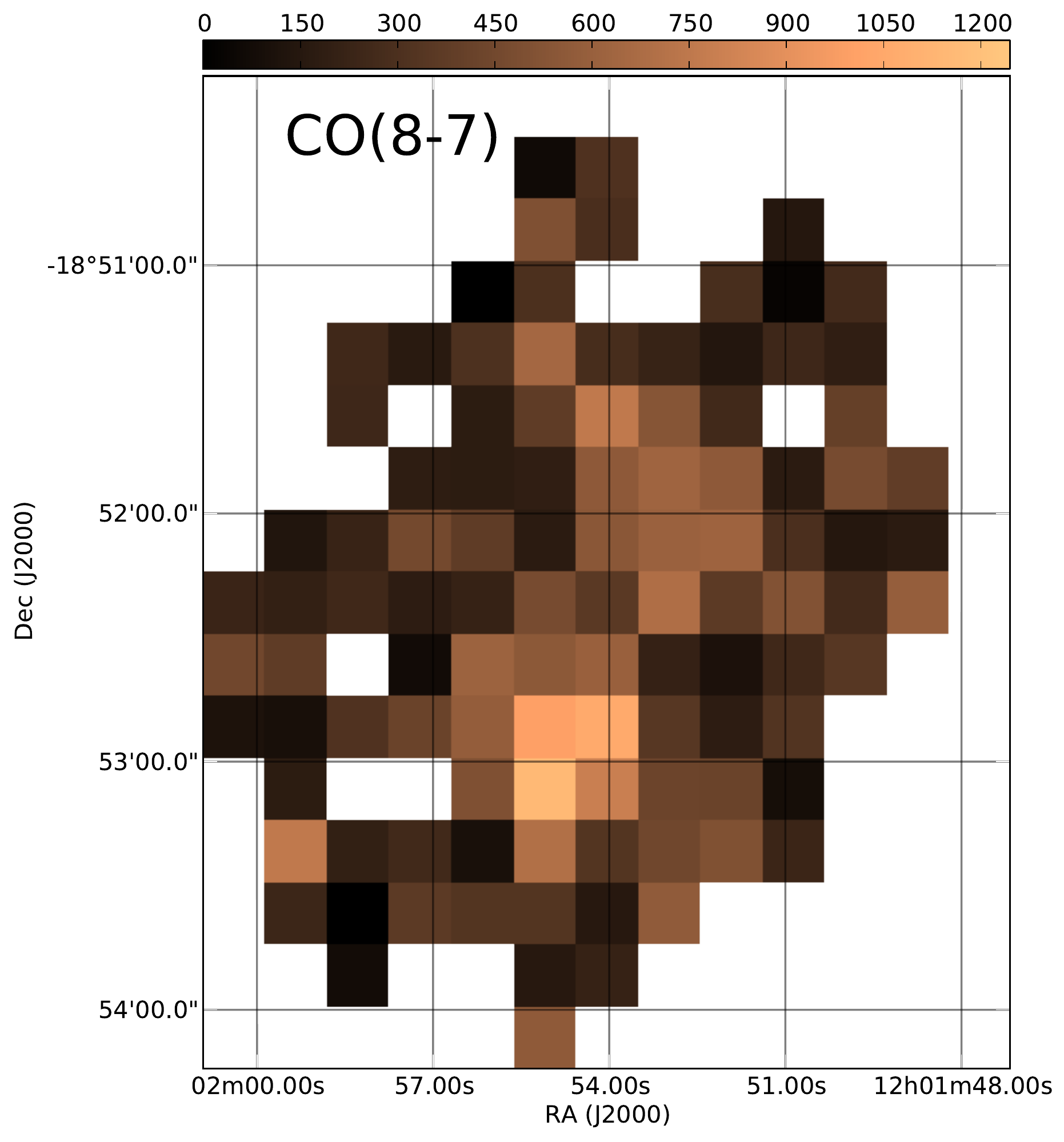}
\end{array}$
\caption[]{Integrated intensity maps for the SPIRE FTS $\mol{CO}$ and $\mol{[CI]}$ emission lines in units of $\unit{Jy \ km \ s^{-1}}$. These maps are at the instrument resolution. The $\mol{[CI]}$ $1-0$ line map is shown in Figure \ref{IMconv}.}
\label{IMunconv}
\end{figure}

\begin{figure}[ht] %%%Convolved maps
\centering
$\begin{array}{@{\hspace{-0.2in}}c@{\hspace{0in}}c@{\hspace{0in}}c}
\includegraphics[height=0.35\linewidth]{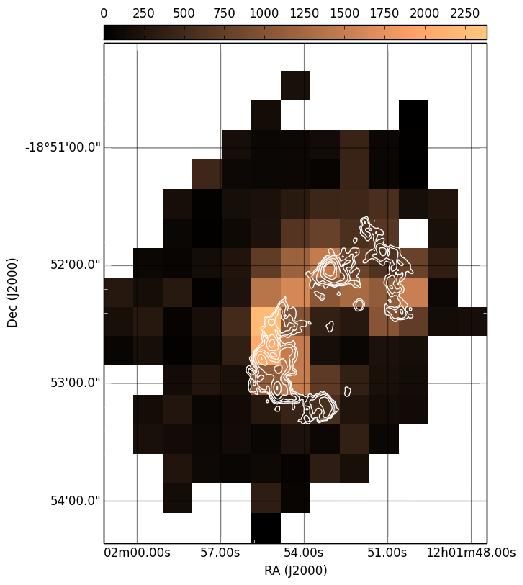} &\includegraphics[height=0.35\linewidth]{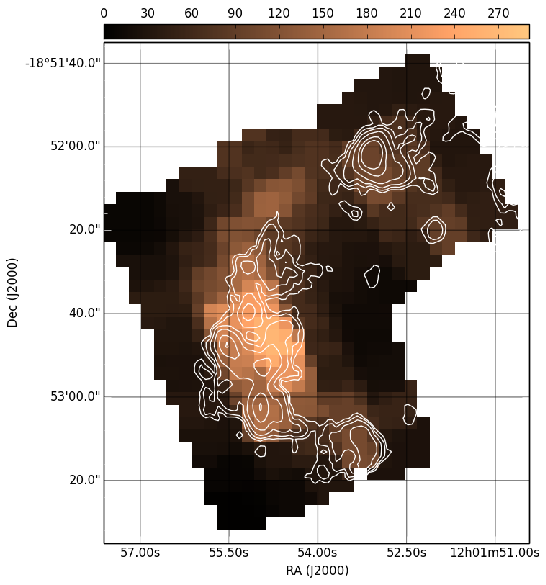} & \includegraphics[height=0.35\linewidth]{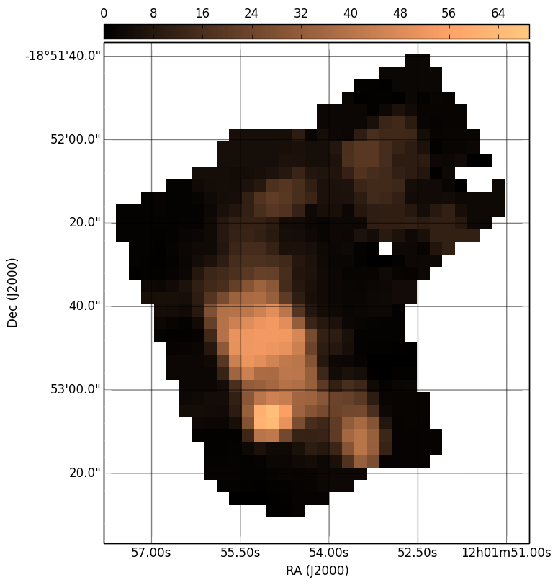} 
\end{array}$ 
\caption[]{ Integrated intensity maps for the $\mol{[NII]}$ $205 \unit{\mu m}$ (left), $\mol{[CII]}$ $158 \unit{\mu m}$ (centre), and $\mol{[OI]}$ $63 \unit{\mu m}$ (right) atomic fine structure lines at the instrument resolution. The $\mol{[NII]}$ map is in units of $\mol{Jy \ beam^{-1} \ km \ s^{-1}}$, while the $\mol{[CII]}$ and $\mol{[OI]}$ maps are in units of $\mol{Jy \ sr^{-1} \ km \ s^{-1}}$.   The $\mol{CO}$ $J=1-0$ contours from \cite{wilson2003} are overlaid in white on the $\mol{[NII]}$ and $\mol{[CII]}$ images, with the contours corresponding to $1 \%$, $2.5 \%$, $6\%$ $15 \%$, $37\%$ and $57\%$ the peak intensity. }
\label{PACSFig}
\end{figure}

We detect 5 $\mol{CO}$ transitions in emission, from $J=4-3$ to $J=8-7$. { We also detect the  $\mol{[CI]}$ $\unit{^3P_1 - ^3P_0}$ and $\unit{^3P_2 - ^3P_1}$ transitions along with the $\unit{[NII]}$ ($\unit{^3P_1 - ^3P_0}$) transition.}   We wrote a custom line fitting routine to measure the integrated intensities for all detected lines in the SLW and the SSW spectra across the entire data cube. In each pixel in the cube, the routine first removes the baseline by masking out all the lines in the spectrum and fitting a high-order polynomial to the remaining spectrum. Next, the routine fits a Sinc function to each line in the spectrum. To calculate the integrated intensity of each line we integrate over the entire Sinc function. In the case of the $\mol{CO} \ \unit{J=7-6}$ and $\mol{[CI]}$ $J=2-1$  lines, the routine fits both lines simultaneously, each with its own Sinc function.  The resulting integrated intensity maps are shown in { Figures \ref{IMunconv} and \ref{PACSFig}}.

\subsection{Ancillary Data}

\subsubsection{$\mol{CO} \ J=1-0$}

We obtained $\mol{CO}$ $J=1-0$ observations of the Antennae from \cite{zhu2003}. They obtained these observations from the Nobeyama Radio Observatory, with a telescope beam size of $15''$. The final map consists of three 64-point maps each covering $68'' \times 68''$, encompassing the $\mol{CO}$ emitting regions in NGC 4038, the overlap region and NGC 4039 in their entirety.  We collapse the data cube using the \verb+Starlink+ software package \citep{2008ASPC..394..650C}.  The total usable bandwidth was $350 \unit{MHz}$, corresponding to $\sim 640 \unit{km \ s^{-1}}$. The FWHM of the $\mol{CO} \ J=1-0$ emission line across the Antennae is $\sim 100 - 200 \unit{km \ s^{-1}}$; therefore, not enough bandwidth was available to estimate the uncertainty in each pixel reliably. We therefore estimate a total uncertainty, calibration included, of $20\%$ in the final collapsed image.

\subsubsection{$\mol{CO} \ J=2-1$ and $\mol{CO} \ J=3-2$}

The $\mol{CO}$ $J=2-1$ transition was observed in the Antennae using the JCMT on 2013 March 25 and 2013 April 3 as part of project M13AC09 (PI: Maximilien Schirm). These data were obtained using Receiver A3 in raster mapping mode. The resulting map is Nyquist sampled and covers the entire $\mol{CO}$ emitting region with a total area of $140'' \times 140''$ and a total integration time of $14,940$ seconds. The main-beam efficiency was $\eta_{MB}=0.69$ and the beam size was $20.8''$ at $230.56 \unit{GHz}$. The JCMT $\mol{CO}$ $J=3-2$ observations were obtained as part of project M09BC05 (PI: Tara Parkin) on 2009 December 10, 16 and 17. The beam size of the telescope was $14.5''$. We obtained a raster map over an area of $159'' \times 186''$ in position-switched mode with a total integration time of $2,111$ seconds, and we used a bandwidth of $1\unit{GHz}$ across $2048$ channels. We reduce both datasets using the methods described in \cite{2010ApJ...714..571W} and \cite{2012MNRAS.422.2291P} using the \verb+Starlink+ software package, with the exception that we convolve the maps using custom convolution kernels described in section \ref{convolution} rather than a Gaussian kernel.

\subsubsection{$\mol{[CII]}$ $158 \unit{\mu m}$ and $\unit{[OI]}$ $63 \unit{\mu m}$}

The $\mol{[CII]}$ $158 \unit{\mu m}$ and $\unit{[OI]}$ $63 \unit{\mu m}$ transitions were observed in the Antennae using the \emph{Herschel} PACS instrument each in three separate pointings: one centered on each of the nuclei of NGC 4038 (Observation IDs 1342199405 and 1342199406), NGC 4039 (Observation IDs 1342210820 and 1342210821), and the overlap region (Observation IDs 1342210822 and 1342210823). { All of these observations were performed in a single pointing in chopping-nodding mode. The field-of-view for each observation is $47'' \times 47''$, and all observations were binned to a $3''$ pixel size. The beam sizes and spectral resolution were $\sim 11''$ and $239 \unit{km/s}$ for $\mol{[CII]}$ $158 \unit{\mu m}$, and $\sim 9''$ and $98 \unit{km/s}$ for $\mol{[OI]}$ $63 \unit{\mu m}$ (PACS observers manual version 2.5.1\footnote{Available at \url{http://herschel.esac.esa.int/Docs/PACS/html/pacs_om.html}}).}

The level 2 data cubes were obtained from the \emph{Herschel} Science Archive on October 10th, 2012 ($\mol{[CII]}158 \unit{\mu m}$) and December 12th, 2012 ($\mol{[OI]}63\unit{\mu m}$). For each data cube, we fit and subtract the baseline in each pixel with a first-order polynomial before fitting the line with a Gaussian. We create an integrated intensity map for each transition by integrating across the fitted Gaussian in each pixel. For each transition, we combine the three maps using \emph{wcsmosaic} in the \verb+Starlink+ software package. { The resulting maps are shown in Figure \ref{PACSFig}.}

\subsection{Convolution}\label{convolution}

\begin{figure}[ht] %%%Convolved maps
\centering
$\begin{array}{@{\hspace{-0.2in}}c@{\hspace{0in}}c@{\hspace{0in}}c}
\includegraphics[width=0.33\linewidth]{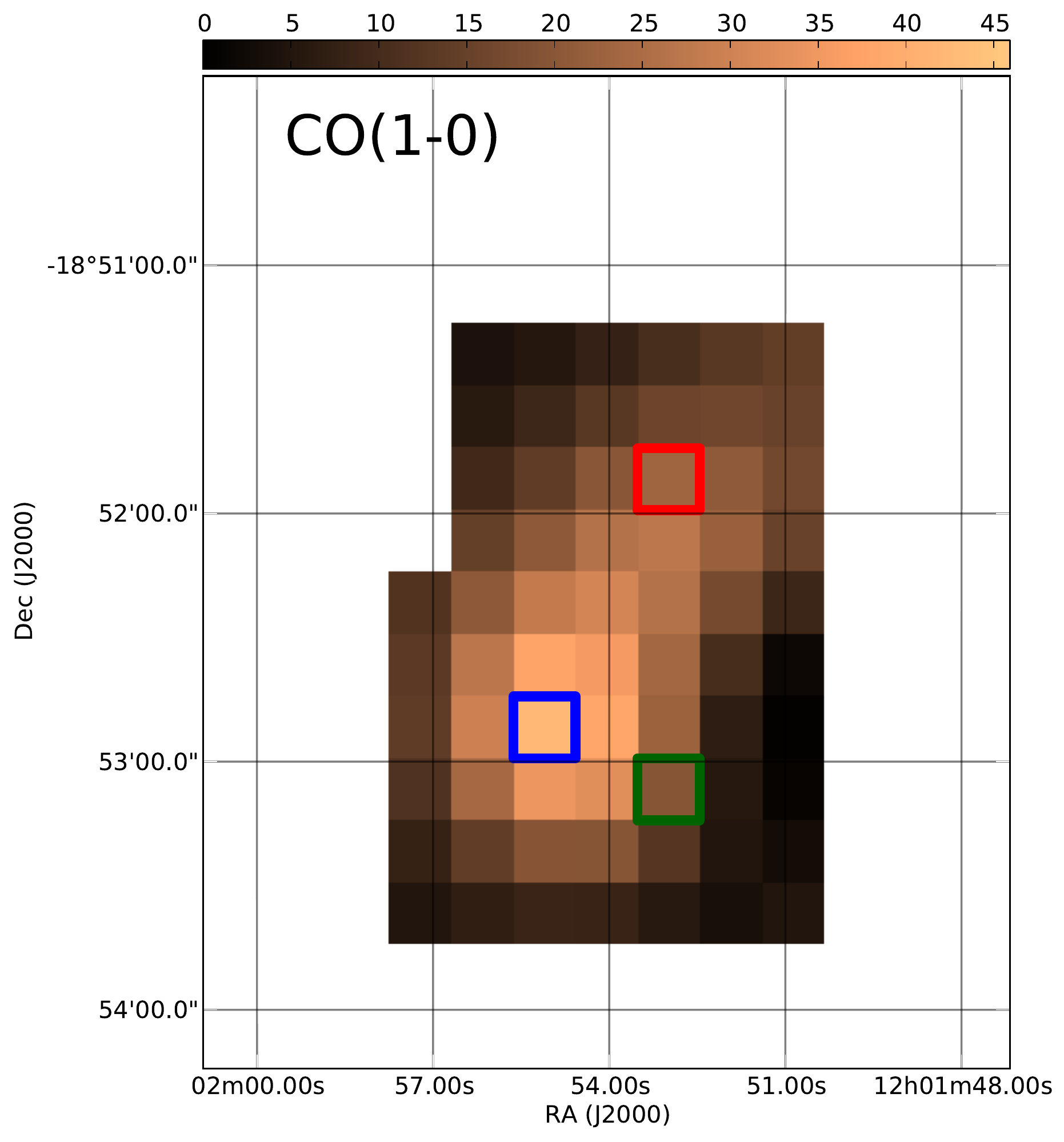} & \includegraphics[width=0.33\linewidth]{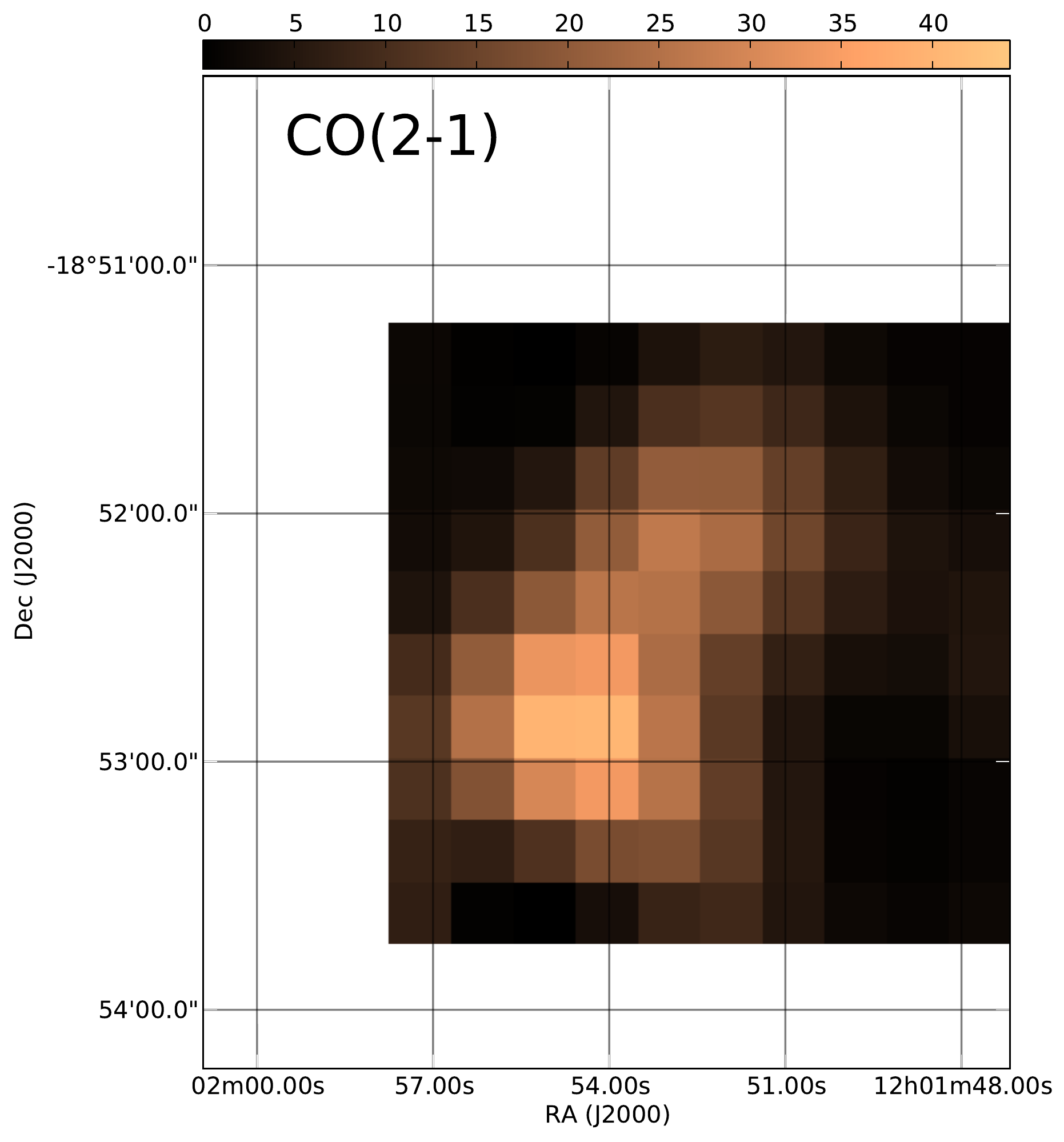} &\includegraphics[width=0.33\linewidth]{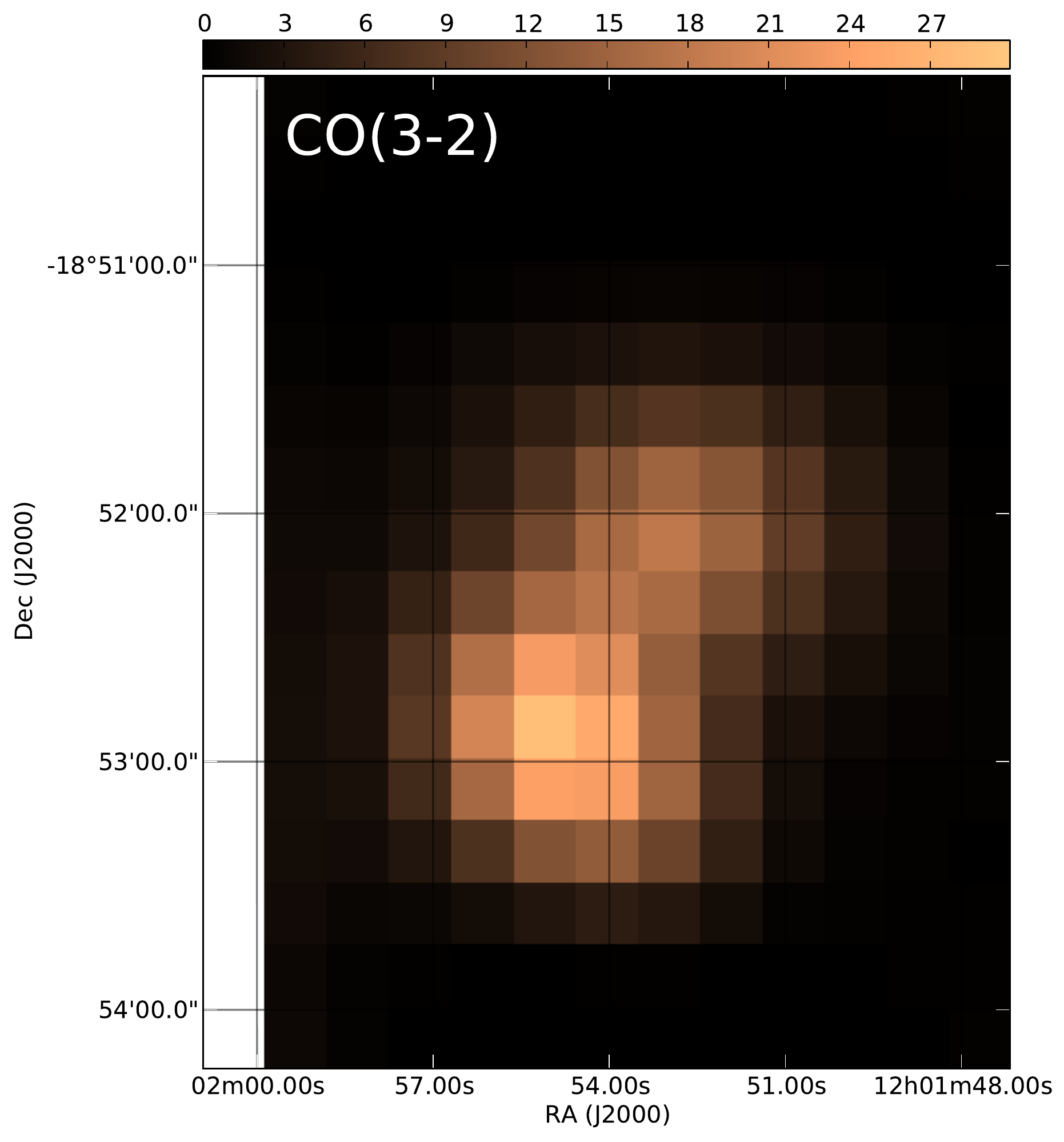} \\
 \includegraphics[width=0.33\linewidth]{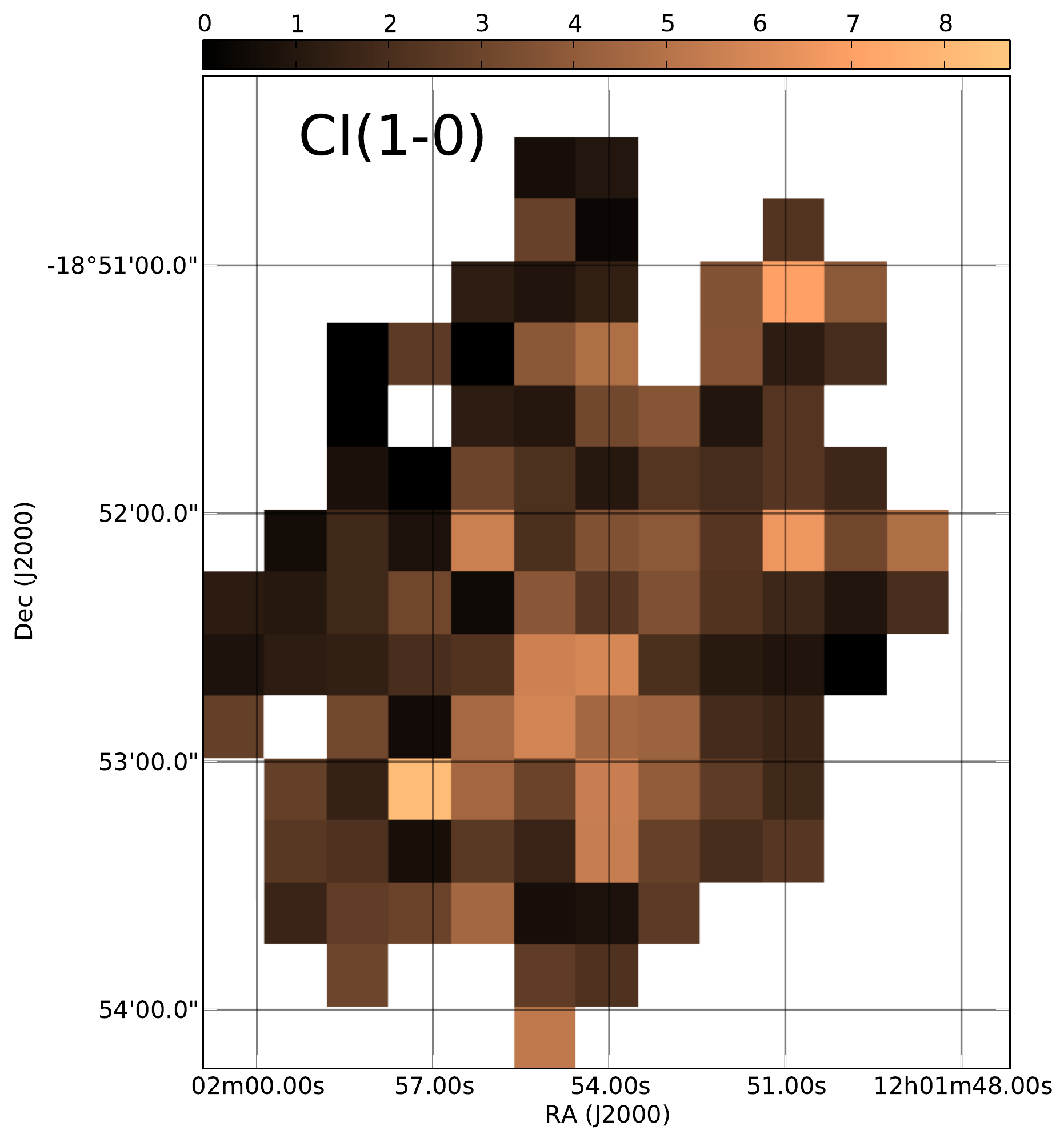} & \includegraphics[width=0.33\linewidth]{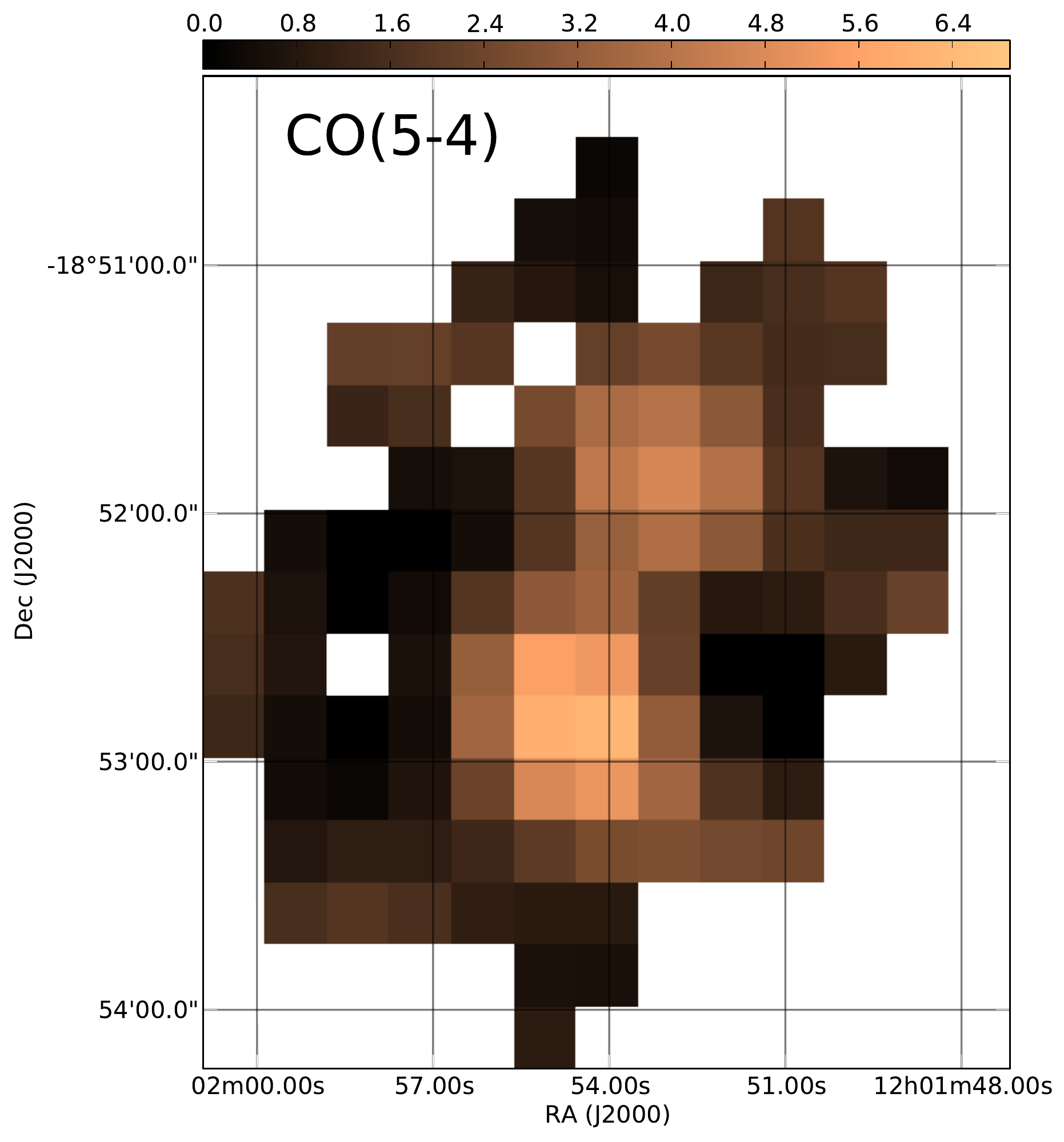} & \includegraphics[width=0.33\linewidth]{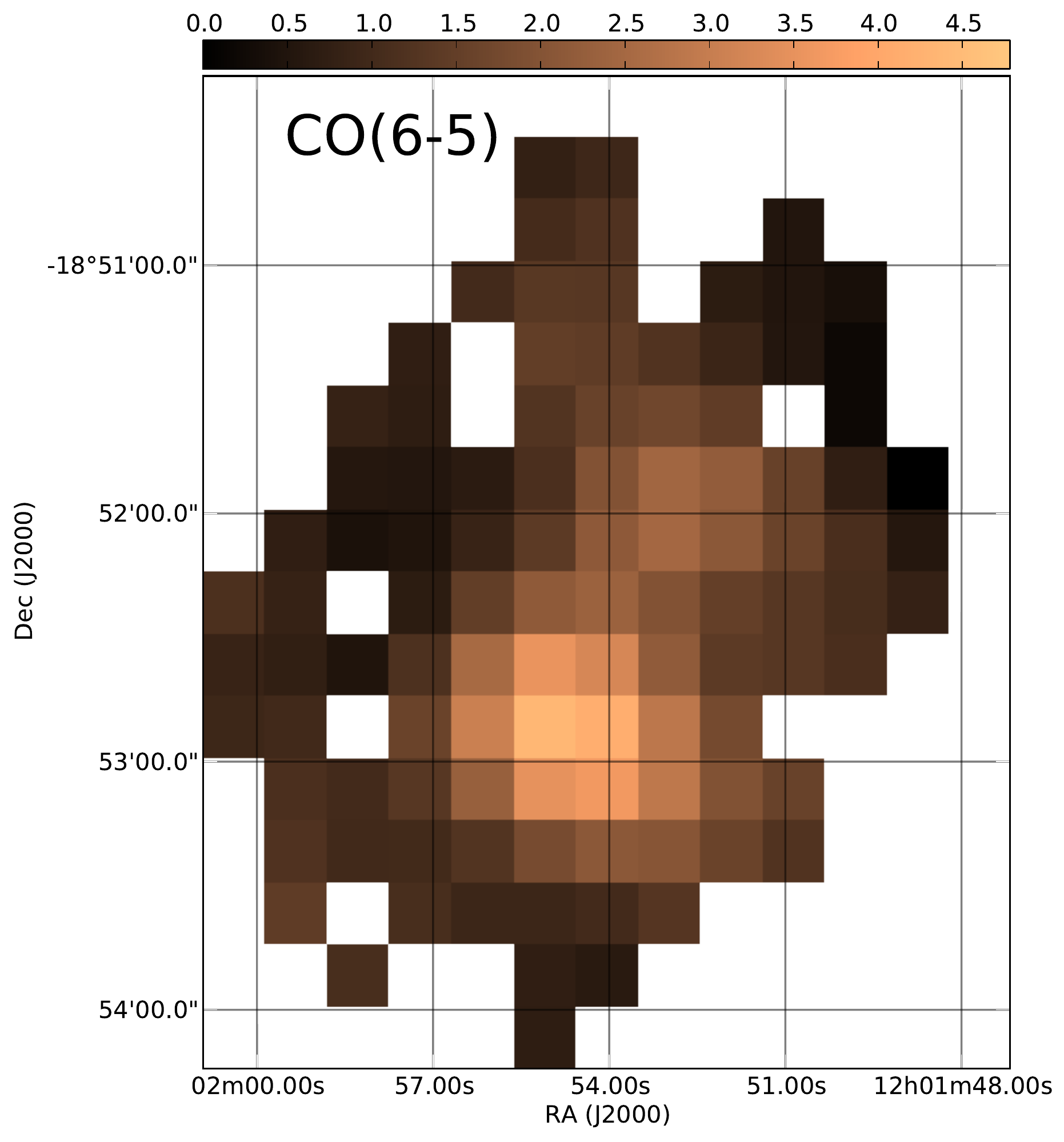} \\\includegraphics[width=0.33\linewidth]{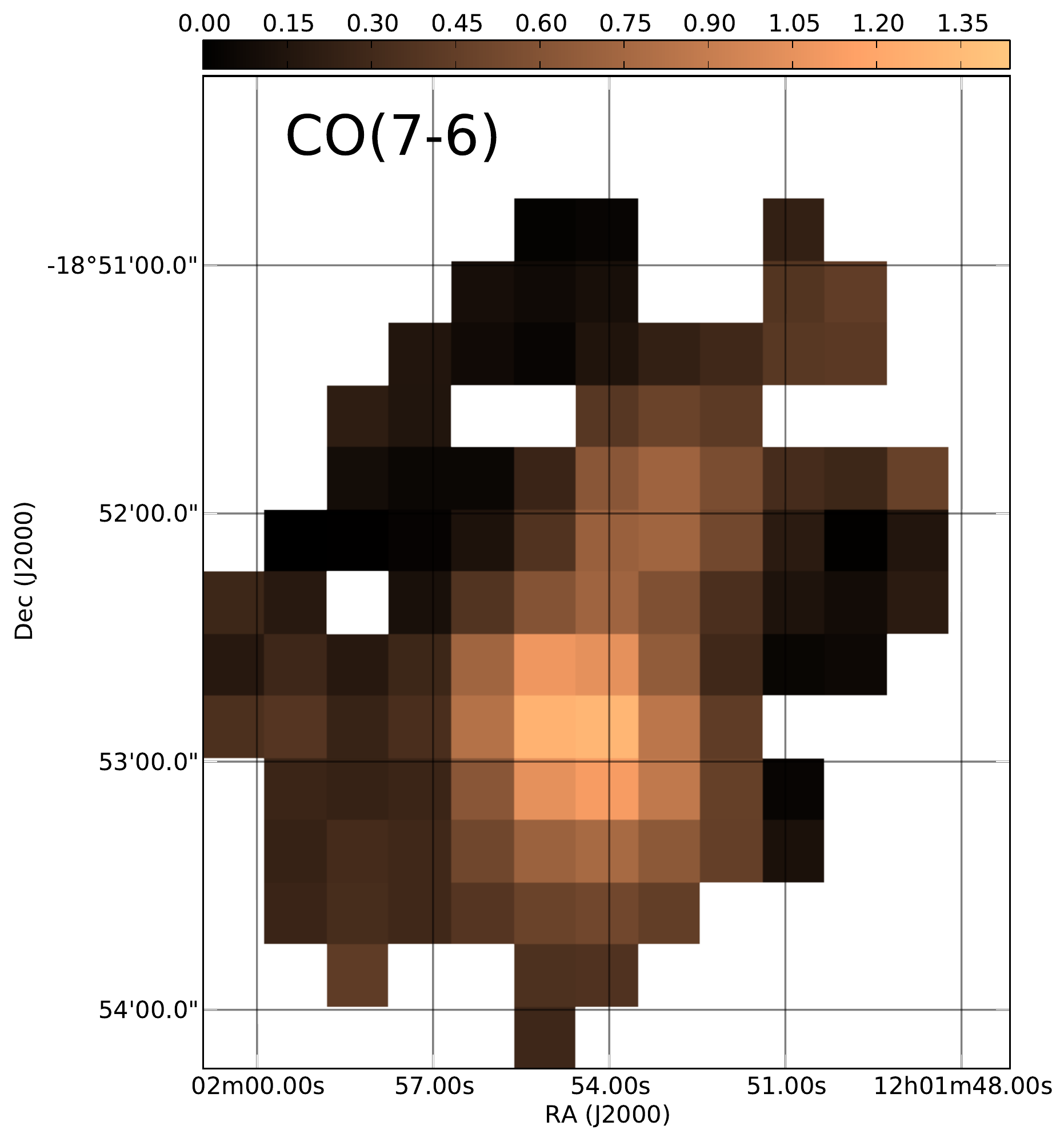} &
\includegraphics[width=0.33\linewidth]{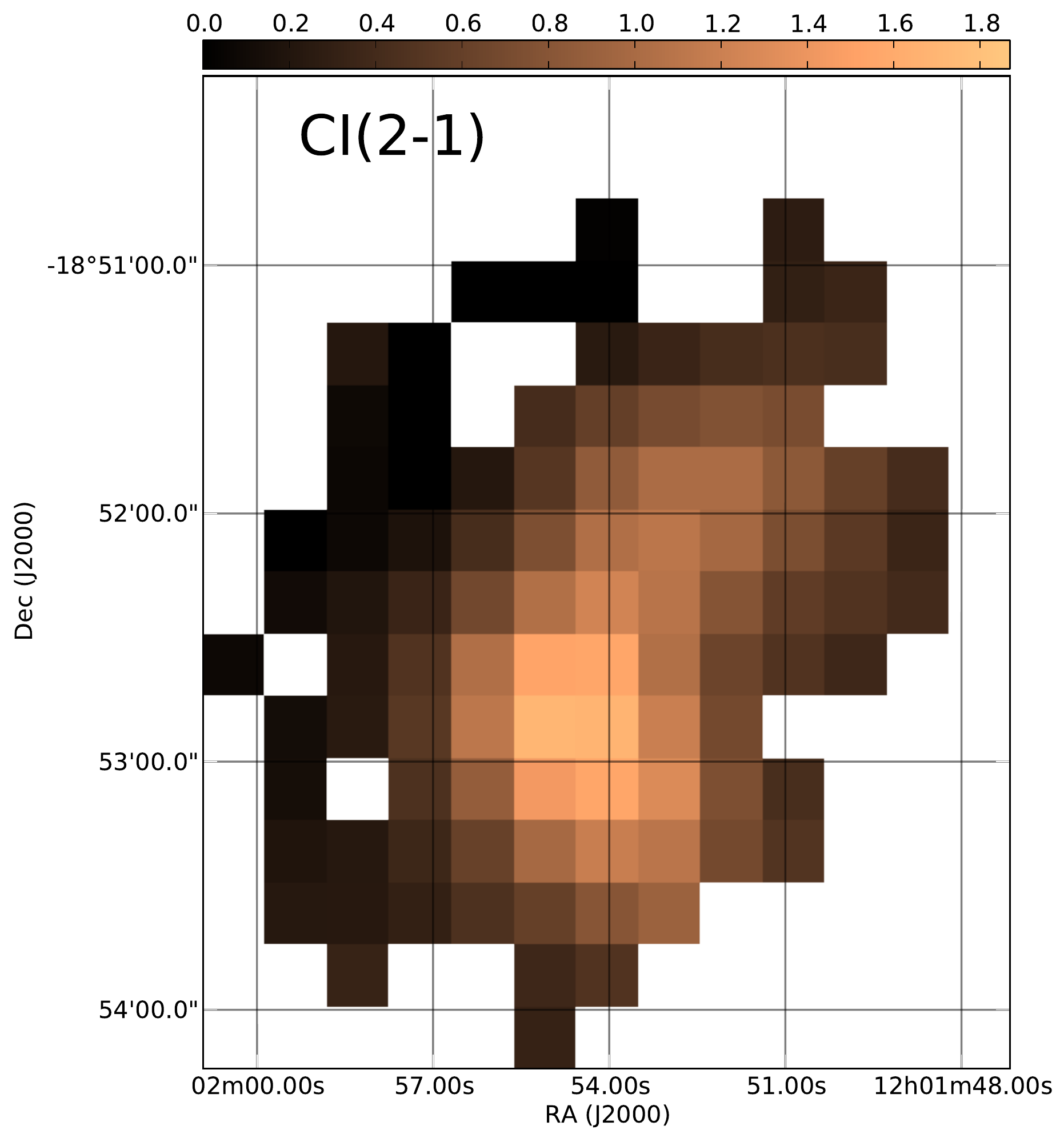} &\includegraphics[width=0.33\linewidth]{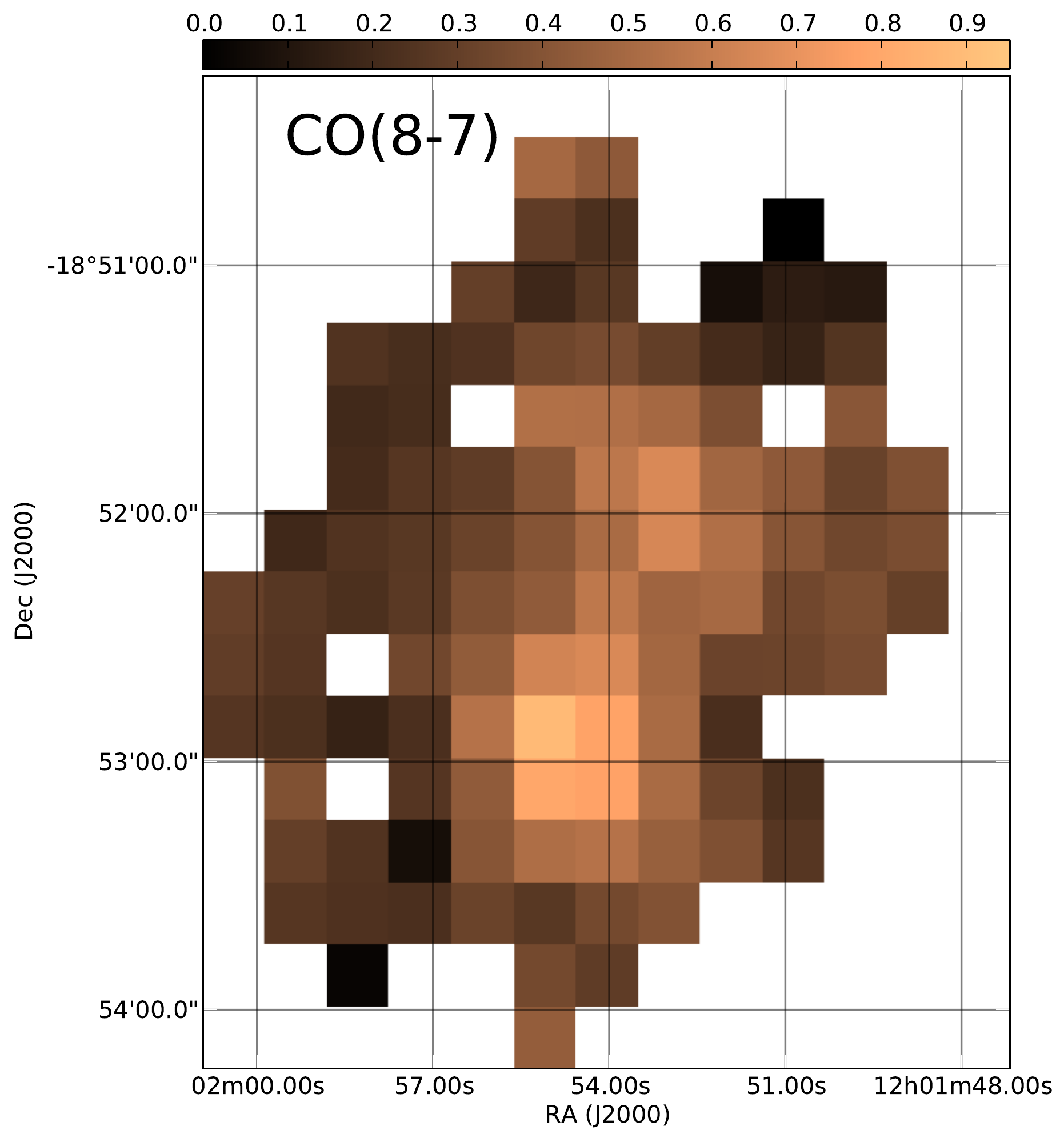}
\end{array}$
\caption[]{Convolved integrated intensity maps for $\mol{CO}$ and $\mol{[CI]}$ in units of $\unit{K \ km \ s^{-1}}$. The $\mol{CO}$ $J=1-0$ observations are from the NRO \citep{zhu2003}, while the $\mol{CO}$ $J=2-1$ and $J=3-2$ observations are from the JCMT. All other observations are from the SPIRE-FTS. All of these maps except for the $\mol{[CI]}$ $J=1-0$ maps have been convolved to match the largest beam size of the FTS at $43''$.  The red, blue and green squares on the $\mol{CO}$ $J=1-0$ map indicate the location of NGC 4038, the overlap region and NGC 4039 (See Figure \ref{PACS70circles} for beam location). { The $\mol{CO}$ $J=4-3$ image is shown in Figure \ref{IMunconv}.}}
\label{IMconv}
\end{figure}

The FTS beam size and shape varies across both the SLW and SSW, from $\sim 17''$ to $\sim 43''$ (SPIRE Observers' Manual version 2.4,   \citealt{swinyard2010}), with the largest beam size occurring at the low frequency end of the SLW.  We developed convolution kernels using the method described in \cite{2012MNRAS.419.1833B} for the ground-based observations of the $\mol{CO}$ $J=1-0$, $J=2-1$ and $J=3-2$ transitions, and for the SPIRE observations of $J=5-4$ to $8-7$ transitions to match the $\mol{CO}$ $J=4-3$ beam ($\sim 43 ''$). The kernels for the SPIRE $\mol{CO}$ transitions are the same kernels used in \cite{kamenetzky2012} and \cite{spinoglio2012}. We use the $\mol{CO}$ $J=7-6$ kernel to convolve the $\mol{[CI]}$ $J=2-1$ map. Finally, we convert from $\mol{Jy \ beam^{-1} \ km \ s^{-1}}$ to $\mol{K \ km \ s^{-1}}$ using 

\begin{equation}
I_{ij} =  S_{ij} \left[0.0109 \theta_{ij}^2 \left(\frac{\nu_{ij}}{115}\right)^2\right]^{-1}
\end{equation}

\noindent where $\nu_{ij}$ is the frequency of the transition in $\mol{GHz}$, $\theta_{ij}$ is the full-width half-maximum beam size in arcseconds, $I_{ij}$ is the integrated intensity in units of $\mol{K \ km \ s^{-1}}$ and $S_{ij}$ is in units of $\mol{Jy \ beam^{-1} \ km \ s^{-1}}$. We use a beam size of $43.4''$ for the convolved maps. The convolved maps are shown in Figure \ref{IMconv} while the integrated intensities for the nuclei of NGC 4038 and NGC 4039, and the overlap region are given in Table \ref{Lineflux} in units of $\mol{K \ km \ s^{-1}}$.

\section{Radiative Transfer Analysis} \label{radtrans}

\subsection{$\mol{[CI]}$ Local Thermodynamic Equilibrium Analysis} \label{CI}

\begin{figure}[ht] %%%FIGURE 3
\centering
\includegraphics[width = \linewidth]{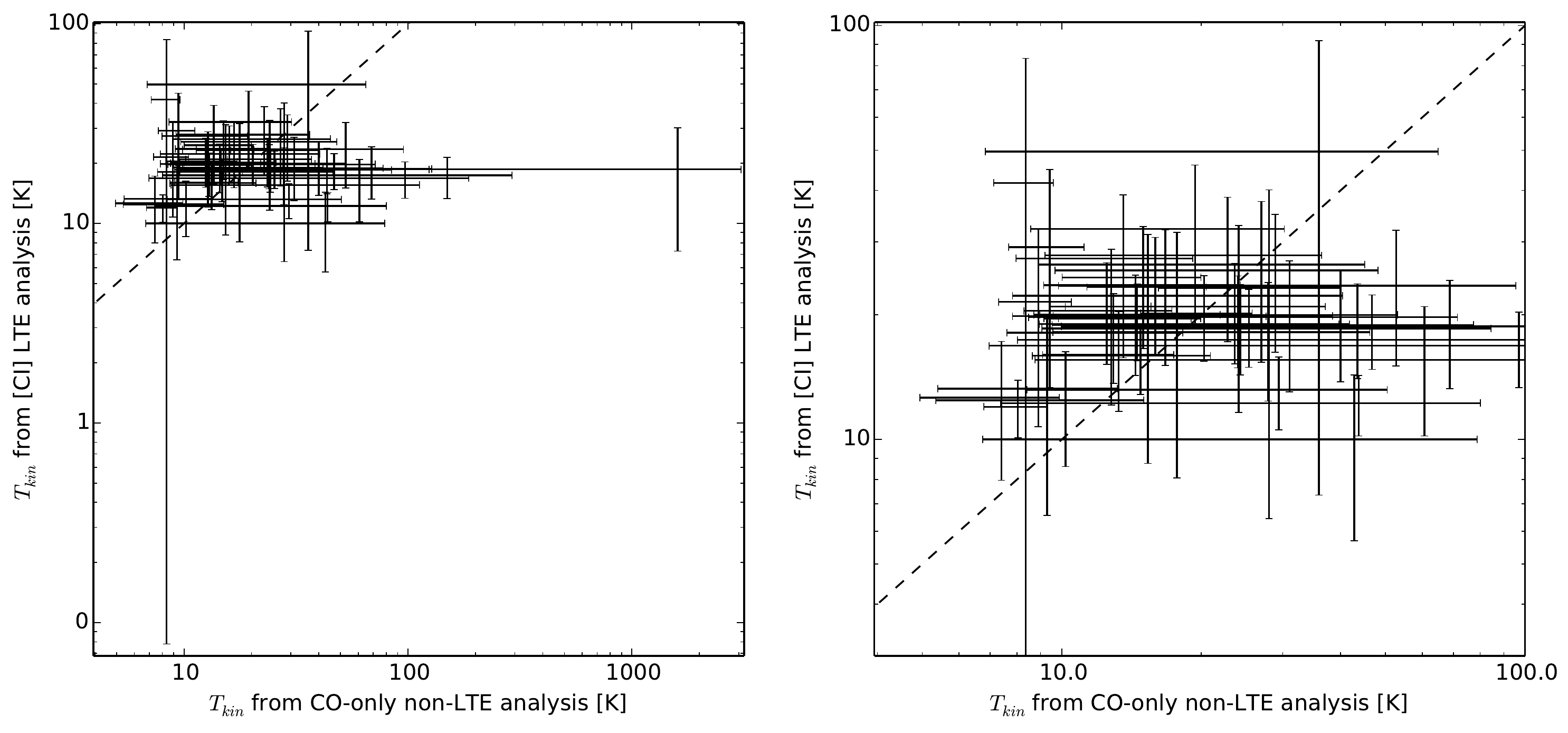} 
\caption[]{Left: Cold molecular gas temperature from the $\mol{[CI]}$ LTE analysis (\emph{y-axis}) and the $\mol{CO}$-only non-LTE radiative transfer analysis (\emph{x-axis}). Right: Zoom in on the upper left portion of the \emph{left} panel. The error bars correspond to the 1-sigma uncertainty in the respective temperatures. Note that where the error bars cross corresponds only to the midpoint of the 1-sigma ranges in log space and not the most probable value. The dashed diagonal line corresponds to where the two temperature are equal.}
\label{CILTE}
\end{figure}

{ In this section we calculate the temperature of the gas using the two $\mol{[CI]}$ lines, $^3P_1 -  ^3P_0$ and $^3P_2- ^3P_1$, assuming local thermodynamic equilibrium (LTE). We calculate the kinetic temperature using a rewritten version of equation 3 from \cite{spinoglio2012}

\begin{equation}
\label{eq:TkF}
T_{kin} = -\frac{E_{21}}{k} \left[\ln{\left(\frac{g_1}{g_2}  \frac{A_{10}\nu_{21}^2 I_{21}}{A_{21}\nu_{10}^2 I_{10}}\right)}\right]^{-1}
\end{equation}

\noindent where $\nu_{ij}$ is the frequency of the transition in $\unit{GHz}$, $I_{ij}$ is the integrated intensity in units of $\unit{K \ km \ s^{-1}}$, $A_{ij}$ is the Einstein coefficient and $\gamma_{ij}$ is the collisional rate. We use the values of $A_{10}=7.93 \times 10^{-8} \unit{s^{-1}}$, $A_{20}=2 \times 10^{-14} \unit{s^{-1}}$, and $A_{21}=2.68 \times 10^{-7} \unit{s^{-1}}$ for the Einstein coefficients \citep{papadopoulos2004}. For the collisional rate coefficients, we assume a temperature of $30\unit{K}$ and an ortho-to-para ratio of 3 and, using the tabulated data from \cite{schroeder1991}, calculate $\gamma_{10}=1.3\times 10^{-10} \unit{cm^3/s}$, $\gamma_{20}=6.9\times 10^{-11} \unit{cm^3/s}$ and $\gamma_{21}=8.3\times 10^{-11} \unit{cm^3/s}$.  

We calculate the temperature in all pixels in our map where we detect both $\mol{[CI]}$ transitions (Figure \ref{CILTE}). Our results suggest that the majority of the $\mol{[CI]}$ emission is associated with cold molecular gas with temperatures of $\sim 10-30 \unit{K}$. }

\subsection{Non-LTE analysis}

We model the $\mol{CO}$ and $\mol{[CI]}$ emission using the non-LTE code RADEX \citep{vandertak2007}, available from the Leiden Atomic and Molecular Database (LAMDA, \citealt{schoier2005}). RADEX iteratively solves for statistical equilibrium based on three input parameters: the molecular gas density ($n(\mol{H_2})$), the column density of the molecular species of interest ($N_{\unit{mol}}$) and the kinetic temperature of the molecular gas ($T_{kin}$). From these three input parameters, RADEX will calculate the line fluxes and optical depths for any molecular or atomic species for which basic molecular data, including the energy levels, Einstein A coefficients and collision rates, are known. Molecular data files are available from LAMDA.  

We calculate a grid of $\mol{CO}$ fluxes and optical depths with RADEX spanning a large parameter space { in density, temperature and column density per unit line width ($N_{\unit{mol}}/\Delta V$) using the uniform sphere approximation.}  In addition, we calculate a secondary grid of $\mol{[CI]}$ fluxes and optical depths based on the same parameter space while varying the $\mol{[CI]}$ abundance relative to $\mol{CO}$ ($x_{\mol{[CI]}}/x_{\mol{CO}}$).  { The Antennae itself is significantly more complex than a simple uniform sphere; however our results are averaged over the entire FTS beam.} The complete list of grid parameters is shown in Table \ref{gridparameter}.  

\subsubsection{Likelihood analysis}

We used a Bayesian likelihood code \citep{ward2003,naylor2010a,panuzzo2010,kamenetzky2011} to determine the most likely solutions for the physical state of the molecular gas for a given set of measured $\mol{CO}$ and $\mol{[CI]}$ line integrated intensities.  We list the highlights of the code here, while further details can be found in \cite{kamenetzky2012}.  The likelihood code includes an area filling factor ($\Phi_A$) with the RADEX grid parameters ($T_{kin}$, $n(\mol{H_2})$, { $N_{\mol{CO}}/\Delta V$)} to create a 4 dimensional parameter space. Assuming Bayes' theorem, the code compares the measured fluxes to those in our RADEX grid to calculate the probability that a given set of parameters produces the observed set of emission lines.  { In addition, the source line width ($\Delta V$) is included as an input parameter in order to properly compare measured and calculated integrated intensities.  We use the convolved $\mol{CO}$ $J=3-2$ second moment map for the source line widths of each pixel in our maps.}

The code calculates three values for each parameter: the median, the 1DMax and the 4DMax. The 1DMax corresponds to the most probable value for the given parameter based upon the 1-dimensional likelihood distribution for that parameter, while the median is also calculated from the 1-dimensional likelihood distribution. The 4DMax is only calculated explicitly for the 4 grid parameters ($T_{kin}$, $n(\mol{H_2})$, $N_{\mol{CO}}$, $\Phi_A$) and is the most probable set of values based upon the 4-dimensional likelihood distribution of the 4 parameters.  Finally, the $1\sigma$ range is calculated from the 1-dimensional likelihood distribution. 

We use three priors to constrain our solutions to those which are physically realizable \citep{ward2003,rangwala2011}. The first prior places a limit on the column density ensuring that the total mass in the column does not exceed the dynamical mass of the system, or

\begin{equation}
N_{\mol{CO}} < \frac{M_{dyn} x_{\mol{CO}}}{\mu m_{\mol{H_2}} A_{\mol{CO}} \Phi_A}
\end{equation}

\noindent where $\mu$ is the mean molecular weight, $M_{dyn}$ is the dynamical mass of the system, $x_{\mol{CO}}$ is the $\mol{CO}$ abundance relative to $\mol{H_2}$ and $A_{\mol{CO}}$ is the area of the CO emitting region. The dynamical mass is calculated as the sum of the virial masses of all of the SGMCs in the overlap region from \cite{wilson2000}, corrected for incompleteness as some of the mass will be found in unresolved SGMCs (e.g. see \citealt{wilson2003}). This incompleteness correction is performed by calculating the fraction of $\mol{CO}$ emission in unresolved SGMCs in the overlap region. The corrected dynamical mass is $3.1 \times 10^9 \unit{M_{\odot}}$. While the dynamical mass in other parts of the galaxy will be less than in the overlap region (e.g. see \citealt{zhu2003}), it is very difficult to determine how much mass is observed by each pixel in our maps. Therefore we conservatively use the highest possible mass we expect in the beam for one pixel.  Using these values along with those listed in Table \ref{lparam}, we limit the product of the column density and filling factor to

\begin{equation}
N_{\mol{CO}} \Phi_A < 10^{18.36} \  \unit{cm^{-2}}
\end{equation}

The second prior limits the total length of the column to be less than the length of the molecular region on the plane of the sky, so that

\begin{equation}
\frac{N_{\mol{CO}}}{\sqrt{\Phi_A} x_{\mol{CO}} n(\mol{H_2})} \le L
\end{equation}

\noindent where $L$ is the size of the $\mol{CO}$ emitting region. We use the diameter of the nucleus of NGC 4038 ($L=1,900 \unit{pc}$) from \cite{wilson2000}, corrected to a distance of $22 \unit{Mpc}$, since it is the largest single molecular complex in the Antennae. As such, this provides an upper limit on the true size of $\mol{CO}$, regardless of which pixel is being considered. All of the physical parameters used to calculate the first two priors are shown in Table \ref{lparam}. 

The third prior limits the optical depth to be $-1 < \tau<100$ as recommended by the RADEX documentation \citep{vandertak2007}. A negative optical depth is indicative of a ``maser'', which is nonlinear amplification of the incoming radiation. RADEX cannot accurately calculate the line intensities when the optical depth is less than $\tau < -1$, and masing is not expected, so these solutions should be disregarded. Conversely, a high optical depth can lead to unrealistic high calculated temperatures and so should be disregarded, and, in any case, our models do not approach that high optical depth limit. 

\subsection{RADEX results}

\subsubsection{\mol{CO} only}
\label{COonly}

We model the $\mol{CO}$ emission for each pixel in our map where we have a detection for all 8 of our observed $\mol{CO}$ transitions ($J=1-0$ to $J=8-7$) and both $\mol{[CI]}$ transitions. The pixels associated with NGC 4038, NGC 4039 and the overlap region are shown on the $\mol{CO}$ $J=1-0$ map in Figure \ref{IMconv}, while the beams associated with these pixels are shown in Figure \ref{PACS70circles}. We assume that all of the molecular gas in each pixel is in one of two distinct components\footnote{ We performed 1-component fits in NGC 4038, NGC 4039 and the overlap region both with and without the two $\mol{[CI]}$ transitions in addition to our 8 $\mol{CO}$ transitions. In all three cases, we recovered only warm  ($T_{kin} \gtrsim 100 \unit{K}$), low density ($n(\mol{H_2}) \lesssim 10^3 \unit{cm^{-3}}$) molecular gas.  Furthermore, we recovered a molecular gas mass of $M_{beam} \sim 10^8 M_\odot$ or less in all three regions, which leads to a total molecular gas mass $\sim$ a few $10^8 M_\odot$ for the entire galaxy (assuming a $\mol{CO}$ abundance of $3 \times 10^{-4}$).  $\mol{CO}$ $J=1-0$ interferometric observations from Wilson et al. (2000) found that, in all three regions, the amount of molecular gas exceeds $5.0 \times 10^8 M_\odot$ using two different methods. Given the warm temperatures and low density of the gas we recovered with our 1-component fit, along with the low molecular gas mass, we feel that the 1-component fit does not represent a physical solution.}: a cold component and a warm component. Studies of the Antennae have revealed that there is both cold gas (e.g. \citealt{wilson2000}, \citealt{zhu2003}) and warm gas (e.g. \citealt{brandl2009}, \citealt{herrera2012}); however the molecular gas likely populates a spectrum of temperature and density ranges. While a two component model  is unlikely to represent the true physical state of the molecular gas, it does provide us with an average, along with a statistical range, of the temperature, density and column density of the molecular gas.  

Under the two-component assumption, the cold component dominates the lower $J$ $\mol{CO}$ emission while the warm component dominates the higher $J$ $\mol{CO}$ emission.  We begin by fitting the cold component to the lower $J$ lines up to some transition $J_{up}\le J_{break}$ and setting the measurements of the higher transitions ($J_{up} > J_{break}$) as upper limits. We subtract the resulting calculated cold component from the line fluxes and fit the residual high $J$ $\mol{CO}$ emission as a ``warm component", while keeping the residuals of the lower $J$ $\mol{CO}$ transitions as upper limits. Following this fit, we subtract the warm component from our measured data and fit the cold component once again. We continue to iterate in this manner until we converge upon a set of solutions. We solve for values of $J_{break}=3$, $4$ and $5$ and present a $\chi^2$ goodness of fit parameter for each solution in Table \ref{chisquared}. For the overlap region and NGC 4038, the $J_{break}=3$ 	solution presents the worst fit, while the differences between the $J_{break}=4$ and $J_{break}=5$ solutions are minimal. Furthermore, for NGC 4039, all three solutions present reasonable fits with the $J_{break}=3$ solution producing the best fit. Therefore, for consistency we report the $J_{break}=4$ solution for each pixel in our map; however, it is important to note that the statistical ranges in the physical parameters for all 3 solutions do not depend appreciably on the value of $J_{break}$.
 
%The best-fit solution for the Overlap region is the $J_{break}=4$ solution
%The best-fit solution for the nuclei of NGC 4038 and NGC 4039 is the $J_{break}=4$ solution. In the overlap region, the $J_{break}=5$ solution presents the ``best fit'' of the measured data; however, the improvement in $\chi^2$ value in the $J_{break}=5$ solution over the $J_{break}=4$ solution is marginal (Table \ref{chisquared}). 

\begin{figure}[ht] %%%2CSLED Jbreak = 4
\centering
\includegraphics[width = 0.8\linewidth]{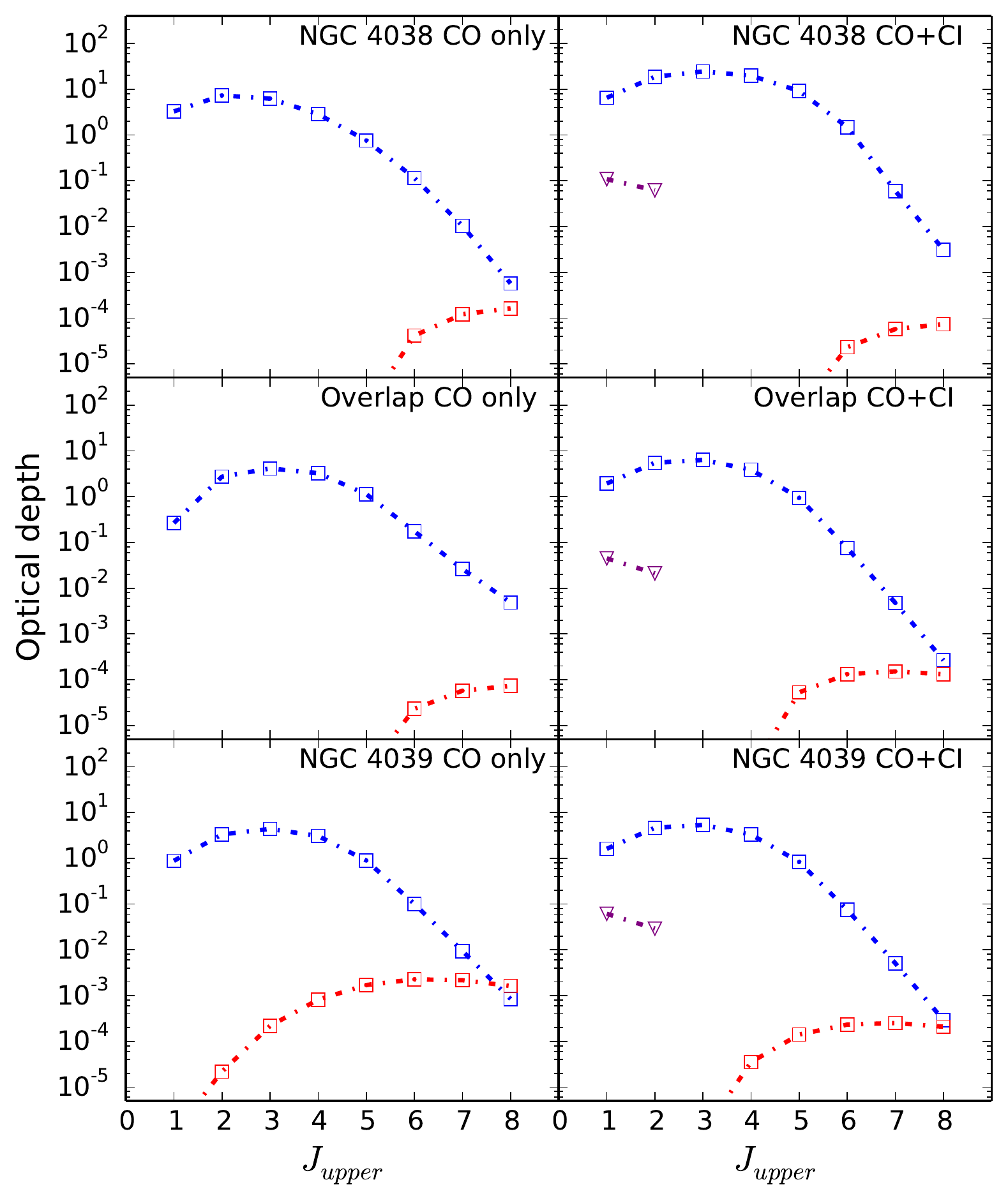} 
\caption[]{Best-fit model $\mol{CO}$ optical depth for the nucleus of NGC 4038 (\emph{top}), the overlap region (\emph{middle}) and the nucleus of NGC 4039 (\emph{bottom}) and for $J_{break} = 4$. The left panels correspond to solutions including only $\mol{CO}$ while the right panels correspond to solutions including both $\mol{CO}$ and $\mol{[CI]}$.  In all panels, the blue dashed-dot line and squares correspond to the cold component, the red dashed-dot line and squares to the warm component. In the right column, the purple dotted line and squares correspond to the calculated $\mol{[CI]}$ optical depth. The warm component $\mol{CO}$ emission and the $\mol{[CI]}$ emission are optically thin.}
\label{Tau43}
\end{figure}

The measured and calculated $\mol{CO}$ spectral line energy distributions (SLEDs) are shown in Figure \ref{COSLED43} while the optical depths are shown in Figure \ref{Tau43}, both calculated from the 4DMax solutions (see Tables \ref{cold_43_UA} and \ref{warm_43_UA}). In NGC 4038 and the overlap region, the cold component dominates the emission for all transitions where $J_{upper}\le 5$, while the warm component is dominant only for the 2 highest $J$ $\mol{CO}$ transitions. In NGC 4039, the cold component dominates only for the $J_{upper} \le 4$ transitions. In  all cases, the warm component is more optically thin than the cold component for almost all of the $\unit{CO}$ transitions (Figure \ref{Tau43}).

\begin{figure}[htp] %%%FIGURE 11a
\centering
\includegraphics[width=0.8\linewidth]{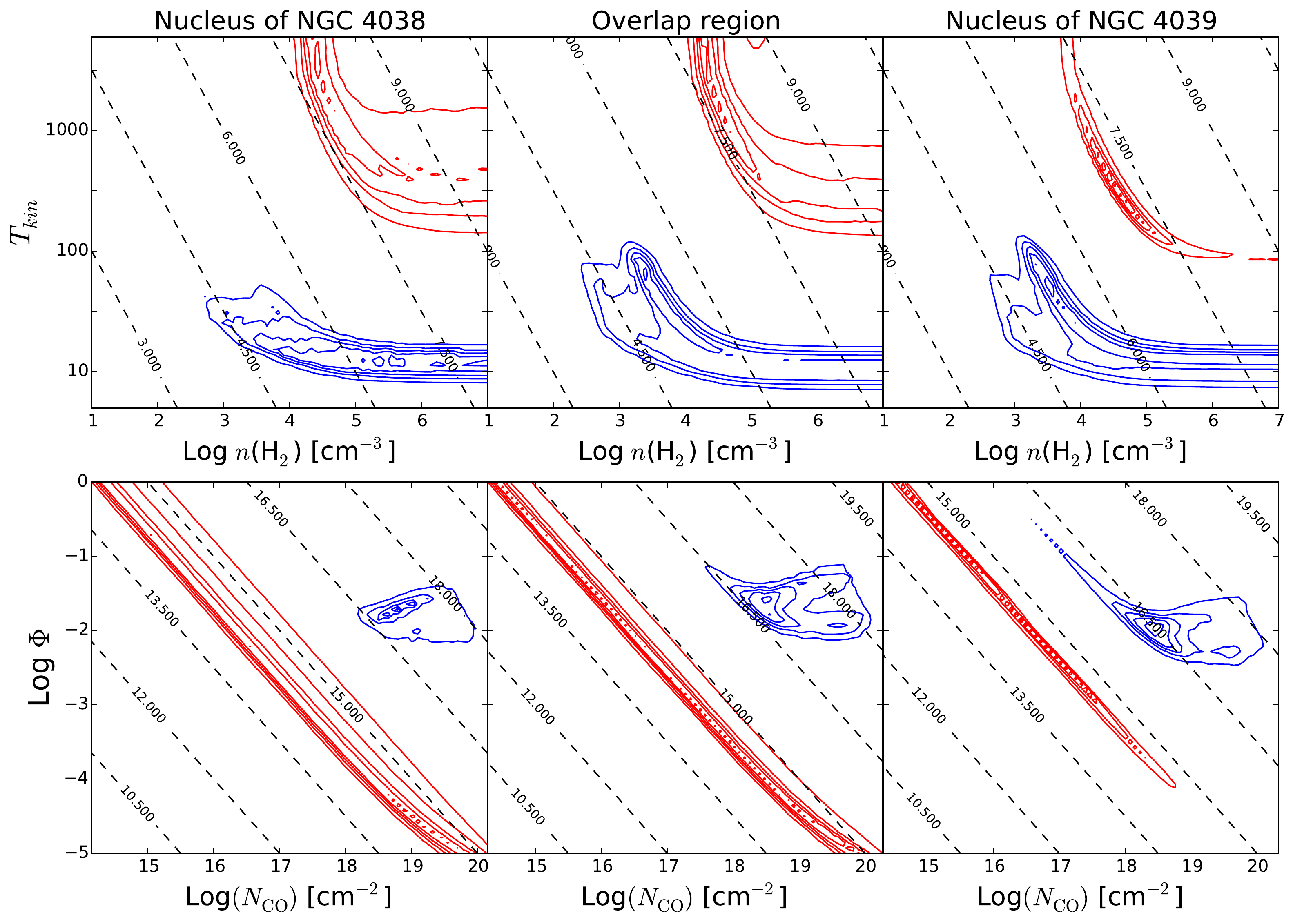} 
\caption[]{Contour probability plots for the $\mol{CO}$-only solutions with $J_{break}=4$ for the warm (red contours) and cold (blue contours) in the nucleus of NGC 4038 (\emph{left}), the overlap region (\emph{middle}) and the nucleus of NGC 4039 (\emph{right}) and for kinetic temperature and density (\emph{top}), and the filling factor and column density (\emph{bottom}). The dashed diagonal lines correspond to the logarithm of the pressure in units of $\mathrm{Log} (\unit{K \ cm^{-2}})$ (\emph{top}) and the beam-averaged column-density, denoted as $\left< N_{\mol{CO}} \right>$ in the tables, in units of $\mathrm{Log} (\unit{cm^{-2}})$ (\emph{bottom}). The contours correspond to $10\%$, $30\%$, $50\%$, $70\%$ and $90\%$ peak probability. }
\label{ContoursCO43}
\end{figure}

The fitted physical parameters for NGC 4038, the overlap region and NGC 4039 are shown in Tables \ref{cold_43_UA} (cold component) and \ref{warm_43_UA} (warm component). In all three regions, the upper limits of the density for both the cold and warm components are not well constrained (Figure \ref{ContoursCO43} \emph{top}). As a result, the upper limit on the pressure, which is the product of the temperature and density, is not well constrained. { The $1\sigma$ range of the temperature of the cold component in all three regions is constrained to being cold ($T_{kin} \lesssim 40 \unit{K}$),} which agrees well with our $\mol{[CI]}$ LTE analysis (Figure \ref{CILTE}). The filling factor and the $\unit{CO}$ column density for the cold component in all three regions is well constrained (Figure \ref{ContoursCO43} \emph{bottom}). The lack of constraint for the warm components of these three regions can be attributed to the degeneracy of $\Phi_A$ and $N_{\mol{CO}}$ (e.g. see \citealt{kamenetzky2012}).  Their product, which is equal to the beam-averaged column-density ($\left<N_{\mol{CO}}\right>$) is well constrained for both the warm and cold component in all three regions (Figure \ref{ContoursCO43} \emph{bottom}). If we assume that the $\mol{CO}$ abundance ($x_{\mol{CO}}$) is the same for both the warm and cold component, the warm component would correspond to $\sim 0.1\%-0.3\%$ of the total molecular gas mass in the nucleus of NGC 4038 and the overlap region, and $\sim 1 \%$ of the total gas mass in the nucleus of NGC 4039.

\begin{figure}[ht] %%%BACD and P for CO only
\centering
\includegraphics[width = \linewidth]{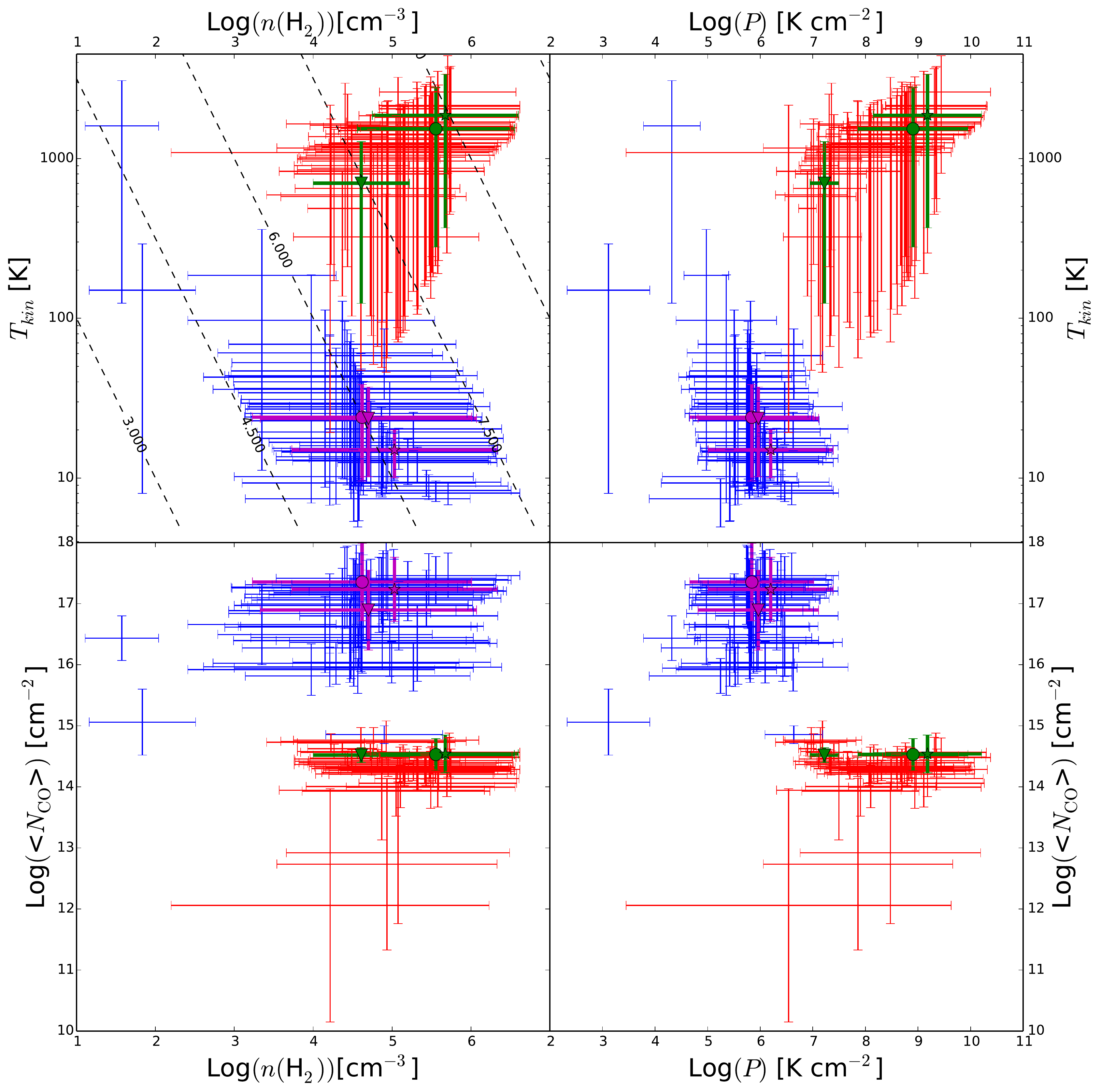} 
\caption[]{$\mol{CO}$-only non-LTE radiative transfer results for every pixel in our maps for which we detect all 8 CO and both [CI] transitions. The temperature (top row) and beam-averaged column density (bottom row) are compared to the molecular gas density (left column) and pressure (right column) for both the cold (\emph{blue}) and warm (\emph{red}) components. The error bars correspond to the 1-sigma range for each physical parameter. { The results for NGC 4038 (star), the overlap region (circle) and NGC 4039 (triangle) are indicated by the thick magenta (cold) and green (warm) symbols and lines.} Note the location where the vertical and horizontal error bars cross corresponds to the midpoint of the $1\sigma$ range and not the most probable value. The dashed diagonal lines in the top-left plot correspond to contours of constant pressure.}
\label{nT43}
\end{figure}

Results for the entire system are shown in Figure \ref{nT43}. In this figure, it is important to note that, since our pixel size ($15 ''$) is less than our beam size ($\sim 43''$), data points on these plots are not entirely independent. We compare the $1\sigma$ ranges for the temperature (\emph{top row}) and beam-averaged column density (\emph{bottom row}) to those for the density (\emph{left column}) and pressure (\emph{right column}). Both a cold and warm component are revealed outside of NGC 4038, NGC 4039, and the overlap region (Figure \ref{nT43} \emph{top row}). Furthermore, no distinction can be made between the density of the cold and warm components (Figure \ref{nT43} \emph{left column}). The pressure in each of the cold and warm components does not vary by more than $\sim 1-2$ orders of magnitude, (Figure \ref{nT43} \emph{right column}) which, given the large $1\sigma$ ranges, may in turn suggest that the conditions under which stars form are nearly constant across the entire system. In addition, the pressure of the warm component is higher than that of the cold component, which is likely attributable to the increased temperature. 

Finally, the beam-averaged column density for both the warm and cold components are well constrained across the entire region, with the beam-averaged column density of the warm component varying only by about an order of magnitude from pixel to pixel (Figure \ref{nT43} \emph{bottom row}). Furthermore, the beam-averaged column density of the warm component is $2-3$ orders of magnitude less than that of the cold component.  Assuming a $\mol{CO}$ abundance of $x_{\mol{CO}} = 3\times10^{-4}$ \citep{kamenetzky2012}, the total mass of the warm component across the entire map ($\log{(M_{warm}/M_{\odot})} = 6.2_{-0.2}^{+0.2}$) is only $\sim 0.1\%$ that of the cold component ($\log{(M_{cold}/M_{\odot})} = 9.1_{-0.9}^{+0.3}$).

\subsubsection{\mol{CO} and \mol{[CI]}}
\label{COCI}

We expand upon our likelihood analysis by assuming both $\mol{[CI]}$ transitions trace the same molecular gas as $\mol{CO}$. \cite{2002ApJS..139..467I} found that in the Orion Giant Molecular Cloud, $\mol{[CI]}$ $^3P_1-^{3}P_0$ and $\unit{^{13}CO}$ $J=1-0$ show structural similarities across the entire cloud both spatially and in velocity, suggesting that both transitions trace much of the same molecular gas especially in the denser regions of GMCs. Furthermore, our ratio of the two $\mol{[CI]}$ transitions along with the $\mol{CO}$-only molecular gas temperature strongly suggests that it originates from cold, rather than warm molecular gas (Section \ref{CI}). 

We fit all of the pixels in our maps where we have a detection in all of the $\mol{CO}$ and both $\mol{[CI]}$ transitions with both a cold and warm component following the same procedure in section \ref{COonly} with the following addition. We include the $\mol{[CI]}$ emission in the cold component only and not the warm component, assuming it does not contribute appreciably to the molecular gas traced by the higher $J$ $\mol{CO}$ transitions. We report the $\chi^2$ goodness of fit parameters for both the $\mol{CO}$ and $\mol{[CI]}$ measured and calculated SLEDs separately in Table \ref{chisquared}. It is important to note that the $\mol{CO}$ SLED is calculated from the 4DMax solutions, while the $\mol{[CI]}$ solutions are calculated from the 1DMax solutions, as only the 1DMax is calculated for the $\mol{[CI]}$ abundance relative to $\mol{CO}$.  The best fit solution to $\mol{CO}$ for NGC 4038 and the overlap region is the $J_{break}=4$ solution, while in NGC 4039, the $J_{break}=3$ solution presents the best solution. Furthermore, the best-fit solutions for $\mol{[CI]}$ in all three regions is the $J_{break}=4$ solution. Therefore, we report the $J_{break}=4$ solution for consistency with our $\mol{CO}$-only results. 

\begin{figure}[tp] %%%FIGURE 11b
\centering
\includegraphics[width=0.8\linewidth]{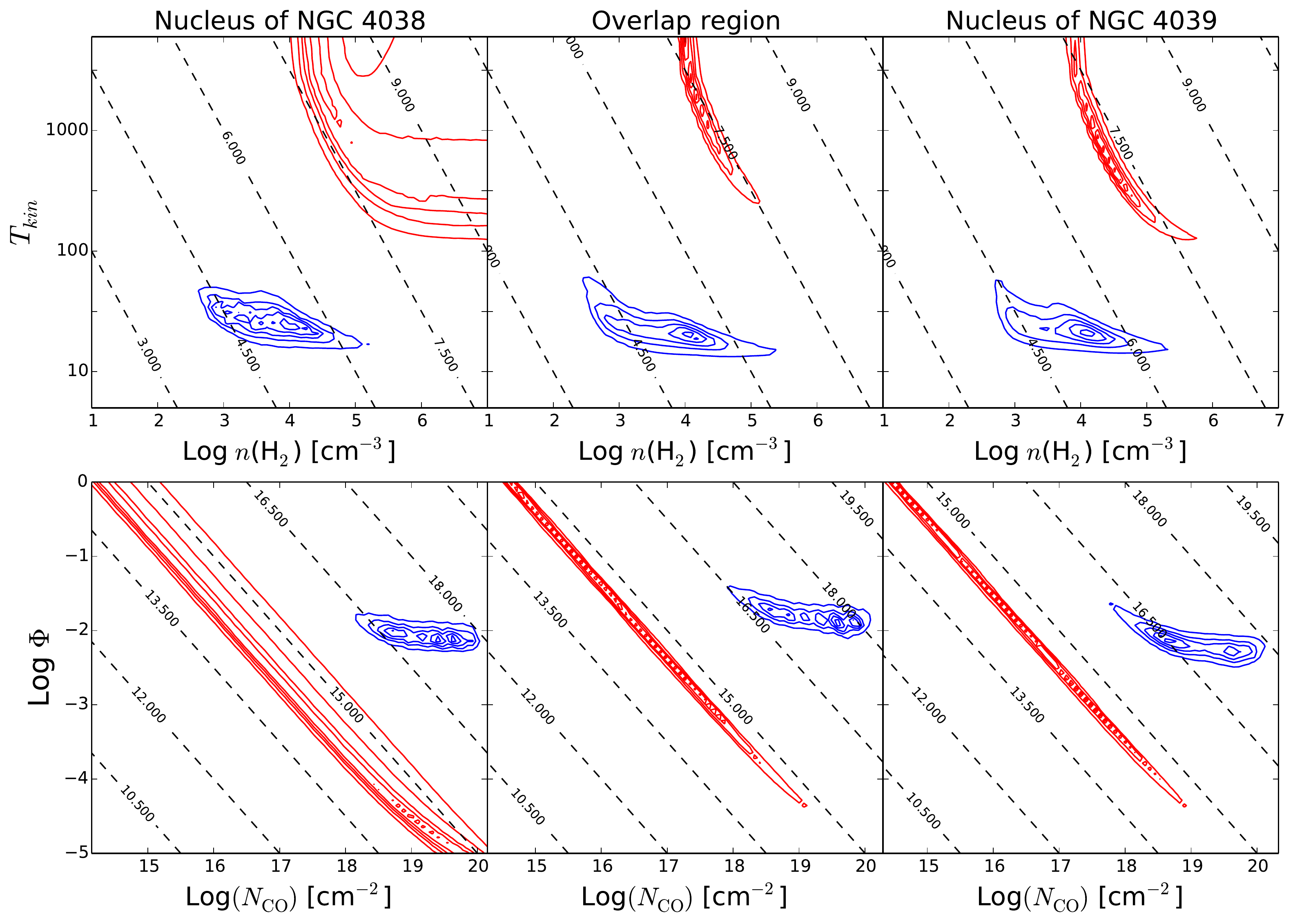} 
\caption[]{Same as figure \ref{ContoursCO43} except for the solutions including both $\mol{CO}$ and $\mol{[CI]}$.}
\label{ContoursCOci}
\end{figure}

The measured and calculated $\mol{CO}$ and $\mol{[CI]}$ SLEDs are shown in Figure \ref{COSLED43} while the optical depths are shown in Figure \ref{Tau43}. 
 The behavior of the cold and warm components of the best fit SLED for all three regions is strikingly similar to the $\mol{CO}$-only results. Furthermore, in all three regions the measured $\mol{[CI]}$ flux is reproduced. Once again, the warm component is more optically thin than the cold component (Figure \ref{Tau43}). In addition, the $\mol{[CI]}$ emission is optically thin, as assumed in our $\mol{[CI]}$ LTE anlaysis. As in the $\mol{CO}$-only solutions, when $\mol{[CI]}$ is included both a cold ($\lesssim 30 \unit{K}$) and a warm ($\gtrsim 200 \unit{K}$) component are recovered (Tables \ref{cold_43_UA_cicold} and \ref{warm_43_UA_cicold}). However, in the $\mol{CO}$ and $\mol{[CI]}$ solution, the density of the cold component in all three components is better constrained than in the $\unit{CO}$ only solution (Figure \ref{ContoursCOci}), along with the density of the warm component in the overlap region and NGC 4039. 

% The behavior of the cold and warm components of the best fit SLED of NGC 4038 and the overlap region are similar to the $\mol{CO}$-only results. In NGC 4039, however, the cold component dominates for $J_{upper} \le 4$, with the warm component dominating the $J_{upper} \ge 6$ transitions. Finally, in NGC 4038 and NGC 4039, the model reproduces the measured $\mol{[CI]}$ fluxes. . %The density of both the warm and cold components remain unconstrained along with the temperature of the warm component (Figure \ref{ContoursCO43}). 

For the remaining pixels in the map, the addition of $\mol{[CI]}$ to the radiative transfer analysis does not change the resulting $1\sigma$ ranges for the various physical parameters (Figure \ref{nT43_ci}). As in the $\mol{CO}$ solution, we find both a warm and cold component with comparable densities. The beam-averaged column density of the warm component is less than in the cold component. The pressure of the warm component is once again higher than in the cold component, while remaining nearly constant for each component separately. Furthermore, we find that the mass of the warm component ($\log{(M_{warm}/M_{\odot})} = 6.2_{-0.2}^{+0.3}$) is only $\sim 0.2 \%$ that of the cold component ($\log{(M_{cold}/M_{\odot})} = 8.9_{-0.7}^{+0.3}$).  

%For the remaining pixels in the map, the addition of $\mol{[CI]}$ to the radiative transfer analysis bears many similarities to the $\unit{[CO]}$ only solution (Figures \ref{ContoursCOci} and \ref{nT43_ci}, and Tables \ref{cold_43_NCO_cicold} and \ref{warm_43_NCO_cicold}). As in the previous solution, we find that on average that the warm component has a significantly higher pressure than the cold component. In addition, the pressure of the cold component is nearly constant across the entire map (Figure \ref{nT43_ci} (d)). Furthermore, as in the $\mol{CO}$-only solution, 

\subsection{Molecular gas mass correction}

The total molecular gas mass calculated from our radiative transfer modeling will be slightly smaller than the true molecular gas mass as we only modeled the $\mol{CO}$ and $\mol{[CI]}$ emission in pixels where we detect both $\mol{[CI]}$ and all 8 $\mol{CO}$ transitions. We can estimate the missing mass using the $\mol{CO}$ $J=3-2$ map as it encompasses the entire $\mol{CO}$ emitting region in the Antennae. The total integrated intensity in the $\mol{CO}$ $J=3-2$ map is $724 \unit{K \ km \ s^{-1}}$, while the integrated intensity of the pixels used in RADEX modeling is $563 \unit{K \ km \ s^{-1}}$, corresponding to only $\sim 78 \%$ of the total integrated intensity. Therefore, we apply a correction of $\sim 22\%$ to the total molecular gas masses calculated in sections \ref{COonly} and \ref{COCI}.

In addition, the $\mol{CO}$ $J=1-0$ map does not extend far beyond the bright $\mol{CO}$ emitting regions in the Antennae, suggesting that there could be missing flux from beyond the edges of the map.  Using the $\mol{CO}$ $J=3-2$ map, we estimate that only $\sim 88 \%$ of the total integrated intensity is within the bounds of the $\mol{CO}$ $J=1-0$ map. We correct for this missing flux when calculating the total cold molecular gas mass from the $\mol{CO}$ $J=1-0$ map in Section \ref{discuss}.

\section{Discussion} \label{discuss}

\subsection{Radiative transfer modeling results}

\subsubsection{Comparison to previous results}

\begin{figure}[ht] %%%2CSLED Jbreak = 4
\centering
\includegraphics[width = 0.8\linewidth]{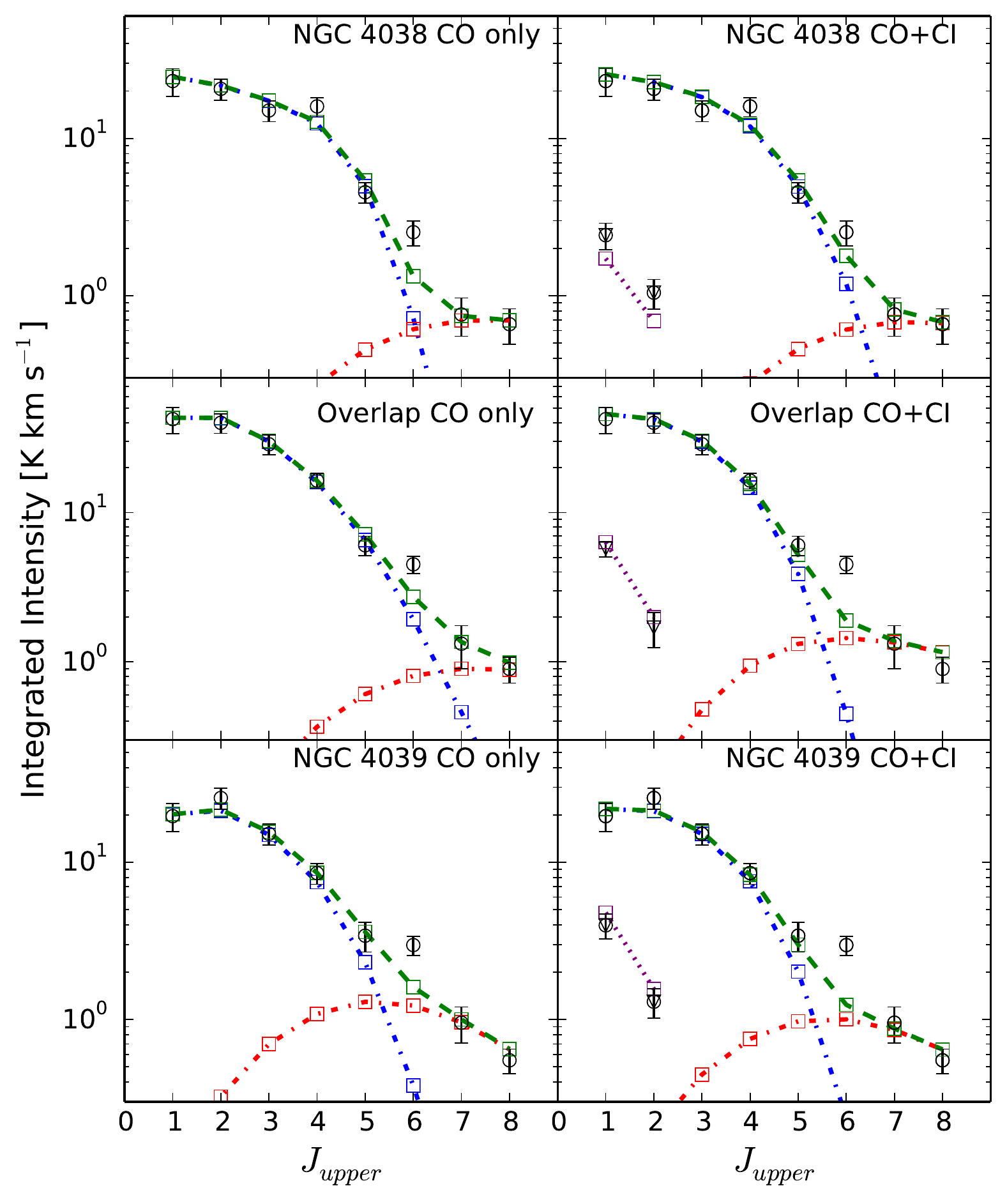} 
\caption[]{Measured and calculated $\mol{CO}$ spectral line energy distributions (SLED) for the nucleus of NGC 4038 (\emph{top}), the overlap region (\emph{middle}) and the nucleus of NGC 4039 (\emph{bottom}) and  for $J_{break} = 4$. The left panels correspond to solutions including only $\mol{CO}$ while the right panels correspond to solutions including both $\mol{CO}$ and $\mol{[CI]}$.  In all panels, the blue dashed-dot line and squares correspond to the cold component, the red dashed-dot line and squares to the warm component, the green dashed line and squares is the sum of the cold and warm component and the black circles correspond to the measured \mol{CO} data.  In the right column, the purple dotted line and squares correspond to the calculated $\mol{[CI]}$ flux while the black triangles correspond to the measured $\mol{[CI]}$ flux.}
\label{COSLED43}
\end{figure}

Both \cite{zhu2003} and \cite{bayet2006} have previously performed a radiative transfer analysis using ground based $\mol{CO}$ data. \cite{bayet2006} used the ratios of $\frac{\unit{^{12}CO} \ J=3-2}{\unit{^{12}CO} \ J=6-5}$, $\frac{\unit{^{12}CO} \ J=2-1}{\unit{^{12}CO} \ J=6-5}$, and $\frac{\unit{^{12}CO} \ J=3-2}{\unit{^{13}CO} \ J=3-2}$ to fit a single warm component in the nucleus of NGC 4038 and the overlap region. The primary difference between their model for a warm component and ours is that we consider the contributions of the cold component to the lower $J$ $\mol{CO}$ transitions while they do not. They found that the temperature and density vary significantly between NGC 4038 and the overlap region. In NGC 4038, the temperature that their model predicts ($T_{kin} = 40 \unit{K}$) does not fall within either the cold or warm component $1\sigma$ range for either the $\mol{CO}$ or the $\mol{CO}$ and $\mol{[CI]}$ solutions, while the density ($n(\mol{H_2}) = 3.5 \times 10^5 \unit{cm^{-2}}$) agrees within $1\sigma$ of the $\mol{CO}$-only cold component and both the $\mol{CO}$-only, and the $\mol{CO}$ and $\mol{[CI]}$ warm components. In the overlap region, their model temperature ($145 \unit{K}$) does not fall within any of our temperature ranges, while their density ($n(\mol{H_2})  = 8.0 \times 10^3 \unit{cm^{-2}}$) agrees with both our cold component densities. Furthermore, their models predict that the $\unit{^{12}CO}$ SLED peaks at the $J=3-2$ transition, while our observations indicate that it instead peaks at the $J=1-0$ transition (Figure \ref{COSLED43}).  In our models we find that the warm component only contributes significantly to the $J=6-5$, $J=7-6$ and $J=8-7$  transitions (Figure \ref{COSLED43}), suggesting that \cite{bayet2006} is, at least in part, modeling a cold component of the molecular gas. 

In comparison, using various ratios of the $\unit{^{12}CO}$ $J=1-0$ to $J=3-2$ and the isotopologue $\unit{^{13}CO}$ $J=2-1$ and $J=3-2$ transitions, \cite{zhu2003} fit both a single and a two-component model in NGC 4038, NGC 4039 and the overlap region. The results from the single component fit suggest that these transitions are tracing a cold component($T_{kin} \sim 20-40 \unit{K}$) with density $n(\mol{H_2}) \sim 10^3-10^4 \unit{cm^{-3}}$, both comparable to our cold component. This agreement is unsurprising, as our models suggest that, in all 3 regions, the cold component dominates the emission from the $\unit{^{12}CO}$ $J=1-0$ to $J=3-2$ transitions. For the two component fit, \cite{zhu2003} argue that there must be a low density component ($n(\mol{H_2}) \sim 10^3 \unit{cm^{-3}}$) which dominates the optically thick $\unit{^{12}CO}$ emission and a high density component ($n(\mol{H_2}) \sim 10^5 \unit{cm^{-3}}$) which dominates the optically thin $\unit{^{13}CO}$ emission. As a result, they find two temperature components, a cold component ($T_{kin} = 36 \unit{K} - 120 \unit{K}$) and a warm component ($T_{kin} = 42 \unit{K} - 220 \unit{K}$), with the density and temperature of the two components depending upon the ratio of $^{12}\mol{CO}/^{13}\mol{CO}$ which varies from $40$ to $70$. In our case, the warm component dominates the high $J$ $\mol{CO}$ transitions, which are optically thin, while the cold component dominates the optically thick lower $J$ $\mol{CO}$ transitions (Figure \ref{Tau43}). Furthermore, while the density is not well constrained, our results suggest that the cold component has a lower density than the warm component, indicating that our lower density component does coincide with the optically thick transitions, as found by \cite{zhu2003}.  

\subsubsection{Comparison across the Antennae}

\begin{figure}[ht] %%%BACD and P for CO only
\centering
\includegraphics[width = \linewidth]{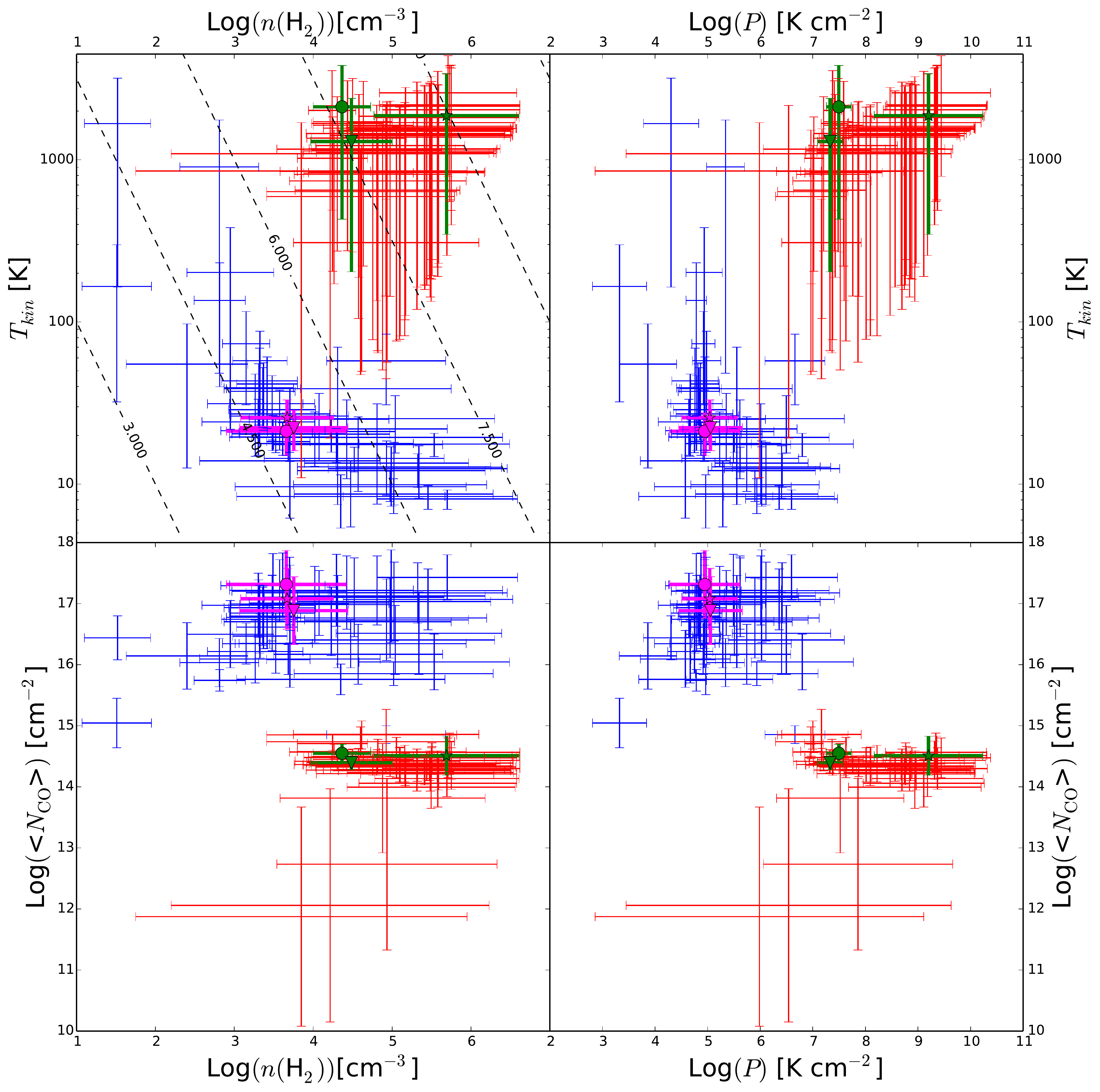} 
\caption[]{Same as figure \ref{nT43} except for solutions including both $\mol{CO}$ and $\mol{[CI]}$.}
\label{nT43_ci}
\end{figure}

Both the LTE and non-LTE radiative transfer analyses suggest there is cold molecular gas across the Antennae with both sets of calculated temperatures agreeing within the uncertainties (Figure \ref{CILTE}).  Furthermore, the non-LTE analysis suggests that warm molecular gas ($T_{kin} \gtrsim 100 \unit{K}$) is prevalent throughout the system; however the warm component temperature is poorly constrained (Figures \ref{nT43} and \ref{nT43_ci}). { Both sets of RADEX solutions for the cold molecular gas suggest that the molecular gas has a similar temperature ($\sim 20\unit{K}$) and density ($\sim 10^3 - 10^4 \unit{cm^{-3}}$) in all three regions.  Similarly, the densities of the warm component in the overlap region and NGC 4039 are similar ($\sim 10^{4.5} \unit{cm^{-3}}$), while it is higher in NGC 4038 ($\gtrsim 10^{4.75} \unit{cm^{-3}}$).  In addition,} the $\mol{CO}$ and $\mol{[CI]}$ RADEX solution suggest the density of the warm component is slightly higher than that of the cold component; however for the overlap region and NGC 4039, the $1 \sigma$ ranges of the densities of the cold and warm components overlap and so no firm conclusions can be drawn concerning these densities. 

\cite{brandl2009} calculated the temperature of the warm molecular gas from Infrared Spectrograph (IRS) Spitzer Space Telescope (\emph{Spitzer}) observations of the $\mol{H_2}$ $\unit{S(1)}$ and $\unit{S(2)}$ transitions and found a warm gas temperature of $~270 - 370 \unit{K}$ in both nuclei and at numerous locations in the overlap region. This temperature range falls within our warm component $1\sigma$ ranges for these three regions.  This suggests that $\mol{H_2}$ is tracing the same warm gas as the upper $J$ $\mol{CO}$ transitions. 

The pressure across the Antennae in each of the cold and warm components is nearly constant within uncertainties (Figures \ref{nT43} and \ref{nT43_ci}), while the pressure in the warm component is $\sim 2-3$ orders of magnitude larger than that in the cold component. The temperature is only $\sim 1-3$ orders of magnitude larger in the warm component over the cold component and may be sufficient to explain the increased pressure.

\subsection{$\mol{CO}$ abundance and the $\mol{CO}$-to-$\mol{H_2}$ conversion factor}\label{xXCO}

%\subsection{$x_{\mol{CO}}$ and $\alpha_{\mol{CO}}$}\label{xXCO}

For the purpose of our RADEX modeling we have assumed a $\mol{CO}$ abundance of $x_{\mol{CO}} = 3\times10^{-4}$ \citep{kamenetzky2012}. Using this abundance, we recover a corrected cold gas mass of $M_{cold} = 1.5_{-1.3}^{+1.7} \times 10^{9} \ M_{\odot}$ which would correspond to a $\mol{CO}$ $J=1-0$ luminosity-to-mass conversion factor of $\alpha_{\mol{CO}} \sim 0.7 \ M_{\odot}\unit{pc^{-2} \ (K \ km \ s^{-1})^{-1}}$. This value is lower than the value for the Milky Way, but agrees with the value of $\alpha_{\mol{CO}} = 0.8 \ M_{\odot}\unit{pc^{-2} \ (K \ km \ s^{-1})^{-1}}$ from \cite{downes1998}, which is the value typically assumed for ULIRGs.  Arp 299, another relatively nearby merger ($D_L = 46 \unit{Mpc}$), is brighter in the infrared than the Antennae ($L_{IR} \sim 7 \times 10^{11} L_{\odot}$). \cite{sliwa2012} found that for a $\mol{CO}$ abundance of $3 \times 10^{-4}$, $\alpha_{\mol{CO}} = 0.4 \ M_{\odot}\unit{pc^{-2} \ (K \ km \ s^{-1})^{-1}}$, which agrees with the value calculated for LIRGs ($\alpha_{\mol{CO}} \sim 0.6 \pm 0.2 \ M_{\odot}\unit{pc^{-2} \ (K \ km \ s^{-1})^{-1}}$, \citealt{papadopoulos2012}).

However, the true $\mol{CO}$ abundance in the Antennae may be smaller \citep{zhu2003}. We investigate the $\mol{CO}$ abundance by comparing the total warm molecular gas mass we calculate in the Antennae to those from previous studies.  \cite{brandl2009} measured the total warm ($\sim 300 \unit{K}$) gas mass in the Antennae using \emph{Spitzer} observations of the $\mol{H_2}$ S(1), S(2) and S(3) transitions to be $2.5 \times 10^{7} M_{\odot}$. Our corrected warm component ($\gtrsim 100 \unit{K}$) mass from the $\mol{CO}$-only RADEX solution is $2.2_{-1.0}^{+1.3} \times 10^6 M_{\odot}$, while for the $\mol{CO}$ and $\mol{[CI]}$ solution it is $2.2_{-1.0}^{+1.5} \times 10^6 M_{\odot}$.  Both of our calculated masses are a factor of $\sim 10$ less than the \cite{brandl2009} mass. Assuming our warm gas solution is tracing the same gas as the $\mol{H_2}$ lines, a $\mol{CO}$ abundance of $x_{\mol{CO}} = 3 \pm 2 \times 10^{-5}$ is required to recover this warm gas mass. This abundance ratio falls in the range of $\sim 10^{-5}-10^{-4}$, which is typically assumed for starburst galaxies \citep{zhu2003,mao2000}. 

Using this abundance ratio of $3 \times 10^{-5}$, we obtain a cold molecular gas mass for the $\mol{CO}$-only solution of $M_{cold} = 1.5_{-1.3}^{+1.7} \times 10^{10} \ M_{\odot}$. In order to recover the same mass from our $\mol{CO}$ $J=1-0$ map, we would require a $\mol{CO}$ $J=1-0$ luminosity-to-mass conversion factor of $\alpha_{\mol{CO}} \sim 7 \ M_{\odot}\unit{pc^{-2} \ (K \ km \ s^{-1})^{-1}}$. This is consistent with the Milky Way value  ($\alpha_{\mol{CO}} \sim 4-9 \ M_{\odot} \unit{pc^{-2} \ (K \ km \ s^{-1})^{-1}}$) and the value for M 31, M 33 and the Large Magellanic Cloud ($\alpha_{\mol{CO}} \sim 3-9 \ M_{\odot} \unit{pc^{-2} \ (K \ km \ s^{-1})^{-1}}$, \citealt{leroy2011}). \cite{wilson2003} calculated the $\mol{CO}$-to-$\mol{H_2}$ conversion factor in the Antennae by comparing the virial mass of resolved SGMCs to the integrated intensity. They found that the conversion factor in the Antennae agrees with the value for the Milky Way ($\alpha_{\mol{CO}} \sim 6.5 \ M_{\odot}\unit{pc^{-2} \ (K \ km \ s^{-1})^{-1}}$). As such, we adopt a value of $\alpha_{\mol{CO}} \sim 7 \ M_{\odot}\unit{pc^{-2} \ (K \ km \ s^{-1})^{-1}}$ and $x_{\mol{CO}} \sim 3 \times 10^{-5}$ for the $\mol{CO}$-to-$\mol{H_2}$ conversion factor and the $\mol{CO}$ abundance respectively.

%The increased $L_{IR}$ in Arp 299 could be associated with increased star formation throughout the system, which in turn could drive the value of $\alpha_{\mol{CO}}$ down. We might expect Arp 299 to be in a more advanced merger state; however this is not necessarily the case \citep{hibbard1997}. It may be that there is more gas near the nuclear regions in Arp 299 as opposed to NGC 4038/39, as both overlap regions are similar \citep{hibbard1997}. This is turn may lead to increased star formation in the nuclear regions which in turn could drive down the $\mol{CO}$ luminosity-to-mass conversion factor. Furthermore, it could be that $\alpha_{\mol{CO}}$ is close to the Milky Way value in both Arp 299 and NGC 4038/39, while the $\mol{CO}$ abundance is lower. This could be evidence of $\mol{CO}$ photo-dissociation from far-ultraviolet (FUV) radiation (e.g. \citealt{tielens1985}). We explore the implications of these photon dominated regions (PDRs) in the overlap region and NGC 4039 in Section \ref{PDR}.

\subsection{Heating and cooling of the molecular gas} \label{heat}

In this section we will discuss the possible heating and cooling mechanisms for the molecular gas in the Antennae.

\subsubsection{Cooling} \label{cooling}

$\mol{CO}$, $\unit{[OI]63\unit{\mu m}}$, $\mol{[CII]}$ and $\mol{H_2}$ will all contribute to the overall cooling budget of both the cold and warm molecular gas; however, which coolant dominates is dependent on the overall state of the molecular gas. In particularly warm molecular gas ($T_{kin} \gtrsim 1000 \unit{K}$), $\mol{H_2}$ will be the dominant coolant (e.g. Arp 220, \citealt{rangwala2011}) while for cooler gas, $\mol{[CII]}$, $\unit{[OI]}63\unit{\mu m}$, and $\mol{CO}$ will also contribute to the total cooling of molecular gas. We investigate the possible cooling mechanisms in order to determine the total rate of cooling of molecular gas in the Antennae. 

$\mol{H_2}$ cooling only becomes important in molecular gas where the temperature is $T_{kin}>100 \unit{K}$. \cite{lebourlot1999} calculated curves for the total $\mol{H_2}$ cooling per unit mass, which is dependent upon the molecular gas density ($n(\mol{H_2})$), the kinetic temperature $T_{kin}$, the $\unit{H}/\mol{H_2}$ ratio and the ratio of ortho-to-para $\mol{H_2}$, { all of which are input parameters for the calculated cooling curves. } \cite{lebourlot1999} have provided a program which interpolates their cooling curves for a given set of input parameters, allowing us to estimate the $\mol{H_2}$ cooling for each pixel in the Antennae using the RADEX calculated $n(\mol{H_2})$ and $T_{kin}$ from the warm component, along with the mass of the warm molecular gas. We assume that $\unit{H}/\mol{H_2}=0.01$ and that the ratio of ortho-to-para is $1$.  It is important to note that changing the ortho-to-para ratio to $3$ does not change the total $\mol{H_2}$ cooling significantly \citep{lebourlot1999}. The $\mol{H_2}$ cooling is highly dependent on the kinetic temperature and given that the temperature of our warm component is not particularly well constrained, we opt to use the $1\sigma$ lower bound on the temperature in each pixel to calculate a lower limit to the $\mol{H_2}$ cooling. There is only a $\sim 10-15\%$ difference between the cooling rate calculated when using either the $1\sigma$ lower bound or the most probable value for the molecular gas density. We opt to use the most probable value as it likely represents a more realistic density (e.g. see Figures \ref{ContoursCO43} and \ref{ContoursCOci}).  

We assume a $\mol{CO}$ abundance of $3\times 10^{-5}$ as determined in Section \ref{xXCO} and apply the same pixel incompleteness correction as we did for the mass. We calculate the total $\mol{H_2}$ cooling to be $\sim 4.9 \times 10^7 L_{\odot}$ from the $\mol{CO}$-only RADEX results, and $\sim 6.5 \times 10^7 L_{\odot}$ from the $\mol{CO}$ and $\mol{[CI]}$ RADEX results. \cite{brandl2009} measured a luminosity of $9.2\times 10^{6} L_{\odot}$ for the  $\mol{H_2}$ $\unit{S(3)}$ transition. Given that the luminosity of the $\unit{S(2)}$ and $\unit{S(1)}$ transitions are comparable to the $\unit{S(3)}$ transition \citep{brandl2009}, our calculated $\mol{H_2}$ cooling is reasonably consistent with this measurement. 

Next, we estimate the total cooling contribution from $\mol{CO}$ ($L_{\mol{CO}}$) by summing the total luminosity for all of our $\mol{CO}$ transitions. { Each individual transition contributes between $1 \%$ and $25 \%$ of the total $\mol{CO}$ cooling, with the $J=1-0$ transition contributing only $1\%$, and $J=4-3$ and $J=6-5$ transitions contributing $>20 \%$. The remaining 5 transitions each contribute between $6\%$ and $14\%$ to the total $\mol{CO}$ cooling. } In addition, we apply the same pixel incompleteness correction as before, and we calculate the contribution of $\mol{CO}$ to the overall cooling to be $L_{\mol{CO}}=1.8 \times 10^7 L_{\odot}$.  

In order to calculate the total contribution from $\mol{[CII]}$ to the total cooling budget, we must correct for the fraction of emission which arises from ionized gas. The ratio of $\mol{[CII]}$ to the $\mol{[NII]}$ transition at $1461 \unit{GHz}$ ($L_\mol{[CII]}/L_{\mol{[NII]}}$) provides a useful diagnostic for determining the contribution from ionized gas \citep{oberst2006}, as emission from $\mol{[NII]}$ arises entirely from ionized gas \citep{malhotra2001}. This ratio depends upon the ionized gas density, $n_e$, with the ratio varying from $\sim 2.4$ to $\sim 4.3$ \citep{oberst2006}, assuming solar abundances for $\mol{C^+}$ and $\mol{N^+}$. We assume $L_\mol{[CII]}/L_{\mol{[NII]}} \sim 3.5$ for the ionized gas as it is near the midpoint between the two extremes for the ratio, and correct the $\mol{[CII]}$ emission by assuming any excess in $L_\mol{[CII]}$ cools the molecular gas. This value is an upper limit as some of the $\mol{[CII]}$ emission will originate from atomic gas.  %Given that the ratio of $L_\mol{[CII]}/L_{\mol{[NII]}}$ does not vary by more than a factor of two, our assumed ratio allows us to calculate the correction to well within a factor of 2 of the true corrected value.

We calculate the ratio of $L_\mol{[CII]}/L_{\mol{[NII]}}$ by first convolving our $\mol{[CII]}$ map to the beam of the $\mol{[NII]}$ map. We approximate the SSW beam at $1461 \unit{GHz}$ as a $17''$ Gaussian and use the kernels from \cite{aniano2011}. We then align our $\mol{[CII]}$ map to the $3''$ pixel scale $\mol{[NII]}$ map, and calculate the ratio of $L_\mol{[CII]}/L_{\mol{[NII]}}$. We linearly interpolate any pixels in which we do not have measurements for $\mol{[NII]}$, due to the large beam size. We correct each pixel in our $\mol{[CII]}$ map before summing over the entire map. After this correction, we calculate that the contribution to the total molecular gas cooling from $\mol{[CII]}$ to be $L_{\mol{[CII]}} = 5.5 \times 10^7 L_\odot$. In comparison, without the correction for $\mol{[CII]}$ from ionized gas, the total $\mol{[CII]}$ luminosity is $L_{\mol{[CII]}} = 9.4\times10^7 L_\odot$ from the PACS observations.   

The total $\mol{[CII]}$ luminosity corresponds to only $\sim 23\%$ of the total $\mol{[CII]}$ luminosity calculated by \cite{nikola1998} using observations from the Kuiper Airborne Observatory (KAO). Their map, however, covered a region of $5' \times 5'$ which is significantly larger than the region mapped by PACS. Furthermore, \cite{nikola1998} compared their KAO observations to those from the Infrared Space Observatory (ISO) and found that the KAO flux is a factor of 2 larger across the same region. The KAO observations also have a calibration uncertainty of $30\%$. { Given the large uncertainties in these previous observations, we elect to estimate the total $\mol{[CII]}$ flux from the PACS observations.}  We estimate the missing flux in our $\mol{[CII]}$ flux by comparing the total PACS $160 \unit{\mu m}$ flux to the PACS $160 \unit{\mu m}$ flux in the region mapped in our $\mol{[CII]}$ observation. { The PACS $160 \unit{\mu m}$ was graciously provided by \cite{Klaas2010}, and covers a total region of approximately $8.5' \times 9.5'$ centered on the Antennae.} We estimate that only $\sim 70 \%$ of the total $\mol{[CII]}$ luminosity is in our PACS map, which gives us an ionized gas corrected luminosity of $L_{\mol{[CII]}} = 7.9 \times10^7 L_\odot$. Due to the uncertainty in the KAO flux, we use this corrected luminosity for the contribution of $\mol{[CII]}$ to the total cooling budget.

 We also calculate the total cooling due to $\unit{[OI]}63\unit{\mu m}$ to be $L_{\unit{[OI]}63} = 3.6 \times10^7 L_\odot$ from the PACS observations. The ratio of $L_{\mol{[CII]}}/L_{\unit{[OI]}63}$ is not constant, typically increasing further away from the nuclear regions of galaxies, as $\mol{[CII]}$ starts to dominate the cooling in more diffuse environments. As such, we do not correct the $\unit{[OI]}63\unit{\mu m}$ emission for missing flux when including it in our total cooling budget.  We estimate the total cooling budget for the molecular gas in the Antennae to be $\sim 2.0 \times 10^8 L_\odot$. Assuming a molecular gas mass of $1.5 \times 10^{10} M_{\odot}$ (see Section \ref{xXCO}), this would correspond to $\sim 0.01 L_{\odot}/M_{\odot}$, with $\mol{[CII]}$ dominating the cooling, followed by $\mol{H_2}$, $\unit{[OI]}63$ and $\mol{CO}$. In comparison, the cooling per unit mass in M82 and Arp220 is $\sim 3 L_{\odot}/M_{\odot}$ and $\sim 20 L_{\odot}/M_{\odot}$ respectively (see Section \ref{CompGal}).

%As for $L_{\mol{[CII]}}$, we correct for lack of coverage using the $160 \unit{\mu m}$ map and calculate a corrected luminosity of $L_{\unit{[OI]}63} = 5.0 \times10^7 L_\odot$.

\subsubsection{Mechanical heating}

We consider two forms of mechanical heating: turbulent heating \citep{bradford2005}, and supernova and stellar wind heating \citep{maloney1999}. Turbulent heating is caused by the turbulent motion of the molecular gas which can be be caused by a strong interaction or ongoing merger. We can calculate the energy per unit mass injected back into the Antennae using \citep{bradford2005}

\begin{equation}
\frac{L}{M} = 1.10 \left(\frac{\vm_{rms}}{25 \unit{km \ s^{-1}}} \right)^3 \left(\frac{1 \unit{pc}}{\Lambda_d} \right) \frac{L_{\odot}}{M_{\odot}}
\end{equation}
\noindent where $\vm_{rms}$ is the turbulent velocity and $\Lambda_d$ is the size scale.  We assume the turbulent heating rate is equal to our calculated cooling rate ($L/M \sim 0.01 {L_{\odot}}/{M_{\odot}}$), and for a size scale of $\Lambda_d = 1 \unit{pc}$ we calculate a turbulent velocity of $\vm_{rms} \sim 5 \unit{km \ s^{-1}}$. For a size scale of $\Lambda_d = 1000 \unit{pc}$ the corresponding turbulent velocity is $\vm_{rms} \sim 52 \unit{km \ s^{-1}}$. The line widths for the resolved SGMCs from \cite{wilson2003} are on the order of $10-50 \unit{km \ s^{-1}}$, which correspond to turbulent velocities on the order of $\vm_{rms} \sim 5-25 \unit{km \ s^{-1}}$ on a size scale of $1 \unit{kpc}$. This is comparable to the values calculated for a $1 \unit{pc}$ size scale, which is on the order of the Jeans length for our warm component.  Furthermore, it is comparable to velocities from simulations of extreme star-forming galaxies ($\sim 30-140 \unit{km \ s^{-1}}$, \citealt{downes1998}). Given that the Antennae is both undergoing an intense starburst \citep{hibbard1997} and is in the process of merging, a turbulent velocity of $\sim 5 \unit{km \ s^{-1}}$ is not unreasonable. Thus, turbulent velocity is a possible contributor to heating in the Antennae. 

The mechanical energy due to supernovae is \citep{maloney1999}
\begin{equation}
L_{SN} \sim 3\times10^{43} \left( \frac{\nu_{SN}}{1 \unit{yr^{-1}}} \right) \left(\frac{E_{SN}}{10^{51} \unit{erg}} \right) \unit{erg \ s^{-1}}
\end{equation}
\noindent where $\unit{\nu_{SN}}$ is the supernova rate and $E_{SN}$ is the energy released per supernova ($\sim 10^{51} \unit{erg}$). In the Antennae, the observed global supernova rate is $\unit{\nu_{SN}} \sim 0.2 - 0.3\unit{yr^{-1}}$ \citep{neff2000}.  This corresponds to a rate of $L_{SN} \sim (1.6 - 2.3) \times 10^{9} \ L_{\odot}$ for the energy released from supernova. If we assume that the contribution from stellar winds is comparable \citep{rangwala2011}, the total mechanical energy injected into the interstellar medium (ISM) from supernovae and stellar winds is $ (3.2-4.6)\times 10^{9} \ L_{\odot}$. Only $\sim 5\%$ of this energy would be required to balance the measured cooling rate of $\sim 2.0 \times 10^8 L_\odot$. This situation corresponds to a supernova heating efficiency of $0.05$. In comparison, in the Milky Way only $\sim 10\%$ of the total energy from supernovae is injected back into the surrounding ISM in the form of kinetic energy, which in turn contributes to both moving and heating the gas \citep{thornton1998}. 

{ By comparing the position of the nonthermal radio sources in \cite{neff2000}, along with their respective derived supernova rates, to the beams in Figure \ref{PACS70circles}, we estimate that $14\%$, $6\%$ and $66\%$ of the supernova originate from NGC 4038, NGC 4039 and the overlap region, respectively. In comparison, we estimate that $9\%$, $15\%$ and $54\%$ of the $\mol{[CII]}$ emission (corrected for the ionized gas fraction), which is the dominant coolant (see Section \ref{cooling}), originates from NGC 4038, NGC 4039 and the overlap region, respectively. The differences between the relative heating and cooling rates could be an indicator of a different balance between the varying sources of heating in the three regions. }

{ Globally,} supernovae and stellar winds are a possible source of heating in the Antennae. Given the turbulent nature of the molecular gas as a result of the ongoing merger, it is likely that both the merger induced turbulent motion as well as supernovae and stellar winds contribute to the heating, with their relative importance dependent on the local environment  within the Antennae.

%while the supernova feedback efficiency in M82 is $\sim 0.3-1$ \citep{strickland2009}.

\subsubsection{Photon dominated regions}\label{PDR}

Photon dominated regions (PDRs) are neutral regions located near the surfaces of molecular clouds which are irradiated by strong far-ultraviolet (FUV) radiation \citep{tielens1985}. The FUV photons are absorbed by dust grains and may liberate electrons through the photoelectric effect; the liberated electrons in turn heat the gas. The strength of the incident FUV field, $G_0$, is measured in units of the Habing interstellar radiation field, which is $1.3 \times 10^{-4} \unit{ergs \ cm^{-2} \ s^{-1} \ sr^{-1}}$ \citep{wolfire2010}, and is the strength of the local interstellar field. This FUV radiation will photo-dissociate the $\mol{CO}$ located near the edge of the molecular cloud where the FUV radiation is the strongest. Typically, massive, young, hot stars are the source of the FUV radiation, and as such can have a profound effect on the chemical and physical state of the entire molecular cloud. 

%presented in \cite{hollenbach2012}

We use PDR models (\citealt{hollenbach2012} and M. Wolfire, private communication) to interpret the observed $\mol{CO}$ SLED for the three regions in the Antennae (Table \ref{Lineflux}). These models consist of a grid of $\mol{CO}$ fluxes for transitions from $J=1-0$ to $J=29-28$ spanning a large range of densities ($n(\mol{H_2})=10^{1.0} \unit{cm^{-3}}$ to $10^{7.0} \unit{cm^{-3}}$) and incident FUV fluxes ($G_0 = 10^{-0.5}$ to $10^{6.5}$). Furthermore, these models typically assume that the FIR flux is a factor of two larger than the incident FUV flux. Using these models along with the densities calculated from our radiative transfer analysis, we can constrain the FUV field strength.  In this section, we model the ratio of two $\mol{CO}$ transitions ($J=3-2$ and $J=6-5$) to the FIR flux along with the ratios of numerous $\mol{CO}$ transitions to each other. 

We estimate the FIR luminosity ($L_\mol{FIR}$) by first calculating the total infrared luminosity ($L_\mol{TIR}$) using equation 4 from \cite{dale2002}. We acquired the Multiband Imaging Photometer for Spitzer (MIPS) $24\unit{\mu m}$ map from the Spitzer Space Telescope (\emph{Spitze}) archive, while we were graciously provided with the PACS $70\unit{\mu m}$ and $160 \unit{\mu m}$ photometric maps by \cite{Klaas2010}. All 3 maps are convolved to the $160 \mu m$ beam size ($12.13''$, PACS Observer's Manual version 2.4\footnote{Available from \url{http://herschel.esac.esa.int/Docs/PACS/html/pacs_om.html}}) using the appropriate convolution kernels and scripts from \cite{aniano2011}\footnote{Available from \url{http://www.astro.princeton.edu/~ganiano/Kernels.html}}. Next, we assume a ratio of $L_\mol{TIR}/L_\mol{FIR}\sim 2$ \citep{dale2001} and calculate a map of $L_\mol{FIR}$. { This $L_{\mol{FIR}}$ map is convolved to the $43''$ beam of the FTS by first convolving it to a $15''$ Gaussian beam using the appropriate kernel from \cite{aniano2011}, and then to the $43''$ FTS beam using the same kernel used to convolve the $\mol{CO}$ $J=3-2$ map.}

\begin{figure}[tp] %%%FIGURE PDR1
\centering
\includegraphics[width=0.75\linewidth]{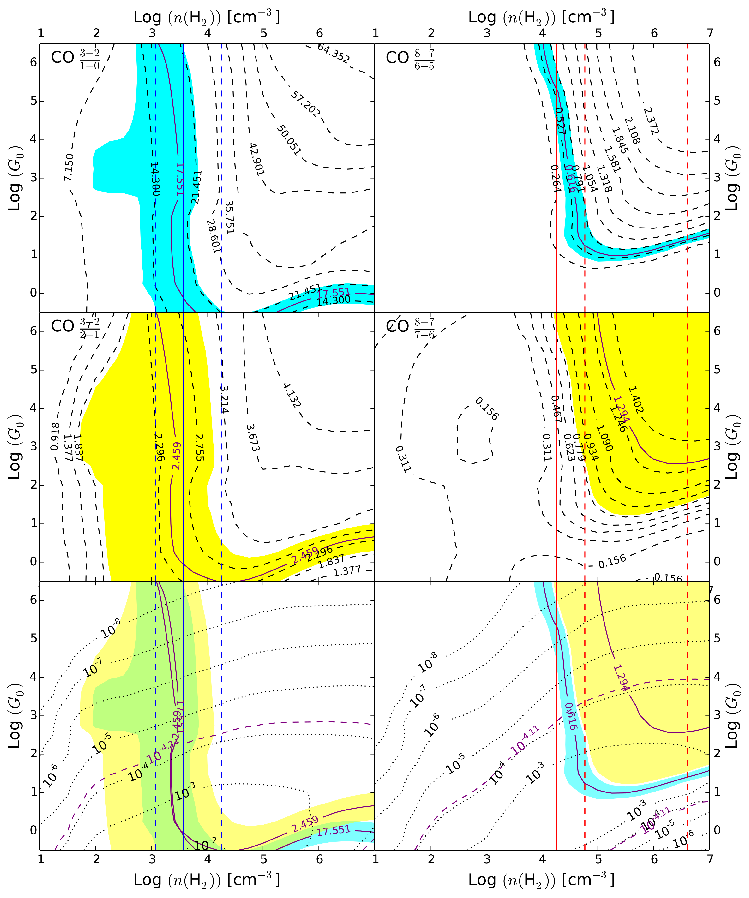} 
\caption[]{PDR models for NGC 4038 comparing $\mol{CO}$ line ratios to the field strength in units of the Habing field ($G_0$) and the gas density ($n_{\mol{H_2}}$).  Each of the  panels in the top two rows correspond to a different $\mol{CO}$ ratio: $J=3-2/1-0$ (\emph{top-left}), $J=3-2/2-1$ (\emph{middle-left}), $J=8-7/6-5$ (\emph{top-right}) and $J=8-7/7-6$ (\emph{middle-right}). The $J=3-2/1-0$ and $J=3-2/2-1$ ratios are combined in the bottom-left panel, while the $J=8-7/6-5$ and $J=8-7/7-6$ ratios are combined in the bottom-right panel.  The dashed black contours correspond to contours of constant ratios, while the purple contour corresponds to the measured ratio of the two lines. The blue and yellow shaded regions indicate the $1\sigma$ uncertainty range in the measured ratio. The blue and red solid lines correspond to the cold and warm component densities as determine from the $J_{break}=4$ radiative transfer solution for both $\mol{CO}$ and $\mol{[CI]}$ while the blue and red dashed lines correspond to the $1\sigma$ range to this density for the cold and warm components respectively. In the bottom panels, the dashed purple contours corresponds to the measured ratio of $\mol{CO}$ $J=3-2/$FIR (\emph{bottom-left}) and $\mol{CO}$ $J=6-5/$FIR (\emph{bottom-right}) in units of $\mol{erg\ cm^{-2}\ s^{-1}\ sr^{-1}}$, while the dotted black contours correspond to the model $\mol{CO}/$FIR values.}
\label{PDR4038Rat}
\end{figure}

\begin{figure}[tp] %%%FIGURE PDR3
\centering
\includegraphics[width=0.75\linewidth]{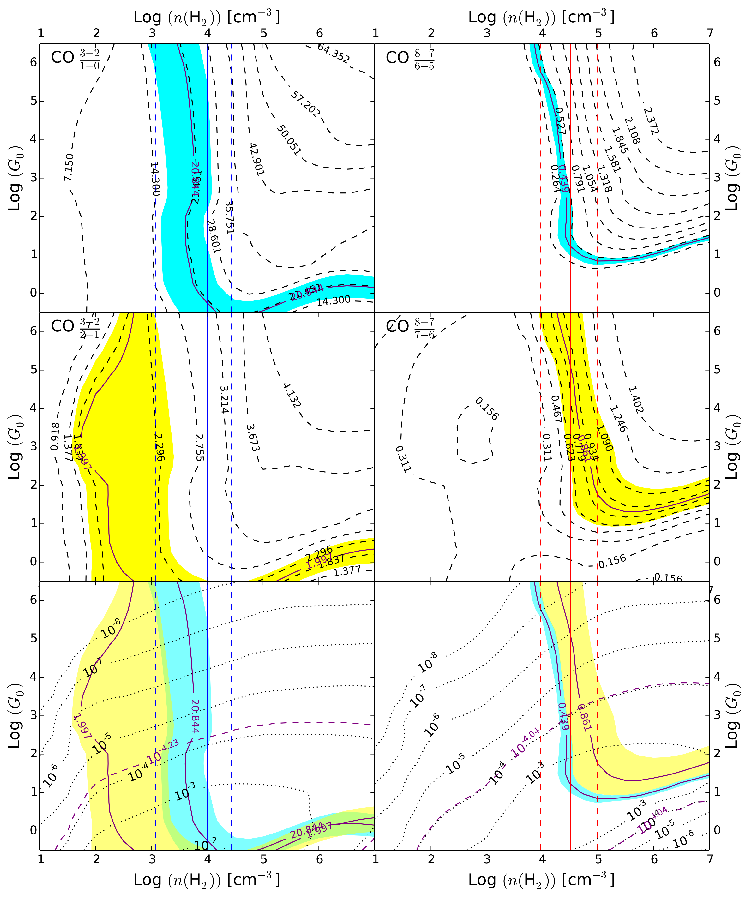} 
\caption[]{Same as figure \ref{PDR4038Rat} except for NGC 4039.}
\label{PDR4039Rat}
\end{figure}

\begin{figure}[tp] %%%FIGURE PDR3
\centering
\includegraphics[width=0.75\linewidth]{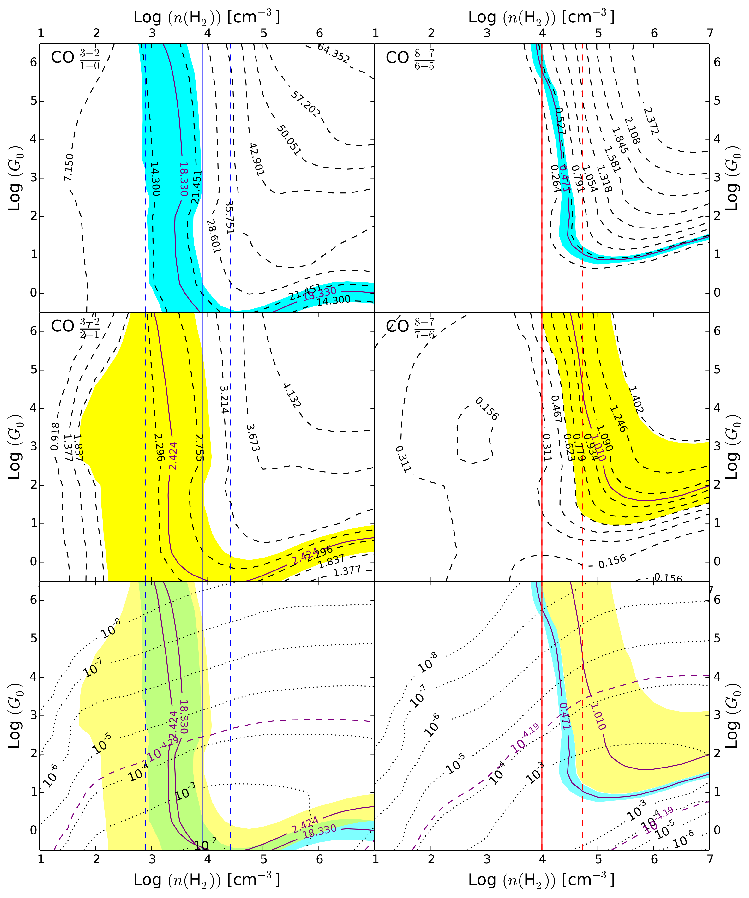} 
\caption[]{Same as figure \ref{PDR4038Rat} except for the overlap region. { Note that in the right column, the solid red line coincides with the lower dashed red line.}}
\label{PDROverlapRat}
\end{figure}

We calculate the ratios of $L_{\mol{CO}}/L_\mol{FIR}$ for the $\mol{CO}$ $J=3-2$ transition and $\mol{CO}$ $J=6-5$ transition for NGC 4038 (Figure \ref{PDR4038Rat} bottom), NGC 4039 (Figure \ref{PDR4039Rat} bottom), and the overlap region (Figure \ref{PDROverlapRat} bottom).  We further constrain the field strength by plotting various ratios of $\mol{CO}$ transitions for NGC 4038 (Figure \ref{PDR4038Rat}), NGC 4039 (Figure \ref{PDR4039Rat}) and the overlap region (Figure \ref{PDROverlapRat}). For all three regions, we plot the ratio of $\mol{CO}$ $\frac{3-2}{1-0}$ (top-left), $\frac{3-2}{2-1}$ (middle-left), $\frac{8-7}{6-5}$ (top-right), and $\frac{8-7}{7-6}$ (middle-right). We associate the $\mol{CO}$ $J=3-2$ transition ratios with the cold component and $J=6-5$ transition ratios with the warm component, and as such we compare these ratios to the densities of the cold ($J=3-2$) and warm ($J=6-5$) components from the $\mol{CO}$ and $\mol{[CI]}$ non-LTE radiative transfer solutions. In the bottom two panels of all 3 figures, we combine the cold (bottom-left) and warm (bottom-right) $\mol{CO}$ transitions with the corresponding $L_\mol{CO}/L_\mol{FIR}$ ratio. 

The results for all three regions are similar: the various ratios for the warm component are consistent with a field strength of $\mol{Log}(G_0)\sim 3$, while the various ratios for the cold component are consistent with a field strength of $\mol{Log}(G_0) \sim 2$. It is important to note that the ratios of $L_\mol{CO}/L_\mol{FIR}$ are lower limits as there will be contributions to $L_\mol{FIR}$ from both the cold and warm components. These lower limits correspond to upper limits in the FUV field strength $G_0$ (see bottom of Figure \ref{PDR4038Rat}, Figure \ref{PDR4039Rat} and Figure \ref{PDROverlapRat}). We are unable to constrain the relative contributions from the warm and cold components to the total FIR luminosity, and so all values for $G_0$ are upper limits. 

%These lower limits correspond to upper limits in $G_0$, since we are unable to constrain the relative contributions from the warm and cold components to the total FIR luminosity (see Figure \ref{CO32FIR} and Figure \ref{CO65FIR}). 

%However, given that the field strength of the cold component is an order of magnitude less than that of the warm component, we do not expect the warm component contributions to the FIR luminosity to change appreciably. As such, we treat the warm component results as the actual field strength, while we make no conclusions based on the cold component.

Given an FUV field strength of $\mol{Log} (G_0) = 3$ for our warm PDR models, and assuming a ratio of $L_\mol{FIR}/L_\mol{FUV}\sim2$, the corresponding FIR flux is $2.6 \times 10^{-1} \unit{ergs \ cm^{-2} \ s^{-1} \ sr^{-1}}$. The peak FIR flux, as estimated from our TIR map, in NGC 4038 is $4.9 \times 10^{-2} \unit{ergs \ cm^{-2} \ s^{-1} \ sr^{-1}}$, NGC 4039 is $2.6 \times 10^{-2} \unit{ergs \ cm^{-2} \ s^{-1} \ sr^{-1}}$ and the overlap region is $9.3 \times 10^{-2} \unit{ergs \ cm^{-2} \ s^{-1} \ sr^{-1}}$. In NGC 4039, the weakest of the three regions, our model PDRs would need to fill only $\sim 10\%$ of the $12''$ ($\sim 1.2 \unit{kpc}$) PACS beam in order to recover the measured peak flux. Given that the typical size scale of GMCs and stellar clusters is $10-100 \unit{pc}$, only a few model PDR regions are required to recover the measured FIR flux.  

In comparison, our cold PDR models have a FUV field strength of $\mol{Log} (G_0) = 2$, corresponding to a FIR flux of $2.6 \times 10^{-2} \unit{ergs \ cm^{-2} \ s^{-1} \ sr^{-1}}$. In NGC 4039, this would require that our model PDRs fill $\sim 100\%$ of the PACS beam. Given the face-on nature of the Antennae coupled with previous interferometric observations of $\mol{CO}$ (e.g. \citealt{wilson2003}), we do not expect PDRs to fill the $12 ''$ PACS beam and thus the warm PDRs must make a significant contribution to $L_{FIR}$. For example, if cold PDRs filled $30\%$ of the beam and so contributed $30\%$ of $L_{FIR}$, then the warm PDRs would have to account for the remaining $70\%$ of the far-infrared luminosity in NGC 4039.

%We might expect that cold PDRs fill up to $\sim 50\%$ of the beam, which in turn corresponds to a maximum of $50\%$ of the $L_{FIR}$ originating from the cold PDRs, with at least $50\%$ originating from warm PDRs. 

\cite{bayet2006} modeled various ratios of the $\mol{CO}$ $J=3-2$, $J=2-1$ and $J=6-5$ transitions, along with the $^{13}\mol{CO}$ $J=3-2$ transition with PDR models in NGC 4038 and the overlap region. They find a FUV field strength, in units of the Habing field, of $\mol{Log}(G_0) \sim 5.4$ and density of $n(\mol{H_2}) = 3.5 \times 10^5 \unit{cm^{-3}}$ for the overlap region, while for NGC 4038 they find that $\mol{Log}(G_0) \sim 5.6$ and $n(\mol{H_2}) = 3.5 \times 10^5 \unit{cm^{-3}}$. In both cases, their field strength does not lie within the fields strengths allowed by our solutions for both regions (Figure \ref{PDROverlapRat} and Figure \ref{PDR4038Rat}). Furthermore, our ratio of $L_\mol{CO}/L_\mol{FIR}$ for the $J=6-5$ transition would need to be two orders of magnitude smaller to recover such a field strength, even at very high densities (e.g. see bottom-right of Figure \ref{PDR4038Rat} and Figure \ref{PDROverlapRat}). Given that our ratio provides an upper limit on $G_0$, we can rule out their solutions. 

In comparison, \cite{schulz2007} modeled various ratios of the peak brightness of the $^{12}\mol{CO}$ $J=1-0$, $J=2-1$, and $J=3-2$ transitions, along with the $^{13}\mol{CO}$ $J=1-0$ and $J=2-1$ transitions. They apply their model to NGC 4038, NGC 4039 and the overlap region (their ``interaction region''). Their findings are consistent with ours: they are able to recover their line ratios with an FUV field strength equivalent to $\mol{Log}(G_0) \sim 3.2$ with densities of $n(\mol{H_2}) = 10^{4.5}$, $n(\mol{H_2}) = 10^{4.3}$ and $n(\mol{H_2}) = 10^{4.4}$ for NGC 4038, NGC 4039 and the overlap region. All of these densities lie either within our $1\sigma$ ranges for the respective cold and warm components, or lie near the boundary, further suggesting that our results are consistent with \cite{schulz2007}. 

In summary, we model the ratios of $\mol{CO}$ $J=3-2$ and $J=6-5$ transitions to the FIR emission, along with various ratios of different $\mol{CO}$ transitions with PDRs. By comparing our densities as calculated from our non-LTE radiative transfer analysis, we find a field strength of $\mol{Log}(G_0) \sim 3$ for PDRs in all three regions. Our field strength and densities for our PDR models are both significantly less than the values from \cite{bayet2006}, but are both consistent with the results from \cite{schulz2007}. Thus, PDRs remain as a possible source of significant heating throughout the Antennae. Further study using transitions from atomic species, such as $\mol{[CII]}$ and $\mol{[OI]}$, will be useful in further constraining not only the physical characteristics of the PDRs throughout the Antennae, but the location of these PDRs.

%Both \cite{bayet2006} and \cite{schulz2007} modeled the $\mol{CO}$ emission with PDRs in the overlap region. \cite{bayet2006} find that a FUV field strength of $1.5 \times 10^5 G_0$ with a density of $n(\mol{H_2}) = 3.5 \times 10^5 \unit{cm^{-3}}$.This model is unlikely to represent the true conditions within the overlap region as this field strength is higher than the FUV field suggested by our $\mol{[CII]}$ map. Furthermore, this combination of density and field strength fails to reproduce all of our measured line rations in figure \ref{PDROverlapRat} except for the $\mol{CO} \frac{8-7}{7-6}$ ratio. 

\subsection{Comparison to other galaxies}\label{CompGal}

The Antennae is the fourth system from the VNGS-FTS sample to be analyzed using a non-LTE radiative transfer analysis and is the only early-stage merger from our sample. Furthermore, of these four systems, it is the only one in which { large scale} structure is resolved in our $43''$ beam. Of the three previously studied galaxies, the Antennae has more similarities to M82 \citep{kamenetzky2012} and Arp 220 \citep{rangwala2011}. The third galaxy, NGC 1068, is a Seyfert type 2 galaxy, whose nuclear physical and chemical state is driven by an active galactic nucleus (AGN) \citep{spinoglio2012}.

M82 is a nearby galaxy ($3.4 \unit{Mpc}$) currently undergoing a starburst \citep{yun1993} due to a recent interaction with the nearby galaxy M81. This starburst has led to an enhanced star formation rate, and as a result an infrared brightness ($L_{IR} = 5.6 \times 10^{10} \unit{L_\odot}$, \citealt{sanders2003}) approaching that of a LIRG. Like in the Antennae, radiative transfer modeling of M82 found that there is both a cold ($\lesssim 100 \unit{K}$) and warm ($\sim 450 \unit{K}$) molecular gas, with the mass of the warm component ($\sim 1.5 \times 10^{6} M_{\odot}$) being on the order of $\sim 10\%$ of the mass of the cold component ($\sim 2 \times 10^{7} M_{\odot}$) \citep{kamenetzky2012}. Arp 220, on the other hand, is a nearby ($77 \unit{Mpc}$, \citealt{scoville1997}) ULIRG with an increased star formation rate that is the result of an ongoing merger in an advanced state.  As in M82 and the Antennae, both a cold ($T_{kin} \sim 50 \unit{K}$) and a warm ($\sim 1300 \unit{K}$) component are recovered from the radiative transfer analysis, albeit significantly warmer in Arp 220 \citep{rangwala2011}.  Similarly to M82, the warm gas mass ($\sim 4.7 \times 10^{8} M_{\odot}$) is about $\sim 10 \%$ that of the cold gas mass ($\sim 5.2 \times 10^{9} M_{\odot}$, \citealt{rangwala2011}). 

Both Arp 220 and M82 have a significantly higher warm gas mass fraction than in the Antennae, where we found a warm gas mass fraction of $\sim 0.2 \%$. This may be due to either Arp 220 and M82 having a larger source of heating or NGC 4038/39 cooling more efficiently. Evidence suggests Arp 220 hosts a central AGN \citep{clements2002,iwasawa2005}; however, the strongest candidates for heating are supernova and stellar winds, which contribute $\sim 200 L_{\odot}/M_{\odot}$ to the overall heating \citep{rangwala2011}. (This value does not account for supernova feedback efficiency.) The majority of the cooling is from $\mol{H_2}$ due to the high temperature of the warm molecular gas ($T_{kin} \sim 1300 \unit{K}$) and the cooling rate is $\sim 20 L_{\odot}/M_{\odot}$. In Arp 220, a supernova heating efficiency $\gtrsim 0.1$ is required to match the cooling; however Arp 220 is compact in comparison to the Antennae with the size of its molecular region only $\sim 400 \unit{pc}$. Therefore, the increase in both the temperature and mass fraction of the warm molecular gas in Arp 220 is likely a result of a larger amount of supernova and stellar wind energy being injected into the surrounding ISM, a higher supernova feedback efficiency and a larger difference between the heating and cooling rates. 

Similarly to Arp 220, turbulent motions due to supernovae and stellar winds are the strongest candidate for molecular gas heating in M82 \citep{panuzzo2010,kamenetzky2012}. The cooling rate in M82 is greater than in NGC 4038/39 by two orders of magnitude ($3 L_{\odot}/M_{\odot}$). The supernova rate in M82 is $\sim 0.09 \unit{yr^{-1}}$ \citep{fenech2010}, which corresponds to $L_{SN} \sim (7 \times 10^8 )L_{\odot}$, or $L_{SN}/M \sim 32 L_{\odot}/M_{\odot}$ assuming a molecular gas mass of $2.2\times 10^7 M_{\odot}$ \citep{kamenetzky2012}. A supernova feedback efficiency of $\sim 0.1$ is required to match the cooling in M82. As such, the higher warm gas fraction and temperature in M82 is possibly the result of a higher supernova heating rate, likely in part due to the increased supernova feedback efficiency.

%Given how efficient the supernova feedback is in M82, it is sensible to assume similar feedback efficiency (e.g. $>0.3$)

\section{Summary and conclusions} \label{conc}

In this paper, we present maps of the $\mol{CO}$ $J=4-3$ to $J=8-7$ and two $\mol{[CI]}$ transitions of the Antennae observed using the \emph{Herschel} SPIRE-FTS. We supplement the SPIRE-FTS maps with observations of $\mol{CO}$ $J=2-1$ and $J=3-2$ from the JCMT, $\mol{CO}$ $J=1-0$ from the NRO, and observations of $\mol{[CII]}$ and $\unit{[OI]}63 \unit{\mu m}$ from the \emph{Herschel} PACS spectrometer. 

\begin{enumerate}
\item We perform a local thermodynamic equilibrium analysis using the two observed $\mol{[CI]}$ transitions across the entire galaxy. We find that throughout the Antennae there is cold molecular gas with temperatures $\sim 10-30 \unit{K}$. Our non-local thermodynamic equilibrium radiative transfer analysis using both $\mol{CO}$ and $\mol{[CI]}$ transitions shows that the $\mol{[CI]}$ emission is optically thin, which suggests that $\mol{[CI]}$ is in local thermodynamic equilibrium. 

\item Using the non-local thermodynamic equilibrium radiative transfer code RADEX, we perform a likelihood analysis using our 8 $\mol{CO}$ transitions, both with and without the two $\mol{[CI]}$ transitions. We find that the molecular gas in the Antennae is in both a cold ($T_{kin} \sim 10-30 \unit{K}$) and a warm ($T_{kin} \gtrsim 100 \unit{K}$) state, with the warm molecular gas comprising only $\sim 0.2\%$ of the total molecular gas fraction in the Antennae. Furthermore, the physical state of the molecular gas does not vary substantially, with the pressure of both the warm and cold components being nearly constant within uncertainties and our angular resolution across the Antennae. %However, the pressure of the warm component is $2-3$ times the pressure in the cold component due to both the increased temperature and density of the molecular gas. 

\item By considering the contributions of $\mol{H_2}$, $\mol{[CII]}$, $\unit{[OI]}63\unit{\mu m}$ and $\mol{CO}$, we calculate a total cooling rate of $\sim 2.0 \times 10^8 L_{\odot}$ for the molecular gas, or $\sim 0.01 L_{\odot}/M_{\odot}$, with $\mol{[CII]}$ as the dominant coolant. The contributions calculated for $\mol{H_2}$ and $\unit{[OI]}63\unit{\mu m}$ are lower limits due to unconstrained temperatures from the radiative transfer analysis ($\mol{H_2}$) and limits in the size of the map ($\unit{[OI]}63\unit{\mu m}$). Furthermore, the contributions from $\mol{[CII]}$ is an upper limit as some of the $\unit{[CII]}$ emission likely originates from atomic gas. Mechanical heating is sufficient to match the total cooling and heat the molecular gas throughout the Antennae, with both turbulent heating due to the ongoing merger, and supernovae and stellar winds contributing to the mechanical heating.

\item  We model the ratio of the $\mol{CO}$ flux to the FIR flux for $\mol{CO}$ $J=3-2$ and $J=6-5$, along with the ratio of various $\mol{CO}$ lines in the nucleus of NGC 4038, the nucleus of NGC 4039 and the overlap region using models of photon dominated regions. Using the densities calculated from our non-LTE radiative transfer analysis, we find that a photon dominated region with a field strength of $G_0 \sim 1000$ can explain the warm component $\mol{CO}$ and $\mol{FIR}$ emission in all three regions. We also find that this field strength is consistent with the observed peak $\mol{FIR}$ flux in all three regions. These results are consistent with a previous study by \cite{schulz2007}. While photon dominated regions are not necessary to heat the molecular gas, they remain as a possible contributor in heating the molecular gas in the star forming regions of the Antennae. 

\item Both the warm gas fraction and temperature are smaller in NGC 4038/39 than in either Arp 220, or M82, both of which are likely heated by turbulent motion due to supernova and stellar winds. We suggest that this is due to increased supernova feedback efficiency in both Arp 220 and M82 due to their compactness. 

\item In the warm molecular gas, we calculate a $\mol{CO}$ abundance of $x_{\mol{CO}} \sim 3 \times 10^{-5}$, corresponding to a warm molecular gas mas of $\sim 2.2 \times 10^{7} M_{\odot}$. If we assume the same $\mol{CO}$ abundance in the cold molecular gas, this corresponds to a cold molecular gas mass of $1.5 \times 10^{10} M_{\odot}$ and a $\mol{CO}$ luminosity-to-mass conversion factor of $\alpha_{\mol{CO}} \sim 7 \ M_{\odot}\unit{pc^{-2} \ (K \ km \ s^{-1})^{-1}}$, comparable to the Milky Way value. This value is consistent with previous results for the Antennae \citep{wilson2003} where the luminosity-to-mass conversion factor was determined using the virial mass of resoled SGMCs.

\end{enumerate}

\acknowledgments

This research was supported by grants from the Canadian Space Agency and the Natural Sciences and Engineering Research Council of Canada (PI: C. D. Wilson). MPS has been funded by the Agenzia Spaziale Italiana (ASI) under contract I/005/11/0. PACS has been developed by a consortium of institutes led by MPE (Germany) and including UVIE (Austria); KU Leuven, CSL, IMEC (Belgium); CEA, LAM (France); MPIA (Germany); INAF-IFSI/OAA/OAP/OAT, LENS, SISSA (Italy); IAC
(Spain). This development has been supported by the funding agencies BMVIT (Austria), ESA-PRODEX (Belgium), CEA/CNES (France), DLR (Germany), ASI/INAF (Italy) and CICYT/MCYT (Spain). SPIRE has been developed by a consortium of institutes led by Cardiff University (UK) and including Univ. Lethbridge (Canada); NAOC (China); CEA, LAM (France); IFSI, Univ. Padua (Italy); IAC (Spain); Stockholm Observatory (Sweden); Imperial College London, RAL, UCL-MSSL, UKATC, Univ. Sussex (UK); and Caltech, JPL, NHSC, Univ. Colorado (USA). This development has been supported by national funding agencies: CSA
(Canada); NAOC (China); CEA, CNES, CNRS (France); ASI (Italy); MCINN
(Spain); SNSB (Sweden); STFC (UK); and NASA (USA). HIPE is a joint
development by the Herschel Science Ground Segment Consortium,
consisting of ESA, the NASA Herschel Science Center, and the HIFI,
PACS and SPIRE consortia. This research has made use of the NASA/IPAC
Extragalactic Database (NED) which is operated by the Jet Propulsion
Laboratory, California Institute of Technology, under contract with
the National Aeronautics and Space Administration. This research made use of the python plotting package matplotlib \citep{hunter2007}. This research made use of APLpy, an open-source plotting package for Python hosted at http://aplpy.github.com. We would like to thank Mark Wolfire for providing the PDR model grids used in this paper. IDL is a postdoctoral researcher of the FWO-Vlaanderen (Belgium).

\bibliographystyle{apj}
\bibliography{NGC4038_Schirm}

\begin{thebibliography}{}
\expandafter\ifx\csname natexlab\endcsname\relax\def\natexlab#1{#1}\fi

\bibitem[{{Aniano} {et~al.}(2011){Aniano}, {Draine}, {Gordon}, \&
  {Sandstrom}}]{aniano2011}
{Aniano}, G., {Draine}, B.~T., {Gordon}, K.~D., \& {Sandstrom}, K. 2011, \pasp,
  123, 1218

\bibitem[{{Bayet} {et~al.}(2006){Bayet}, {Gerin}, {Phillips}, \&
  {Contursi}}]{bayet2006}
{Bayet}, E., {Gerin}, M., {Phillips}, T.~G., \& {Contursi}, A. 2006, \aap, 460,
  467

\bibitem[{{Bendo} {et~al.}(2012){Bendo}, {Boselli}, {Dariush}, {Pohlen},
  {Roussel}, {Sauvage}, {Smith}, {Wilson}, {Baes}, {Cooray}, {Clements},
  {Cortese}, {Foyle}, {Galametz}, {Gomez}, {Lebouteiller}, {Lu}, {Madden},
  {Mentuch}, {O'Halloran}, {Page}, {Remy}, {Schulz}, \&
  {Spinoglio}}]{2012MNRAS.419.1833B}
{Bendo}, G.~J., {Boselli}, A., {Dariush}, A., {et~al.} 2012, \mnras, 419, 1833

\bibitem[{{Bradford} {et~al.}(2005){Bradford}, {Stacey}, {Nikola}, {Bolatto},
  {Jackson}, {Savage}, \& {Davidson}}]{bradford2005}
{Bradford}, C.~M., {Stacey}, G.~J., {Nikola}, T., {et~al.} 2005, \apj, 623, 866

\bibitem[{{Brandl} {et~al.}(2009){Brandl}, {Snijders}, {den Brok}, {Whelan},
  {Groves}, {van der Werf}, {Charmandaris}, {Smith}, {Armus}, {Kennicutt}, \&
  {Houck}}]{brandl2009}
{Brandl}, B.~R., {Snijders}, L., {den Brok}, M., {et~al.} 2009, \apj, 699, 1982

\bibitem[{{Clements} {et~al.}(2002){Clements}, {McDowell}, {Shaked}, {Baker},
  {Borne}, {Colina}, {Lamb}, \& {Mundell}}]{clements2002}
{Clements}, D.~L., {McDowell}, J.~C., {Shaked}, S., {et~al.} 2002, \apj, 581,
  974

\bibitem[{{Clements} {et~al.}(1996){Clements}, {Sutherland}, {McMahon}, \&
  {Saunders}}]{clements1996}
{Clements}, D.~L., {Sutherland}, W.~J., {McMahon}, R.~G., \& {Saunders}, W.
  1996, \mnras, 279, 477

\bibitem[{{Cluver} {et~al.}(2010){Cluver}, {Jarrett}, {Kraan-Korteweg},
  {Koribalski}, {Appleton}, {Melbourne}, {Emonts}, \& {Woudt}}]{cluver2010}
{Cluver}, M.~E., {Jarrett}, T.~H., {Kraan-Korteweg}, R.~C., {et~al.} 2010,
  \apj, 725, 1550

\bibitem[{{Currie} {et~al.}(2008){Currie}, {Draper}, {Berry}, {Jenness},
  {Cavanagh}, \& {Economou}}]{2008ASPC..394..650C}
{Currie}, M.~J., {Draper}, P.~W., {Berry}, D.~S., {et~al.} 2008, in
  Astronomical Society of the Pacific Conference Series, Vol. 394, Astronomical
  Data Analysis Software and Systems XVII, ed. R.~W. {Argyle}, P.~S.
  {Bunclark}, \& J.~R. {Lewis}, 650

\bibitem[{{Dale} \& {Helou}(2002)}]{dale2002}
{Dale}, D.~A., \& {Helou}, G. 2002, \apj, 576, 159

\bibitem[{{Dale} {et~al.}(2001){Dale}, {Helou}, {Contursi}, {Silbermann}, \&
  {Kolhatkar}}]{dale2001}
{Dale}, D.~A., {Helou}, G., {Contursi}, A., {Silbermann}, N.~A., \&
  {Kolhatkar}, S. 2001, \apj, 549, 215

\bibitem[{{Downes} \& {Solomon}(1998)}]{downes1998}
{Downes}, D., \& {Solomon}, P.~M. 1998, \apj, 507, 615

\bibitem[{{Fenech} {et~al.}(2010){Fenech}, {Beswick}, {Muxlow}, {Pedlar}, \&
  {Argo}}]{fenech2010}
{Fenech}, D., {Beswick}, R., {Muxlow}, T.~W.~B., {Pedlar}, A., \& {Argo}, M.~K.
  2010, \mnras, 408, 607

\bibitem[{{Fischer} {et~al.}(1996){Fischer}, {Shier}, {Luhman}, {Satyapal},
  {Smith}, {Stacey}, {Unger}, {Greenhouse}, {Spinoglio}, {Malkan}, {Lord},
  {Miles}, {Shure}, {Clegg}, {Ade}, {Armand}, {Burgdorf}, {Church}, {Davis},
  {di Giorgio}, {Ewart}, {Furniss}, {Glencross}, {Gry}, {Lim}, {Molinari},
  {Nguyen-Q-Rieu}, {Price}, {Sidher}, {Smith}, {Swinyard}, {Texier}, {Trams},
  \& {Wolfire}}]{1996A&A...315L..97F}
{Fischer}, J., {Shier}, L.~M., {Luhman}, M.~L., {et~al.} 1996, \aap, 315, L97

\bibitem[{{Fulton} {et~al.}(2010){Fulton}, {Baluteau}, {Bendo}, {Benielli},
  {Gastaud}, {Griffin}, {Guest}, {Imhof}, {Lim}, {Lu}, {Naylor}, {Panuzzo},
  {Polehampton}, {Schwartz}, {Surace}, {Swinyard}, \& {Xu}}]{fulton2010}
{Fulton}, T.~R., {Baluteau}, J., {Bendo}, G., {et~al.} 2010, in Society of
  Photo-Optical Instrumentation Engineers (SPIE) Conference Series, Vol. 7731,
  Society of Photo-Optical Instrumentation Engineers (SPIE) Conference Series

\bibitem[{{Gao} {et~al.}(2001){Gao}, {Lo}, {Lee}, \& {Lee}}]{gao2001}
{Gao}, Y., {Lo}, K.~Y., {Lee}, S.-W., \& {Lee}, T.-H. 2001, \apj, 548, 172

\bibitem[{{Gehrz} {et~al.}(1983){Gehrz}, {Sramek}, \& {Weedman}}]{gehrz1983}
{Gehrz}, R.~D., {Sramek}, R.~A., \& {Weedman}, D.~W. 1983, \apj, 267, 551

\bibitem[{{Genzel} {et~al.}(2010){Genzel}, {Tacconi}, {Gracia-Carpio},
  {Sternberg}, {Cooper}, {Shapiro}, {Bolatto}, {Bouch{\'e}}, {Bournaud},
  {Burkert}, {Combes}, {Comerford}, {Cox}, {Davis}, {Schreiber},
  {Garcia-Burillo}, {Lutz}, {Naab}, {Neri}, {Omont}, {Shapley}, \&
  {Weiner}}]{genzel2010}
{Genzel}, R., {Tacconi}, L.~J., {Gracia-Carpio}, J., {et~al.} 2010, \mnras,
  407, 2091

\bibitem[{{Griffin} {et~al.}(2010){Griffin}, {Abergel}, {Abreu}, {Ade},
  {Andr{\'e}}, {Augueres}, {Babbedge}, {Bae}, {Baillie}, {Baluteau}, {Barlow},
  {Bendo}, {Benielli}, {Bock}, {Bonhomme}, {Brisbin}, {Brockley-Blatt},
  {Caldwell}, {Cara}, {Castro-Rodriguez}, {Cerulli}, {Chanial}, {Chen},
  {Clark}, {Clements}, {Clerc}, {Coker}, {Communal}, {Conversi}, {Cox},
  {Crumb}, {Cunningham}, {Daly}, {Davis}, {de Antoni}, {Delderfield}, {Devin},
  {di Giorgio}, {Didschuns}, {Dohlen}, {Donati}, {Dowell}, {Dowell}, {Duband},
  {Dumaye}, {Emery}, {Ferlet}, {Ferrand}, {Fontignie}, {Fox}, {Franceschini},
  {Frerking}, {Fulton}, {Garcia}, {Gastaud}, {Gear}, {Glenn}, {Goizel},
  {Griffin}, {Grundy}, {Guest}, {Guillemet}, {Hargrave}, {Harwit}, {Hastings},
  {Hatziminaoglou}, {Herman}, {Hinde}, {Hristov}, {Huang}, {Imhof}, {Isaak},
  {Israelsson}, {Ivison}, {Jennings}, {Kiernan}, {King}, {Lange}, {Latter},
  {Laurent}, {Laurent}, {Leeks}, {Lellouch}, {Levenson}, {Li}, {Li},
  {Lilienthal}, {Lim}, {Liu}, {Lu}, {Madden}, {Mainetti}, {Marliani}, {McKay},
  {Mercier}, {Molinari}, {Morris}, {Moseley}, {Mulder}, {Mur}, {Naylor},
  {Nguyen}, {O'Halloran}, {Oliver}, {Olofsson}, {Olofsson}, {Orfei}, {Page},
  {Pain}, {Panuzzo}, {Papageorgiou}, {Parks}, {Parr-Burman}, {Pearce},
  {Pearson}, {P{\'e}rez-Fournon}, {Pinsard}, {Pisano}, {Podosek}, {Pohlen},
  {Polehampton}, {Pouliquen}, {Rigopoulou}, {Rizzo}, {Roseboom}, {Roussel},
  {Rowan-Robinson}, {Rownd}, {Saraceno}, {Sauvage}, {Savage}, {Savini},
  {Sawyer}, {Scharmberg}, {Schmitt}, {Schneider}, {Schulz}, {Schwartz},
  {Shafer}, {Shupe}, {Sibthorpe}, {Sidher}, {Smith}, {Smith}, {Smith},
  {Spencer}, {Stobie}, {Sudiwala}, {Sukhatme}, {Surace}, {Stevens}, {Swinyard},
  {Trichas}, {Tourette}, {Triou}, {Tseng}, {Tucker}, {Turner}, {Vaccari},
  {Valtchanov}, {Vigroux}, {Virique}, {Voellmer}, {Walker}, {Ward}, {Waskett},
  {Weilert}, {Wesson}, {White}, {Whitehouse}, {Wilson}, {Winter}, {Woodcraft},
  {Wright}, {Xu}, {Zavagno}, {Zemcov}, {Zhang}, \& {Zonca}}]{griffin2010}
{Griffin}, M.~J., {Abergel}, A., {Abreu}, A., {et~al.} 2010, \aap, 518, L3

\bibitem[{{Herrera} {et~al.}(2012){Herrera}, {Boulanger}, {Nesvadba}, \&
  {Falgarone}}]{herrera2012}
{Herrera}, C.~N., {Boulanger}, F., {Nesvadba}, N.~P.~H., \& {Falgarone}, E.
  2012, \aap, 538, L9

\bibitem[{{Hibbard}(1997)}]{hibbard1997}
{Hibbard}, J.~E. 1997, in American Institute of Physics Conference Series, Vol.
  393, American Institute of Physics Conference Series, ed. {S.~S.~Holt \&
  L.~G.~Mundy}, 259--270

\bibitem[{{Hollenbach} {et~al.}(2012){Hollenbach}, {Kaufman}, {Neufeld},
  {Wolfire}, \& {Goicoechea}}]{hollenbach2012}
{Hollenbach}, D., {Kaufman}, M.~J., {Neufeld}, D., {Wolfire}, M., \&
  {Goicoechea}, J.~R. 2012, \apj, 754, 105

\bibitem[{Hunter(2007)}]{hunter2007}
Hunter, J.~D. 2007, Computing In Science \& Engineering, 9, 90

\bibitem[{{Ikeda} {et~al.}(2002){Ikeda}, {Oka}, {Tatematsu}, {Sekimoto}, \&
  {Yamamoto}}]{2002ApJS..139..467I}
{Ikeda}, M., {Oka}, T., {Tatematsu}, K., {Sekimoto}, Y., \& {Yamamoto}, S.
  2002, \apjs, 139, 467

\bibitem[{{Iwasawa} {et~al.}(2005){Iwasawa}, {Sanders}, {Evans}, {Trentham},
  {Miniutti}, \& {Spoon}}]{iwasawa2005}
{Iwasawa}, K., {Sanders}, D.~B., {Evans}, A.~S., {et~al.} 2005, \mnras, 357,
  565

\bibitem[{{Joseph} \& {Wright}(1985)}]{joseph1985}
{Joseph}, R.~D., \& {Wright}, G.~S. 1985, \mnras, 214, 87

\bibitem[{{Kamenetzky} {et~al.}(2011){Kamenetzky}, {Glenn}, {Maloney},
  {Aguirre}, {Bock}, {Bradford}, {Earle}, {Inami}, {Matsuhara}, {Murphy},
  {Naylor}, {Nguyen}, \& {Zmuidzinas}}]{kamenetzky2011}
{Kamenetzky}, J., {Glenn}, J., {Maloney}, P.~R., {et~al.} 2011, \apj, 731, 83

\bibitem[{{Kamenetzky} {et~al.}(2012){Kamenetzky}, {Glenn}, {Rangwala},
  {Maloney}, {Bradford}, {Wilson}, {Bendo}, {Baes}, {Boselli}, {Cooray},
  {Isaak}, {Lebouteiller}, {Madden}, {Panuzzo}, {Schirm}, {Spinoglio}, \&
  {Wu}}]{kamenetzky2012}
{Kamenetzky}, J., {Glenn}, J., {Rangwala}, N., {et~al.} 2012, \apj, 753, 70

\bibitem[{{Klaas} {et~al.}(2010){Klaas}, {Nielbock}, {Haas}, {Krause}, \&
  {Schreiber}}]{Klaas2010}
{Klaas}, U., {Nielbock}, M., {Haas}, M., {Krause}, O., \& {Schreiber}, J. 2010,
  \aap, 518, L44

\bibitem[{{Le Bourlot} {et~al.}(1999){Le Bourlot}, {Pineau des For{\^e}ts}, \&
  {Flower}}]{lebourlot1999}
{Le Bourlot}, J., {Pineau des For{\^e}ts}, G., \& {Flower}, D.~R. 1999, \mnras,
  305, 802

\bibitem[{{Leroy} {et~al.}(2011){Leroy}, {Bolatto}, {Gordon}, {Sandstrom},
  {Gratier}, {Rosolowsky}, {Engelbracht}, {Mizuno}, {Corbelli}, {Fukui}, \&
  {Kawamura}}]{leroy2011}
{Leroy}, A.~K., {Bolatto}, A., {Gordon}, K., {et~al.} 2011, \apj, 737, 12

\bibitem[{{Malhotra} {et~al.}(2001){Malhotra}, {Kaufman}, {Hollenbach},
  {Helou}, {Rubin}, {Brauher}, {Dale}, {Lu}, {Lord}, {Stacey}, {Contursi},
  {Hunter}, \& {Dinerstein}}]{malhotra2001}
{Malhotra}, S., {Kaufman}, M.~J., {Hollenbach}, D., {et~al.} 2001, \apj, 561,
  766

\bibitem[{{Maloney}(1999)}]{maloney1999}
{Maloney}, P.~R. 1999, \apss, 266, 207

\bibitem[{{Mao} {et~al.}(2000){Mao}, {Henkel}, {Schulz}, {Zielinsky},
  {Mauersberger}, {St{\"o}rzer}, {Wilson}, \& {Gensheimer}}]{mao2000}
{Mao}, R.~Q., {Henkel}, C., {Schulz}, A., {et~al.} 2000, \aap, 358, 433

\bibitem[{{Meijerink} {et~al.}(2013){Meijerink}, {Kristensen}, {Wei{\ss}}, {van
  der Werf}, {Walter}, {Spaans}, {Loenen}, {Fischer}, {Israel}, {Isaak},
  {Papadopoulos}, {Aalto}, {Armus}, {Charmandaris}, {Dasyra}, {Diaz-Santos},
  {Evans}, {Gao}, {Gonz{\'a}lez-Alfonso}, {G{\"u}sten}, {Henkel}, {Kramer},
  {Lord}, {Mart{\'{\i}}n-Pintado}, {Naylor}, {Sanders}, {Smith}, {Spinoglio},
  {Stacey}, {Veilleux}, \& {Wiedner}}]{meijerink2013}
{Meijerink}, R., {Kristensen}, L.~E., {Wei{\ss}}, A., {et~al.} 2013, \apjl,
  762, L16

\bibitem[{{Naylor} {et~al.}(2010{\natexlab{a}}){Naylor}, {Bradford}, {Aguirre},
  {Bock}, {Earle}, {Glenn}, {Inami}, {Kamenetzky}, {Maloney}, {Matsuhara},
  {Nguyen}, \& {Zmuidzinas}}]{naylor2010a}
{Naylor}, B.~J., {Bradford}, C.~M., {Aguirre}, J.~E., {et~al.}
  2010{\natexlab{a}}, \apj, 722, 668

\bibitem[{{Naylor} {et~al.}(2010{\natexlab{b}}){Naylor}, {Baluteau}, {Barlow},
  {Benielli}, {Ferlet}, {Fulton}, {Griffin}, {Grundy}, {Imhof}, {Jones},
  {King}, {Leeks}, {Lim}, {Lu}, {Makiwa}, {Polehampton}, {Savini}, {Sidher},
  {Spencer}, {Surace}, {Swinyard}, \& {Wesson}}]{naylor2010}
{Naylor}, D.~A., {Baluteau}, J.-P., {Barlow}, M.~J., {et~al.}
  2010{\natexlab{b}}, in Presented at the Society of Photo-Optical
  Instrumentation Engineers (SPIE) Conference, Vol. 7731, Society of
  Photo-Optical Instrumentation Engineers (SPIE) Conference Series

\bibitem[{{Neff} \& {Ulvestad}(2000)}]{neff2000}
{Neff}, S.~G., \& {Ulvestad}, J.~S. 2000, \aj, 120, 670

\bibitem[{{Nikola} {et~al.}(1998){Nikola}, {Genzel}, {Herrmann}, {Madden},
  {Poglitsch}, {Geis}, {Townes}, \& {Stacey}}]{nikola1998}
{Nikola}, T., {Genzel}, R., {Herrmann}, F., {et~al.} 1998, \apj, 504, 749

\bibitem[{{Oberst} {et~al.}(2006){Oberst}, {Parshley}, {Stacey}, {Nikola},
  {L{\"o}hr}, {Harnett}, {Tothill}, {Lane}, {Stark}, \& {Tucker}}]{oberst2006}
{Oberst}, T.~E., {Parshley}, S.~C., {Stacey}, G.~J., {et~al.} 2006, \apjl, 652,
  L125

\bibitem[{{Panuzzo} {et~al.}(2010){Panuzzo}, {Rangwala}, {Rykala}, {Isaak},
  {Glenn}, {Wilson}, {Auld}, {Baes}, {Barlow}, {Bendo}, {Bock}, {Boselli},
  {Bradford}, {Buat}, {Castro-Rodr{\'{\i}}guez}, {Chanial}, {Charlot},
  {Ciesla}, {Clements}, {Cooray}, {Cormier}, {Cortese}, {Davies}, {Dwek},
  {Eales}, {Elbaz}, {Fulton}, {Galametz}, {Galliano}, {Gear}, {Gomez},
  {Griffin}, {Hony}, {Levenson}, {Lu}, {Madden}, {O'Halloran}, {Okumura},
  {Oliver}, {Page}, {Papageorgiou}, {Parkin}, {P{\'e}rez-Fournon}, {Pohlen},
  {Polehampton}, {Rigby}, {Roussel}, {Sacchi}, {Sauvage}, {Schulz}, {Schirm},
  {Smith}, {Spinoglio}, {Stevens}, {Srinivasan}, {Symeonidis}, {Swinyard},
  {Trichas}, {Vaccari}, {Vigroux}, {Wozniak}, {Wright}, \&
  {Zeilinger}}]{panuzzo2010}
{Panuzzo}, P., {Rangwala}, N., {Rykala}, A., {et~al.} 2010, \aap, 518, L37

\bibitem[{{Papadopoulos} {et~al.}(2004){Papadopoulos}, {Thi}, \&
  {Viti}}]{papadopoulos2004}
{Papadopoulos}, P.~P., {Thi}, W.-F., \& {Viti}, S. 2004, \mnras, 351, 147

\bibitem[{{Papadopoulos} {et~al.}(2012){Papadopoulos}, {van der Werf},
  {Xilouris}, {Isaak}, \& {Gao}}]{papadopoulos2012}
{Papadopoulos}, P.~P., {van der Werf}, P., {Xilouris}, E., {Isaak}, K.~G., \&
  {Gao}, Y. 2012, \apj, 751, 10

\bibitem[{{Parkin} {et~al.}(2012){Parkin}, {Wilson}, {Foyle}, {Baes}, {Bendo},
  {Boselli}, {Boquien}, {Cooray}, {Cormier}, {Davies}, {Eales}, {Galametz},
  {Gomez}, {Lebouteiller}, {Madden}, {Mentuch}, {Page}, {Pohlen}, {Remy},
  {Roussel}, {Sauvage}, {Smith}, \& {Spinoglio}}]{2012MNRAS.422.2291P}
{Parkin}, T.~J., {Wilson}, C.~D., {Foyle}, K., {et~al.} 2012, \mnras, 422, 2291

\bibitem[{{Pereira-Santaella} {et~al.}(2013){Pereira-Santaella}, {Spinoglio},
  {Busquet}, {Wilson}, {Glenn}, {Isaak}, {Kamenetzky}, {Rangwala}, {Schirm},
  {Baes}, {Barlow}, {Boselli}, {Cooray}, \& {Cormier}}]{pereirasantaella2013}
{Pereira-Santaella}, M., {Spinoglio}, L., {Busquet}, G., {et~al.} 2013, \apj,
  768, 55

\bibitem[{{Pilbratt} {et~al.}(2010){Pilbratt}, {Riedinger}, {Passvogel},
  {Crone}, {Doyle}, {Gageur}, {Heras}, {Jewell}, {Metcalfe}, {Ott}, \&
  {Schmidt}}]{pilbratt2010}
{Pilbratt}, G.~L., {Riedinger}, J.~R., {Passvogel}, T., {et~al.} 2010, \aap,
  518, L1

\bibitem[{{Rangwala} {et~al.}(2011){Rangwala}, {Maloney}, {Glenn}, {Wilson},
  {Rykala}, {Isaak}, {Baes}, {Bendo}, {Boselli}, {Bradford}, {Clements},
  {Cooray}, {Fulton}, {Imhof}, {Kamenetzky}, {Madden}, {Mentuch}, {Sacchi},
  {Sauvage}, {Schirm}, {Smith}, {Spinoglio}, \& {Wolfire}}]{rangwala2011}
{Rangwala}, N., {Maloney}, P.~R., {Glenn}, J., {et~al.} 2011, ArXiv e-prints,
  arXiv:1106.5054

\bibitem[{{Read} {et~al.}(1995){Read}, {Ponman}, \& {Wolstencroft}}]{read1995}
{Read}, A.~M., {Ponman}, T.~J., \& {Wolstencroft}, R.~D. 1995, \mnras, 277, 397

\bibitem[{{Sanders} {et~al.}(2003){Sanders}, {Mazzarella}, {Kim}, {Surace}, \&
  {Soifer}}]{sanders2003}
{Sanders}, D.~B., {Mazzarella}, J.~M., {Kim}, D.-C., {Surace}, J.~A., \&
  {Soifer}, B.~T. 2003, \aj, 126, 1607

\bibitem[{{Sanders} \& {Mirabel}(1996)}]{sanders1996}
{Sanders}, D.~B., \& {Mirabel}, I.~F. 1996, \araa, 34, 749

\bibitem[{{Sch{\"o}ier} {et~al.}(2005){Sch{\"o}ier}, {van der Tak}, {van
  Dishoeck}, \& {Black}}]{schoier2005}
{Sch{\"o}ier}, F.~L., {van der Tak}, F.~F.~S., {van Dishoeck}, E.~F., \&
  {Black}, J.~H. 2005, \aap, 432, 369

\bibitem[{{Schroder} {et~al.}(1991){Schroder}, {Staemmler}, {Smith}, {Flower},
  \& {Jaquet}}]{schroeder1991}
{Schroder}, K., {Staemmler}, V., {Smith}, M.~D., {Flower}, D.~R., \& {Jaquet},
  R. 1991, Journal of Physics B Atomic Molecular Physics, 24, 2487

\bibitem[{{Schulz} {et~al.}(2007){Schulz}, {Henkel}, {Muders}, {Mao},
  {R{\"o}llig}, \& {Mauersberger}}]{schulz2007}
{Schulz}, A., {Henkel}, C., {Muders}, D., {et~al.} 2007, \aap, 466, 467

\bibitem[{{Schweizer} {et~al.}(2008){Schweizer}, {Burns}, {Madore}, {Mager},
  {Phillips}, {Freedman}, {Boldt}, {Contreras}, {Folatelli}, {Gonz{\'a}lez},
  {Hamuy}, {Krzeminski}, {Morrell}, {Persson}, {Roth}, \&
  {Stritzinger}}]{schweizer2008}
{Schweizer}, F., {Burns}, C.~R., {Madore}, B.~F., {et~al.} 2008, \aj, 136, 1482

\bibitem[{{Scoville} {et~al.}(1997){Scoville}, {Yun}, \&
  {Bryant}}]{scoville1997}
{Scoville}, N.~Z., {Yun}, M.~S., \& {Bryant}, P.~M. 1997, \apj, 484, 702

\bibitem[{{Sliwa} {et~al.}(2012){Sliwa}, {Wilson}, {Petitpas}, {Armus},
  {Juvela}, {Matsushita}, {Peck}, \& {Yun}}]{sliwa2012}
{Sliwa}, K., {Wilson}, C.~D., {Petitpas}, G.~R., {et~al.} 2012, \apj, 753, 46

\bibitem[{{Spinoglio} {et~al.}(2012){Spinoglio}, {Pereira-Santaella},
  {Busquet}, {Schirm}, {Wilson}, {Glenn}, {Kamenetzky}, {Rangwala}, {Maloney},
  {Parkin}, {Bendo}, {Madden}, {Wolfire}, {Boselli}, {Cooray}, \&
  {Page}}]{spinoglio2012}
{Spinoglio}, L., {Pereira-Santaella}, M., {Busquet}, G., {et~al.} 2012, ArXiv
  e-prints, arXiv:1208.6132

\bibitem[{{Stanford} {et~al.}(1990){Stanford}, {Sargent}, {Sanders}, \&
  {Scoville}}]{stanford1990}
{Stanford}, S.~A., {Sargent}, A.~I., {Sanders}, D.~B., \& {Scoville}, N.~Z.
  1990, \apj, 349, 492

\bibitem[{{Swinyard} {et~al.}(2010){Swinyard}, {Ade}, {Baluteau}, {Aussel},
  {Barlow}, {Bendo}, {Benielli}, {Bock}, {Brisbin}, {Conley}, {Conversi},
  {Dowell}, {Dowell}, {Ferlet}, {Fulton}, {Glenn}, {Glauser}, {Griffin},
  {Griffin}, {Guest}, {Imhof}, {Isaak}, {Jones}, {King}, {Leeks}, {Levenson},
  {Lim}, {Lu}, {Makiwa}, {Naylor}, {Nguyen}, {Oliver}, {Panuzzo},
  {Papageorgiou}, {Pearson}, {Pohlen}, {Polehampton}, {Pouliquen},
  {Rigopoulou}, {Ronayette}, {Roussel}, {Rykala}, {Savini}, {Schulz},
  {Schwartz}, {Shupe}, {Sibthorpe}, {Sidher}, {Smith}, {Spencer}, {Trichas},
  {Triou}, {Valtchanov}, {Wesson}, {Woodcraft}, {Xu}, {Zemcov}, \&
  {Zhang}}]{swinyard2010}
{Swinyard}, B.~M., {Ade}, P., {Baluteau}, J.-P., {et~al.} 2010, \aap, 518, L4

\bibitem[{{Thornton} {et~al.}(1998){Thornton}, {Gaudlitz}, {Janka}, \&
  {Steinmetz}}]{thornton1998}
{Thornton}, K., {Gaudlitz}, M., {Janka}, H.-T., \& {Steinmetz}, M. 1998, \apj,
  500, 95

\bibitem[{{Tielens} \& {Hollenbach}(1985)}]{tielens1985}
{Tielens}, A.~G.~G.~M., \& {Hollenbach}, D. 1985, \apj, 291, 722

\bibitem[{{Toomre} \& {Toomre}(1972)}]{toomre1972}
{Toomre}, A., \& {Toomre}, J. 1972, \apj, 178, 623

\bibitem[{{Ueda} {et~al.}(2012){Ueda}, {Iono}, {Petitpas}, {Yun}, {Ho},
  {Kawabe}, {Mao}, {Mart{\'{\i}}n}, {Matsushita}, {Peck}, {Tamura}, {Wang},
  {Wang}, {Wilson}, \& {Zhang}}]{2012ApJ...745...65U}
{Ueda}, J., {Iono}, D., {Petitpas}, G., {et~al.} 2012, \apj, 745, 65

\bibitem[{{van der Tak} {et~al.}(2007){van der Tak}, {Black}, {Sch{\"o}ier},
  {Jansen}, \& {van Dishoeck}}]{vandertak2007}
{van der Tak}, F.~F.~S., {Black}, J.~H., {Sch{\"o}ier}, F.~L., {Jansen}, D.~J.,
  \& {van Dishoeck}, E.~F. 2007, \aap, 468, 627

\bibitem[{{van der Werf} {et~al.}(2010){van der Werf}, {Isaak}, {Meijerink},
  {Spaans}, {Rykala}, {Fulton}, {Loenen}, {Walter}, {Wei{\ss}}, {Armus},
  {Fischer}, {Israel}, {Harris}, {Veilleux}, {Henkel}, {Savini}, {Lord},
  {Smith}, {Gonz{\'a}lez-Alfonso}, {Naylor}, {Aalto}, {Charmandaris}, {Dasyra},
  {Evans}, {Gao}, {Greve}, {G{\"u}sten}, {Kramer}, {Mart{\'{\i}}n-Pintado},
  {Mazzarella}, {Papadopoulos}, {Sanders}, {Spinoglio}, {Stacey}, {Vlahakis},
  {Wiedner}, \& {Xilouris}}]{vanderwerf2010}
{van der Werf}, P.~P., {Isaak}, K.~G., {Meijerink}, R., {et~al.} 2010, \aap,
  518, L42

\bibitem[{{Ward} {et~al.}(2003){Ward}, {Zmuidzinas}, {Harris}, \&
  {Isaak}}]{ward2003}
{Ward}, J.~S., {Zmuidzinas}, J., {Harris}, A.~I., \& {Isaak}, K.~G. 2003, \apj,
  587, 171

\bibitem[{{Warren} {et~al.}(2010){Warren}, {Wilson}, {Israel}, {Serjeant},
  {Bendo}, {Brinks}, {Clements}, {Irwin}, {Knapen}, {Leech}, {Matthews},
  {M{\"u}hle}, {Mortimer}, {Petitpas}, {Sinukoff}, {Spekkens}, {Tan},
  {Tilanus}, {Usero}, {van der Werf}, {Vlahakis}, {Wiegert}, \&
  {Zhu}}]{2010ApJ...714..571W}
{Warren}, B.~E., {Wilson}, C.~D., {Israel}, F.~P., {et~al.} 2010, \apj, 714,
  571

\bibitem[{{Wei} {et~al.}(2012){Wei}, {Keto}, \& {Ho}}]{2012ApJ...750..136W}
{Wei}, L.~H., {Keto}, E., \& {Ho}, L.~C. 2012, \apj, 750, 136

\bibitem[{{Whitmore} \& {Schweizer}(1995)}]{whitmore1995}
{Whitmore}, B.~C., \& {Schweizer}, F. 1995, \aj, 109, 960

\bibitem[{{Whitmore} {et~al.}(1999){Whitmore}, {Zhang}, {Leitherer}, {Fall},
  {Schweizer}, \& {Miller}}]{whitmore1999}
{Whitmore}, B.~C., {Zhang}, Q., {Leitherer}, C., {et~al.} 1999, \aj, 118, 1551

\bibitem[{{Whitmore} {et~al.}(2010){Whitmore}, {Chandar}, {Schweizer},
  {Rothberg}, {Leitherer}, {Rieke}, {Rieke}, {Blair}, {Mengel}, \&
  {Alonso-Herrero}}]{whitmore2010}
{Whitmore}, B.~C., {Chandar}, R., {Schweizer}, F., {et~al.} 2010, \aj, 140, 75

\bibitem[{{Wilson} {et~al.}(2000){Wilson}, {Scoville}, {Madden}, \&
  {Charmandaris}}]{wilson2000}
{Wilson}, C.~D., {Scoville}, N., {Madden}, S.~C., \& {Charmandaris}, V. 2000,
  \apj, 542, 120

\bibitem[{{Wilson} {et~al.}(2003){Wilson}, {Scoville}, {Madden}, \&
  {Charmandaris}}]{wilson2003}
---. 2003, \apj, 599, 1049

\bibitem[{{Wolfire}(2010)}]{wolfire2010}
{Wolfire}, M.~G. 2010, \apss, 380

\bibitem[{{Yun} {et~al.}(1993){Yun}, {Ho}, \& {Lo}}]{yun1993}
{Yun}, M.~S., {Ho}, P.~T.~P., \& {Lo}, K.~Y. 1993, \apjl, 411, L17

\bibitem[{{Zhang} {et~al.}(2010){Zhang}, {Gao}, \& {Kong}}]{zhang2010}
{Zhang}, H.-X., {Gao}, Y., \& {Kong}, X. 2010, \mnras, 401, 1839

\bibitem[{{Zhu} {et~al.}(2003){Zhu}, {Seaquist}, \& {Kuno}}]{zhu2003}
{Zhu}, M., {Seaquist}, E.~R., \& {Kuno}, N. 2003, \apj, 588, 243

\end{thebibliography}

\begin{deluxetable}{ccccccccc}
\tablecolumns{6}
\tabletypesize{\footnotesize}
\tablewidth{0pt}
\tablecaption{Line flux measurements}
\tablehead{\colhead{Line} &
	\colhead{Rest frequency} &
	\colhead{NGC 4038} &
	\colhead{Overlap region} & 
	\colhead{NGC 4039} & 
	\colhead{Calibration uncertainty} \\
	\colhead{} & 
	\colhead{$\unit{GHz}$} &
%	\colhead{$\unit{[Jy \ km/s]}$} &
	\colhead{$\unit{[K \ km \ s^{-1}]}$} &
%	\colhead{$\unit{[Jy \ km/s]}$} &
	\colhead{$\unit{[K \ km \ s^{-1}]}$} &
%	\colhead{$\unit{[Jy \ km/s]}$} &
	\colhead{$\unit{[K \ km \ s^{-1}]}$} &
	\colhead{$\%$}			}
\startdata
$\unit{CO}\ J=1-0$ & $115.27$ &  $23.1$ &  $42.3$ &  $19.6$ & $20$\\
$\unit{CO}\ J=2-1$ & $230.54$ & $20.6 \pm 0.4$ & $40.0 \pm 0.4$ & $25.6 \pm 0.4$ & $15$\\
$\unit{CO}\ J=3-2$ & $345.80$&  $15.0 \pm 0.3$ & $28.7 \pm 0.4$ &  $15 \pm 1$ & $15$\\
$\unit{CO}\ J=4-3$ & $461.04$ &  $16 \pm 1$ &  $16.4 \pm 0.2$ &  $8.5 \pm 0.8$ & $12$\\
$\unit{CO}\ J=5-4$ & $576.28$ &  $4.5 \pm 0.4$ &  $6.0 \pm 0.5$ &  $3.4 \pm 0.6$ & $12$\\
$\unit{CO}\ J=6-5$& $691.47$&  $2.5 \pm 0.3$ &  $4.5 \pm 0.3$ &  $3.0 \pm 0.2$ & $12$\\
$\unit{CO}\ J=7-6$& $806.65$&  $0.8 \pm 0.2$ & $1.3 \pm 0.4$  & $1.0 \pm 0.2$ & $12$\\
$\unit{CO}\ J=8-7$& $921.80$&  $0.7 \pm 0.1$ & $0.9 \pm 0.1$  & $0.55 \pm 0.07$ & $12$\\
$\unit{CO}\ J=9-8$\tablenotemark{a} & $1036.91$& $<0.8$ & $<0.9$ & $<1.2$ & $12$\\
$\unit{[CI]}\ J=1-0$& $492.16$&  $2.4 \pm 0.4$ & $5.7 \pm 0.1$ &  $4.0 \pm 0.5$ & $12$\\ 
$\unit{[CI]}\ J=2-1$ & $809.34$&  $1.0 \pm 0.2$ & $1.7 \pm 0.4$ &  $1.3 \pm 0.2$ & $12$ \\
$\unit{[NII]}$\tablenotemark{b}  & $1461.13$ & $2.9 \pm 0.2$ & $2.20 \pm 0.04$ & $1.09 \pm 0.03$ & $12$
\enddata
\tablenotetext{a}{{\bfseries $3\sigma$ upper limit}}
\tablenotetext{b}{Measurements for $\mol{[NII]}$ are from the unconvolved map (beam size $\sim 17''$). All other measurements are at a beam size of $43''$.}
\label{Lineflux}
\end{deluxetable}

%\tablehead{\colhead{Line} &
%	\colhead{Rest frequency} &
%	\multicolumn{2}{c}{NGC 4038} &
%	\multicolumn{2}{c}{Overlap region} & 
%	\multicolumn{2}{c}{NGC 4039} & 
%	\colhead{Calibration error} \\
%	\colhead{} & 
%	\colhead{$\unit{GHz}$} &
%	\colhead{$\unit{Jy \ km/s$}} &
%	\colhead{$\unit{K \ km/s$}} &
%	\colhead{$\unit{Jy \ km/s$}} &
%	\colhead{$\unit{K \ km/s$}} &
%	\colhead{$\unit{Jy \ km/s$}} &
%	\colhead{$\unit{K \ km/s$}} &
%	\colhead{$\%$}			}

\begin{deluxetable}{lcccc} %%TABLE4
\tablecolumns{3}
\tablewidth{0pt}
\tabletypesize{\scriptsize}
\tablecaption{RADEX grid parameters}
\tablehead{\colhead{Parameter} & \colhead{Range} & \colhead{\# of Points}}
\startdata
$T_{kin} \unit{[K]}$ & $10^{0.7} - 10^{3.8}$ & 71\\
$n(\mol{H_2}) \unit{[cm^{-3}]}$ & $10^{1.0}-10^{7.0}$ & 71 \\
$\Phi_A$ & $10^{-5.0}-1$ & 71 \\
$N_{\mol{CO}}/\Delta V \unit{[cm^{-2}]}$ & $10^{12.0}-10^{18.0}$ & 81\\
$N_{\mol{[CI]}}/N_{\mol{CO}}$ & $10^{-2.0} - 10^{2.0}$ & 20 \\
$\Delta V \unit{[km \ s^{-1}]} $ & $1.0$ &
\enddata
\tablecomments{{Column density is calculated per unit linewidth, while the linewidth is held fixed at $1 \unit{km \ s^{-1}}$ in the grid calculations (see text).}}
\label{gridparameter}
\end{deluxetable}

\begin{deluxetable}{lcc} 
\tablecolumns{3}
\tablewidth{0pt}
\tabletypesize{\scriptsize}
\tablecaption{Model Constraints}
\tablehead{\colhead{Parameter} & \colhead{Value} & \colhead{Units}}
\startdata
$\mol{CO}$ abundance ($x_{\mol{CO}}$) &$3\times 10^{-4}$ & ... \\
Mean molecular weight ($\mu$) & $1.5$ & ...\\
Angular size scale &$107$ &$\unit{pc}/''$ \\
Source size & $43$&$''$ \\
Length  ($L$)& $\le 1,930$\tablenotemark{a} & $\unit{pc}$ \\
Dynamical Mass ($M_{dyn}$) & $\le 3.1\times 10^9$\tablenotemark{b} & $M_\odot $
\enddata
\tablenotetext{a}{Physical size of the nucleus of NGC 4038 from \cite{wilson2000} corrected to a distance of $22 \unit{pc}$}
\tablenotetext{b}{Sum of the virial masses of all SGMCs in overlap region from \cite{wilson2003} corrected for incompleteness; see text.}
\label{lparam}
\end{deluxetable}

\begin{deluxetable}{cccccccc}
\tablecolumns{5}
\tabletypesize{\footnotesize}
\tablewidth{0pt}
\tablecaption{$\chi^2$ values of 2 component fit for various $J_{break}$}
\tablehead{
	\colhead{} &
	\colhead{} &
	\multicolumn{2}{c}{NGC 4038} &
	\multicolumn{2}{c}{Overlap region} & 
	\multicolumn{2}{c}{NGC 4039}  \\
	\colhead{Model} &
	\colhead{$J_{break}$} &

	\colhead{$\chi^2_{\unit{CO}}$} &
	\colhead{$\chi^2_{\unit{[CI]}}$} &
	\colhead{$\chi^2_{\unit{CO}}$} &
	\colhead{$\chi^2_{\unit{[CI]}}$} &
	\colhead{$\chi^2_{\unit{CO}}$} &
	\colhead{$\chi^2_{\unit{[CI]}}$}  }
\startdata
$\unit{CO}$ only %& all & $11.36$ & & $4.93$ & & $6.98$\\
& 3 & $31.2$ &$-$ & $27.7$ & $-$& $\mathbf{5.5}$\\
& 4 & $12.2$ & $-$& $\mathbf{10.7}$ & $-$& $13.1$\\
& 5 & $\mathbf{11.9}$ & $-$& $16.0$ & $-$& $15.3$\\
\hline 
$\unit{CO}$ and $\unit{[CI]}$ 
%& all & $39.42$ & $3180$ & $18.63$ & $117.7$ & $18.16$ & $20.21$\\
& 3 & $37.6$ & $663.5$ & $30.5$ & $2431.9$ & $\mathbf{5.4}$ & $969.7$  \\
& 4  & $\mathbf{10.2}$ & $\mathbf{4.8}$ & $\mathbf{22.2}$ & $\mathbf{1.2}$ & $20.4$ & $\mathbf{2.2}$\\
& 5 & $10.4$ & $\mathbf{4.8}$ & $25.5$ & $3.9$ & $10.1$ & $563.6$
\enddata
\tablecomments{The best $\chi^2$ for each position is highlighted in bold}
\label{chisquared}
\end{deluxetable}

%\tablehead{\colhead{Line} &
%	\colhead{Rest frequency} &
%	\multicolumn{2}{c}{NGC 4038} &
%	\multicolumn{2}{c}{Overlap region} & 
%	\multicolumn{2}{c}{NGC 4039} & 
%	\colhead{Calibration error} \\
%	\colhead{} & 
%	\colhead{$\unit{GHz}$} &
%	\colhead{$\unit{Jy \ km/s$}} &
%	\colhead{$\unit{K \ km/s$}} &
%	\colhead{$\unit{Jy \ km/s$}} &
%	\colhead{$\unit{K \ km/s$}} &
%	\colhead{$\unit{Jy \ km/s$}} &
%	\colhead{$\unit{K \ km/s$}} &
%	\colhead{$\%$}			} %%%ChiSquared

\begin{deluxetable}{ccccccc}
\tablecolumns{7}
\tabletypesize{\scriptsize}
\tablewidth{0pt}
\tablecaption{Cold Component Likelihood Results: $\unit{CO}$ only}
\tablehead{\colhead{Source}&\colhead{Parameter}&\colhead{Median} & \colhead{$1\sigma$ Range} & \colhead{1D Max} &\colhead{4D Max} & \colhead{Unit}}
\startdata
NGC 4038 &$T_{kin}$ & $13$ & $10-20$  &$12$ & $14$ & $\unit{(K)}$ \\
&$n(\unit{H_2})$ & $10^{4.97}$ & $10^{3.73}-10^{6.33}$  &$10^{4.00}$ & $10^{5.54}$ & $\unit{(cm^{-3})}$ \\
&$N_{\unit{CO}}$ & $10^{18.95}$ & $10^{18.53}-10^{19.49}$  &$10^{18.80}$ & $10^{18.72}$ & $\unit{(cm^{-2})}$ \\
&$\Phi_A$ & $10^{-1.80}$ & $10^{-2.02}-10^{-1.62}$  &$10^{-1.71}$ & $10^{-1.79}$ & \\
&$P$ & $10^{6.05}$ & $10^{5.01}-10^{7.39}$  &$10^{5.47}$ & $10^{5.47}$ &  $\unit{(K \ cm^{-2})}$ \\
&$\left<N_{\unit{CO}}\right>$ & $10^{17.24}$ & $10^{16.71}-10^{17.76}$  &$10^{17.33}$ & $10^{17.33}$ & $\unit{(cm^{-2})}$ \\
\hline
Overlap region & $T_{kin}$ & $15$ & $10-38$  &$12$ & $63$ & $\unit{(K)}$ \\
&$n(\unit{H_2})$ & $10^{4.16}$ &$10^{3.23}-10^{6.01}$ & $10^{3.40}$ &$10^{3.40}$ & $\unit{(cm^{-3})}$ \\
&$N_{\unit{CO}}$ &  $10^{18.91}$ &$10^{18.35}-10^{19.55}$ & $10^{18.62}$ &$10^{18.70}$ & $\unit{(cm^{-2})}$ \\
&$\Phi_A$ &  $10^{-1.66}$ &$10^{-1.94}-10^{-1.42}$ & $10^{-1.50}$ &$10^{-1.93}$ & \\
&$P$ & $10^{5.37}$ &$10^{4.66}-10^{7.02}$ & $10^{5.22}$ &$10^{5.22}$ & $\unit{(K \ cm^{-2})}$ \\
&$\left<N_{\unit{CO}}\right>$ &  $10^{17.34}$ &$10^{16.72}-10^{17.99}$ & $10^{16.87}$ &$10^{16.87}$ & $\unit{(cm^{-2})}$ \\
\hline
NGC 4039 & $T_{kin}$ & $15$ & $10-37$  &$12$ & $28$ & $\unit{(K)}$ \\
&$n(\unit{H_2})$ & $10^{4.27}$ &$10^{3.33}-10^{6.07}$ & $10^{3.49}$ &$10^{3.83}$ & $\unit{(cm^{-3})}$ \\
&$N_{\unit{CO}}$ &  $10^{18.72}$ &$10^{18.13}-10^{19.48}$ & $10^{18.61}$ &$10^{18.76}$ & $\unit{(cm^{-2})}$ \\
&$\Phi_A$ &  $10^{-2.00}$ &$10^{-2.26}-10^{-1.75}$ & $10^{-1.93}$ &$10^{-2.21}$ & \\
&$P$ & $10^{5.46}$ &$10^{4.82}-10^{7.11}$ & $10^{5.22}$ &$10^{5.22}$ & $\unit{(K \ cm^{-2})}$ \\
&$\left<N_{\unit{CO}}\right>$ &  $10^{16.75}$ &$10^{16.24}-10^{17.55}$ & $10^{16.49}$ &$10^{16.49}$ & $\unit{(cm^{-2})}$ \\
\enddata
\label{cold_43_UA}
\end{deluxetable}%%%%TABLE
\begin{deluxetable}{cccccccccc}
\tablecolumns{10}
\tabletypesize{\scriptsize}
\tablewidth{0pt}
\tablecaption{Warm Component Likelihood Results: $\unit{CO}$ only}
\tablehead{\colhead{Source}&\colhead{Parameter}&\colhead{Median} & \colhead{$1\sigma$ Range} & \colhead{1D Max} &\colhead{4D Max} & \colhead{Unit}}
\startdata
NGC 4038 &$T_{kin}$ & $1071$ & $368-3361$  &$647$ & $5997$ & $\unit{(K)}$ \\
&$n(\unit{H_2})$ & $10^{5.73}$ & $10^{4.75}-10^{6.60}$  &$10^{7.00}$ & $10^{4.26}$ & $\unit{(cm^{-3})}$ \\
&$N_{\unit{CO}}$ & $10^{17.04}$ & $10^{15.30}-10^{18.83}$  &$10^{16.17}$ & $10^{15.05}$ & $\unit{(cm^{-2})}$ \\
&$\Phi_A$ & $10^{-2.55}$ & $10^{-4.27}-10^{-0.81}$  &$10^{-4.57}$ & $10^{-0.71}$ & \\
&$P$ & $10^{9.28}$ & $10^{8.15}-10^{10.21}$  &$10^{9.44}$ & $10^{9.44}$ &  $\unit{(K \ cm^{-2})}$ \\
&$\left<N_{\unit{CO}}\right>$ & $10^{14.47}$ & $10^{14.23}-10^{14.85}$  &$10^{14.43}$ & $10^{14.43}$ & $\unit{(cm^{-2})}$ \\
\hline
Overlap region & $T_{kin}$ & $732$ & $279-2790$  &$389$ & $3999$ & $\unit{(K)}$ \\
&$n(\unit{H_2})$ & $10^{5.57}$ &$10^{4.56}-10^{6.55}$ & $10^{4.43}$ &$10^{4.26}$ & $\unit{(cm^{-3})}$ \\
&$N_{\unit{CO}}$ &  $10^{17.08}$ &$10^{15.32}-10^{18.90}$ & $10^{15.92}$ &$10^{14.80}$ & $\unit{(cm^{-2})}$ \\
&$\Phi_A$ &  $10^{-2.59}$ &$10^{-4.32}-10^{-0.82}$ & $10^{-4.79}$ &$10^{-0.36}$ & \\
&$P$ & $10^{8.92}$ &$10^{7.86}-10^{9.95}$ & $10^{7.90}$ &$10^{7.90}$ & $\unit{(K \ cm^{-2})}$ \\
&$\left<N_{\unit{CO}}\right>$ &  $10^{14.48}$ &$10^{14.27}-10^{14.79}$ & $10^{14.41}$ &$10^{14.41}$ & $\unit{(cm^{-2})}$ \\
\hline
NGC 4039 & $T_{kin}$ & $324$ & $123-1276$  &$212$ & $234$ & $\unit{(K)}$ \\
&$n(\unit{H_2})$ & $10^{4.51}$ &$10^{4.00}-10^{5.22}$ & $10^{4.34}$ &$10^{4.69}$ & $\unit{(cm^{-3})}$ \\
&$N_{\unit{CO}}$ &  $10^{16.34}$ &$10^{15.11}-10^{17.66}$ & $10^{15.68}$ &$10^{16.06}$ & $\unit{(cm^{-2})}$ \\
&$\Phi_A$ &  $10^{-1.86}$ &$10^{-3.16}-10^{-0.63}$ & $10^{-1.07}$ &$10^{-1.43}$ & \\
&$P$ & $10^{7.13}$ &$10^{6.95}-10^{7.49}$ & $10^{7.13}$ &$10^{7.13}$ & $\unit{(K \ cm^{-2})}$ \\
&$\left<N_{\unit{CO}}\right>$ &  $10^{14.49}$ &$10^{14.40}-10^{14.64}$ & $10^{14.61}$ &$10^{14.61}$ & $\unit{(cm^{-2})}$ \\
\enddata
\label{warm_43_UA}
\end{deluxetable}
\begin{deluxetable}{ccccccc}
\tablecolumns{7}
\tabletypesize{\scriptsize}
\tablewidth{0pt}
\tablecaption{Cold Component Likelihood Results: $\unit{CO}$ and $\unit{[CI]}$}
\tablehead{\colhead{Source}&\colhead{Parameter}&\colhead{Median} & \colhead{$1\sigma$ Range} & \colhead{1D Max} &\colhead{4D Max} & \colhead{Unit}}
\startdata
NGC 4038 &$T_{kin}$ & $24$ & $18-33$  &$25$ & $25$ & $\unit{(K)}$ \\
&$n(\unit{H_2})$ & $10^{3.66}$ & $10^{3.08}-10^{4.26}$  &$10^{3.74}$ & $10^{3.57}$ & $\unit{(cm^{-3})}$ \\
&$N_{\unit{CO}}$ & $10^{19.12}$ & $10^{18.58}-10^{19.62}$  &$10^{19.40}$ & $10^{19.40}$ & $\unit{(cm^{-2})}$ \\
&$\Phi_A$ & $10^{-2.09}$ & $10^{-2.21}-10^{-1.96}$  &$10^{-2.07}$ & $10^{-2.07}$ & \\
&$P$ & $10^{5.03}$ & $10^{4.51}-10^{5.57}$  &$10^{5.22}$ & $10^{5.22}$ &  $\unit{(K \ cm^{-2})}$ \\
&$\left<N_{\unit{CO}}\right>$ & $10^{17.09}$ & $10^{16.58}-10^{17.58}$  &$10^{17.62}$ & $10^{17.62}$ & $\unit{(cm^{-2})}$ \\
&$X_{\unit{[CI]}}/X_{\unit{CO}}$ & $10^{-0.65}$ & $10^{-1.11}-10^{-0.16}$  &$10^{-0.95}$ & $10^{-0.95}$ &  \\
&$N_{\unit{[CI]}}$ & $10^{18.50}$ & $10^{18.34}-10^{18.71}$  &$10^{18.65}$ & $10^{18.65}$ & \\
\hline
Overlap region & $T_{kin}$ & $20$ & $15-27$  &$21$ & $21$ & $\unit{(K)}$ \\
&$n(\unit{H_2})$ & $10^{3.67}$ &$10^{2.90}-10^{4.42}$ & $10^{4.00}$ &$10^{3.91}$ & $\unit{(cm^{-3})}$ \\
&$N_{\unit{CO}}$ &  $10^{19.13}$ &$10^{18.44}-10^{19.69}$ & $10^{19.59}$ &$10^{18.84}$ & $\unit{(cm^{-2})}$ \\
&$\Phi_A$ &  $10^{-1.83}$ &$10^{-1.96}-10^{-1.66}$ & $10^{-1.86}$ &$10^{-1.79}$ & \\
&$P$ & $10^{4.91}$ &$10^{4.28}-10^{5.61}$ & $10^{4.32}$ &$10^{4.32}$ & $\unit{(K \ cm^{-2})}$ \\
&$\left<N_{\unit{CO}}\right>$ &  $10^{17.38}$ &$10^{16.76}-10^{17.87}$ & $10^{17.59}$ &$10^{17.59}$ & $\unit{(cm^{-2})}$ \\
&$X_{\unit{[CI]}}/X_{\unit{CO}}$ &  $10^{-0.49}$ &$10^{-0.98}-10^{0.10}$ & $10^{-0.74}$ &$10^{-0.74}$ &  \\
&$N_{\unit{[CI]}}$ &  $10^{18.65}$ &$10^{18.47}-10^{18.86}$ & $10^{18.77}$ &$10^{18.77}$ &  \\
\hline
NGC 4039 & $T_{kin}$ & $21$ & $16-28$  &$21$ & $21$ & $\unit{(K)}$ \\
&$n(\unit{H_2})$ & $10^{3.79}$ &$10^{3.07}-10^{4.44}$ & $10^{4.00}$ &$10^{4.00}$ & $\unit{(cm^{-3})}$ \\
&$N_{\unit{CO}}$ &  $10^{19.01}$ &$10^{18.38}-10^{19.65}$ & $10^{19.65}$ &$10^{18.83}$ & $\unit{(cm^{-2})}$ \\
&$\Phi_A$ &  $10^{-2.21}$ &$10^{-2.36}-10^{-2.03}$ & $10^{-2.21}$ &$10^{-2.14}$ & \\
&$P$ & $10^{5.09}$ &$10^{4.45}-10^{5.66}$ & $10^{5.34}$ &$10^{5.34}$ & $\unit{(K \ cm^{-2})}$ \\
&$\left<N_{\unit{CO}}\right>$ &  $10^{16.86}$ &$10^{16.33}-10^{17.44}$ & $10^{16.49}$ &$10^{16.49}$ & $\unit{(cm^{-2})}$ \\
&$X_{\unit{[CI]}}/X_{\unit{CO}}$ &  $10^{-0.17}$ &$10^{-0.72}-10^{0.36}$ & $10^{-0.53}$ &$10^{-0.53}$ &  \\
&$N_{\unit{[CI]}}$ &  $10^{18.87}$ &$10^{18.63}-10^{19.10}$ & $10^{19.03}$ &$10^{19.03}$ &  \\
\enddata
\label{cold_43_UA_cicold}
\end{deluxetable}%%%%TABLE
\begin{deluxetable}{cccccccccc}
\tablecolumns{10}
\tabletypesize{\scriptsize}
\tablewidth{0pt}
\tablecaption{Warm Component Likelihood Results: $\unit{CO}$ and $\unit{[CI]}$}
\tablehead{\colhead{Source}&\colhead{Parameter}&\colhead{Median} & \colhead{$1\sigma$ Range} & \colhead{1D Max} &\colhead{4D Max} & \colhead{Unit}}
\startdata
NGC 4038 &$T_{kin}$ & $1065$ & $347-3397$  &$647$ & $3999$ & $\unit{(K)}$ \\
&$n(\unit{H_2})$ & $10^{5.75}$ & $10^{4.77}-10^{6.61}$  &$10^{7.00}$ & $10^{4.26}$ & $\unit{(cm^{-3})}$ \\
&$N_{\unit{CO}}$ & $10^{17.04}$ & $10^{15.29}-10^{18.83}$  &$10^{16.10}$ & $10^{14.68}$ & $\unit{(cm^{-2})}$ \\
&$\Phi_A$ & $10^{-2.57}$ & $10^{-4.30}-10^{-0.81}$  &$10^{-4.57}$ & $10^{-0.36}$ & \\
&$P$ & $10^{9.29}$ & $10^{8.17}-10^{10.23}$  &$10^{9.44}$ & $10^{9.44}$ &  $\unit{(K \ cm^{-2})}$ \\
&$\left<N_{\unit{CO}}\right>$ & $10^{14.45}$ & $10^{14.19}-10^{14.83}$  &$10^{14.29}$ & $10^{14.29}$ & $\unit{(cm^{-2})}$ \\
\hline
Overlap region & $T_{kin}$ & $1425$ & $430-3811$  &$2666$ & $4425$ & $\unit{(K)}$ \\
&$n(\unit{H_2})$ & $10^{4.23}$ &$10^{4.00}-10^{4.73}$ & $10^{4.09}$ &$10^{4.00}$ & $\unit{(cm^{-3})}$ \\
&$N_{\unit{CO}}$ &  $10^{16.39}$ &$10^{15.12}-10^{17.80}$ & $10^{14.80}$ &$10^{14.95}$ & $\unit{(cm^{-2})}$ \\
&$\Phi_A$ &  $10^{-1.86}$ &$10^{-3.25}-10^{-0.58}$ & $10^{-0.14}$ &$10^{-0.29}$ & \\
&$P$ & $10^{7.45}$ &$10^{7.26}-10^{7.73}$ & $10^{7.39}$ &$10^{7.39}$ & $\unit{(K \ cm^{-2})}$ \\
&$\left<N_{\unit{CO}}\right>$ &  $10^{14.55}$ &$10^{14.40}-10^{14.70}$ & $10^{14.55}$ &$10^{14.55}$ & $\unit{(cm^{-2})}$ \\
\hline
NGC 4039 & $T_{kin}$ & $629$ & $203-2390$  &$431$ & $431$ & $\unit{(K)}$ \\
&$n(\unit{H_2})$ & $10^{4.35}$ &$10^{3.97}-10^{5.00}$ & $10^{3.91}$ &$10^{4.51}$ & $\unit{(cm^{-3})}$ \\
&$N_{\unit{CO}}$ &  $10^{16.30}$ &$10^{14.99}-10^{17.73}$ & $10^{14.78}$ &$10^{15.15}$ & $\unit{(cm^{-2})}$ \\
&$\Phi_A$ &  $10^{-1.91}$ &$10^{-3.33}-10^{-0.59}$ & $10^{-0.14}$ &$10^{-0.64}$ & \\
&$P$ & $10^{7.27}$ &$10^{7.09}-10^{7.57}$ & $10^{7.26}$ &$10^{7.26}$ & $\unit{(K \ cm^{-2})}$ \\
&$\left<N_{\unit{CO}}\right>$ &  $10^{14.42}$ &$10^{14.28}-10^{14.51}$ & $10^{14.47}$ &$10^{14.47}$ & $\unit{(cm^{-2})}$ \\
\enddata
\label{warm_43_UA_cicold}
\end{deluxetable}

%%%%%%%%%%FIGURES%%%%%%%%%%%%%

%%%%%%%%%%%PHYSICAL PLOTS%%%%%%%%%%%%%

%%%CI LTE stuff

%%%LIKELIHOOD ANALYSIS

%%%CO only results
%%1D contour plots

 \clearpage
 
%CI and CO solutions

\clearpage

%%%%Begin PDR figures%%%%

%\begin{figure}[tp] %%%FIGURE PDR3
%\centering
%\includegraphics[width=\linewidth]{../Figures/PDRplots/CO32FIRRatio.pdf} 
%\caption[]{PDR models for the ratio of the $\mol{CO}$ $J=3-2$ transition to the FIR flux, calculated in units of comparing the ratio of $\mol{CO}$ $J=6-5$ and the FIR flux in units $\mol{erg\ cm^{-2}\ s^{-1}\ sr^{-1}}$. This ratio is calculated for NGC 4038 (\emph{left}), the overlap region (\emph{middle}) and NGC 4039 (\emph{right}). The dashed black lines correspond to the model, while the solid purple line corresponds to the measured ratio. The blue solid lines corresponds to the cold component densities as determine from the $J_{break}=4$ radiative transfer solution for both $\mol{CO}$ and $\mol{[CI]}$ while the blue dashed lines correspond to the $1\sigma$ range to the density for the cold components.}
%\label{CO32FIR}
%\end{figure}
%
%\begin{figure}[tp] %%%FIGURE PDR3
%\centering
%\includegraphics[width=\linewidth]{../Figures/PDRplots/CO65FIRRatio.pdf} 
%\caption[]{Same as Figure \ref{CO32FIR} except for the ratio of $\mol{CO}$ $J=6-5$ and the FIR flux. Here, the red solid lines corresponds to the warm component densities as determine from the $J_{break}=4$ radiative transfer solution for both $\mol{CO}$ and $\mol{[CI]}$ while the red dashed lines correspond to the $1\sigma$ range to the density for the warm components.}
%\label{CO65FIR}
%\end{figure}

%\begin{figure}[ht] %%%Beam Image
%\centering
%\includegraphics[width = \linewidth]{/Users/mschirm/Documents/Grad/MyPapers/NGC4038/Figures/FluxImages/CII/CIITIR_CO10Cont.pdf} 
%\caption[]{Ratio map of $L_{\mol{[CII]}}/L_{\mol{TIR}}$ with the $\mol{CO}$ $J=1-0$ interferometric map from \cite{wilson2003} contours overlaid. The contours correspond to $1\%$, $2.5\%$, $6\%$, $15\%$, $37\%$ and $57 \%$ the peak $\mol{CO}$ flux. Regions where $L_{\mol{[CII]}}/L_{\mol{TIR}} > 0.001$ are consistent with PDR activity (see text). }
%\label{CIITIR_Ratio}
%\end{figure}

\appendix

\end{document}